\documentclass[a4paper,11pt,notitlepage]{report}
\usepackage[english]{babel}
\usepackage[latin1]{inputenc}
\usepackage{amsmath,amssymb}
\usepackage{bbm}
\usepackage{bm}
\usepackage{graphicx}
\usepackage{fancyhdr}
\usepackage[left=3cm,right=3cm,top=3.5cm,bottom=3.5cm,bindingoffset=0.5cm,
twoside]{geometry}
\usepackage[]{hyperref}
\hypersetup{pdftitle={From 3d gravity to the Deformed Special Relativity},pdfauthor={Tomasz Trze\'{s}niewski},pdfkeywords={quantum gravity, 3d gravity, deformed relativistic symmetries}}
\usepackage[all]{hypcap}
\def\lb{\left(}
\def\rb{\right)}
\def\bcp{{\blacktriangleright\!\!\vartriangleleft}}
\def\dcp{{\vartriangleright\!\!\vartriangleleft}}
\def\lsp{{\vartriangleright\!\!<}}
\def\rsp{{>\!\!\vartriangleleft}}
\title{Three-dimensional gravity and deformations of relativistic symmetries}
\author{Tomasz Trze\'{s}niewski}
\date{}
\begin{document}
\thispagestyle{empty}
\begin{center}
\LARGE\textbf{Three-dimensional gravity and deformations of relativistic symmetries\footnote{A thesis for the PhD degree in Physics, defended in October 2015.\\ Supervisor prof. dr hab. Jerzy Kowalski-Glikman.}} \\ 
\vspace{0.75cm}
\large Tomasz Trze\'{s}niewski \\ 
\vspace{0.5cm}
\normalsize{\it Institute for Theoretical Physics, \\ 
University of Wroc\l aw, Wroc\l aw, Poland} \\ 
\vspace{0.25cm}
{\tt e-mail:\ tbwbt@ift.uni.wroc.pl}
\vspace{0.75cm}
\end{center}
\begin{abstract}
\noindent It is possible that relativistic symmetries become deformed in the semiclassical regime of quantum gravity. Mathematically, such deformations lead to the noncommutativity of spacetime geometry and non-vanishing curvature of momentum space. The best studied example is given by the $\kappa$-Poincar\'{e} Hopf algebra, associated with $\kappa$-Minkowski space. On the other hand, the curved momentum space is a well-known feature of particles coupled to three-dimensional gravity. The purpose of this thesis was to explore some properties and mutual relations of the above two models. In particular, I study extensively the spectral dimension of $\kappa$-Minkowski space. I also present an alternative limit of the Chern-Simons theory describing three-dimensional gravity with particles. Then I discuss the spaces of momenta corresponding to conical defects in higher dimensional spacetimes. Finally, I consider the Fock space construction for the quantum theory of particles in three-dimensional gravity.
\end{abstract}

\newpage
\thispagestyle{empty}
~

\tableofcontents

\newpage
\thispagestyle{empty}
~

\chapter*{Preface}
\addcontentsline{toc}{chapter}{Preface}
\markboth{PREFACE}{PREFACE}
The quest for the quantum theory of gravitation faces numerous conceptual and calculational problems, among which we can mention the different roles of time in general relativity and quantum field theory, quantum fluctuations of the background, nonlocality of gravitational observables and nonrenormalizability of the perturbative approach. Nevertheless, the greatest obstacle is actually the almost complete absence of experimental or observational results that would be not explained by classical gravity and standard model and thus could guide us in the construction of quantum gravity models. It is obviously related to the fact that the expected quantum effects of gravity have an extremely small magnitude, proportional to (the power of) the Planck length. The only solution is to look for the phenomena in which these tiny signals could be amplified enough to become measurable by means of either the currently existing or conceivable technology. In particular, such possibilities may be considered in the semiclassical regime in which quantum gravity amounts to some corrections to the structure of classical spacetime. This is usually assumed to result in deformed dispersion relations, which modify the relativistic kinematics of particles. An example of a testable prediction in this context are time delays of photons coming from astrophysical events, e.g.\! gamma-ray bursts, where the sufficient amplification can be achieved due to extremely large distances to the sources. A different possibility is a modification of the threshold energy for the production of pions in the interaction of cosmic rays with the cosmic microwave background radiation, where the ultrahigh energies of the rays give us the advantage. Other examples include anomalies in the production of electron-positron pairs or decay of pions, vacuum Cherenkov radiation, etc.\! (for a recent review see \cite{Amelino:2013qy}). 

Deformed dispersion relations may be a consequence of deformations (or breaking) of relativistic symmetries as well as a nontrivial geometry of momentum space. The idea that quantum gravity should combine general relativistic spacetime with the dynamical, curved momentum space was actually suggested for the first time by M.\! Born \cite{Born:1938ay}. Later it was observed \cite{Snyder:1947qe} that the curvature in momentum space has to be accompanied by the noncommutativity of spacetime coordinates. This intuition was subsequently formalized in the language of quantum groups, i.e.\! nontrivial Hopf algebras, applied to the description of deformations of relativistic symmetries. The best known case is the $\kappa$-deformed Poincar\'{e} algebra, associated with noncommutative $\kappa$-Minkowski space. Such mathematical structures were used \cite{Amelino:2002re,Amelino:2002dy,Magueijo:2002le,Girelli:2005dy} in the family of models called doubly (or deformed) special relativity, whose objective was to represent the semiclassical regime characterized by two invariant scales that are the speed of light and the Planck mass. On the other hand, in the context of more fundamental approaches to quantum gravity it was shown that, e.g.\! \cite{Girelli:2010fs} the $\kappa$-Poincar\'{e} symmetry may arise as a semiclassical symmetry in the formalism of group field theory (which is related to spin foam models). Finally, doubly special relativity was recently recast \cite{Amelino:2011py,Amelino:2011re} as relative locality approach, whose basic principle is that spacetime becomes observer dependent and only the whole phase space is an absolute entity, which leads to the relativity of locality of events in spacetime. There is also an attempt \cite{Freidel:2014be,Freidel:2014qy} to include relative locality in the framework of string theory. 

Meanwhile, there exists a working physical example of the above concepts, which is given by gravity in 2+1 spacetime dimensions. This simpler counterpart of general relativity in 3+1 dimensions has a peculiar feature that momentum space of particles coupled to the (classical) gravitational field is the three-dimensional Lorentz group and thus a curved manifold. Furthermore, a result of the quantization of such particles is that spacetime becomes noncommutative. For this reason three-dimensional gravity attracts interest from the perspective of deformed relativistic symmetries, both as a toy model and a testing ground for different approaches \cite{Batista:2003nu,Freidel:2006py,Noui:2006ty,Joung:2009ty}. In particular, it was argued \cite{Freidel:2006dy} that symmetries of matter fields coupled to three-dimensional quantum gravity in a certain semiclassical limit are described by the $\kappa$-Poincar\'{e} algebra. The essential purpose of this thesis was to explore some properties, consequences and mutual relations of the deformed momentum spaces associated with the $\kappa$-Poincar\'{e} algebra on one side and particles in three-dimensional gravity or analogous objects in higher dimensions (i.e.\! conical defects) on the other. 

We begin Chapter 1 with a general introduction to the formalism of quantum groups and present the structure of the four-dimensional $\kappa$-Poincar\'{e} algebra. Then we discuss the associated noncommutative $\kappa$-Minkowski spacetime and momentum space, which in $n\!+\!1$ dimensions is given by a Lie group ${\rm AN}(n)$. In particular, we show how ${\rm AN}(n)$ can be mapped on a (curved) Lorentzian manifold as well as the corresponding Euclidean manifold, which was obtained by us in \cite{Arzano:2014de}. In Chapter 2, reporting the results from \cite{Arzano:2014de}, we study in the context of $\kappa$-Minkowski space one of the common predictions of the quantum gravity research, which is a scale dependence of the (spectral) dimension of spacetime. Chapter 3 is devoted to point particles coupled to classical gravity in three dimensions. We start from an overall discussion of three-dimensional gravity and describe its point particle solution in terms of geometrical variables. Then we switch to the Chern-Simons formalism and re-derive the effective action of a single particle and a system of multiple particles \cite{Kowalski:2014dy}. In Chapter 4, which contains the results from \cite{Kowalski:2014dy}, we consider an alternative contraction of the de Sitter gauge group in the Chern-Simons theory with point particles and derive a new effective particle action, which may be described as the model of $\kappa$-deformed Carroll particles. Chapter 5, based on our work \cite{Arzano:2015se}, is concerned with conical defects in classical gravity. We discuss massive and massless defects in Minkowski space and explain how their momenta can be characterized by gravitational holonomies. We particularly explore the connection between massless defects and the ${\rm AN}(n)$ groups and we also consider massless conical defects in de Sitter space. In final Chapter 6, using the material of \cite{Arzano:2014by}, we discuss some partial results for the Fock space arising in the quantization of particles coupled to three-dimensional gravity, which is influenced by the deformed symmetry of the theory. 

Everywhere below the speed of light is set as $c = 1$.

\newpage
\thispagestyle{empty}
~

\chapter{$\kappa$-Poincar\'{e} and $\kappa$-Minkowski}
\section{Mathematical preliminaries}
A generalization of ordinary symmetry groups of physical systems are the so-called quantum groups (for a review see \cite{Tjin:1992as,Majid:1995fy}). In order to introduce their structures let us first give a definition of the algebra in the manner that does not explicitly refer to its elements. Namely, an (unital associative) algebra $(A,\nabla,\eta)$ is a vector space $A$ over a field $K$ equipped with the following linear maps $\nabla$ and $\eta$. The multiplication $\nabla: A \otimes A \rightarrow A$ has to satisfy the property of associativity
\begin{align}\label{eq:11.01}
\nabla \circ (\nabla \otimes {\rm id}) = 
\nabla \circ ({\rm id} \otimes \nabla)\,,
\end{align}
where ${\rm id}$ denotes the identity on $A$. In the usual language it translates to $\nabla(a \otimes b) \equiv a \cdot b$ and $(a\cdot b) \cdot c = a \cdot (b \cdot c)$, where $a,b,c \in A$. The map $\eta: K \rightarrow A$, called the unit, is specified by the relation
\begin{align}\label{eq:11.02}
\nabla \circ (\eta \otimes {\rm id}) = 
\nabla \circ ({\rm id} \otimes \eta) = {\rm id}\,.
\end{align}
It means that $\eta(\lambda) \cdot a = a \cdot \eta(\lambda) = \lambda a$ and hence $\eta(\lambda) = \lambda {\bf 1}$, where $\lambda \in K$, $a \in A$ and ${\bf 1}$ is the unit element of $A$. In other words, $\eta$ expresses the presence of the unit element in an algebra. 

In the physical applications we are using specific representations of such abstract algebras. By a representation $(\rho,V)$ of an algebra $A$ we mean a vector space $V$ and linear map $\rho: A \rightarrow {\rm GL}(V)$, where ${\rm GL}(V)$ is the space of linear operators on $V$, which satisfies the homomorphism property
\begin{align}\label{eq:11.03}
\rho(a \cdot b) = \rho(a) \rho(b)\,.
\end{align}
If we want to describe a system composed of two subsystems (e.g.\! two spinning particles) we need to combine two representations into a representation of the tensor product of algebra elements, without violating the definition given above. It turns out that in order to construct such tensor product representations we actually need a generalization of the algebra to a new object, called the bialgebra, which has both the structure of algebra and coalgebra. A coalgebra $(A,\Delta,\epsilon)$ is defined as the vector space equipped with the following linear maps $\Delta$ and $\epsilon$. The comultiplication (or coproduct) $\Delta: A \rightarrow A \otimes A$ is satisfying the property of coassociativity
\begin{align}\label{eq:11.04}
(\Delta \otimes {\rm id}) \circ \Delta = 
({\rm id} \otimes \Delta) \circ \Delta\,.
\end{align}
For $a \in A$ the coproduct can be succinctly written in the form $\Delta(a) = \sum_i a_{(1)}^i \otimes a_{(2)}^i \equiv a_{(1)} \otimes a_{(2)}$, which is known as the Sweedler notation. The map $\epsilon: A \rightarrow K$, called the counit, is specified by the relation
\begin{align}\label{eq:11.05}
(\epsilon \otimes {\rm id}) \circ \Delta = 
({\rm id} \otimes \epsilon) \circ \Delta = 
{\rm id}\,.
\end{align}
Thus the coalgebra can be seen as a dual structure with respect to the algebra. Then $(A,\nabla,\eta,\Delta,\epsilon)$ is a bialgebra if it also satisfies the appropriate compatibility conditions, which may be described as the homomorphism properties. Let us substitute bialgebra elements $a,b \in A$ into these conditions and use the simplified notation for the multiplication and unit. Then the homomorphism conditions have the form
\begin{align}\label{eq:11.06}
\Delta(a \cdot b) = \Delta(a) \Delta(b)\,, \qquad 
\Delta({\bf 1}) = {\bf 1} \otimes {\bf 1}\,, \qquad 
\epsilon(a \cdot b) = \epsilon(a) \epsilon(b)\,, \qquad 
\epsilon({\bf 1}) = 1\,,
\end{align}
where $\Delta(a) \Delta(b) = (a_{(1)} \cdot b_{(1)}) \otimes (a_{(2)} \cdot b_{(2)})$. Using the coalgebra structure we can finally construct a correct tensor product representation of the algebra $A$. For a pair of representations $(\rho_1,V_1)$, $(\rho_2,V_2)$ it is given by $(\rho,V_1 \otimes V_2)$ such that
\begin{align}\label{eq:11.07}
\rho(a) (v_1 \otimes v_2) = (\rho_1 \otimes \rho_2) (\Delta(a)) 
(v_1 \otimes v_2)\,,
\end{align}
where $a \in A$, $v_1 \in V_1$, $v_2 \in V_2$. 

We also observe that if we have a tensor product representation (\ref{eq:11.07}) of a bialgebra $H$ then the representation $(\rho^\prime,V_2 \otimes V_1)$ obtained by the flip of the tensor product $\sigma: a \otimes b \rightarrow b \otimes a$ can be expressed via
\begin{align}\label{eq:11.16}
\rho^\prime(h) = \sigma \circ ((\rho_1 \otimes \rho_2) (\Delta(h))) = 
(\rho_2 \otimes \rho_1) (\Delta^\prime(h))\,,
\end{align}
where $\Delta^\prime \equiv \sigma \circ \Delta$ is called the opposite coproduct. In general $\Delta^\prime(h) \neq \Delta(h)$ and the representations on $V_1 \otimes V_2$ and $V_2 \otimes V_1$ are inequivalent. However, we can find a connection between them if $H$ is a quasitriangular bialgebra, which means that there exists an invertible element $R \in H \otimes H$, $R = \sum_i R_{(1)}^i \otimes R_{(2)}^i \equiv R_{(1)} \otimes R_{(2)}$ such that
\begin{align}\label{eq:11.17}
(\Delta \otimes {\rm id}) R & = R_{13} R_{23}\,, \qquad ({\rm id} \otimes \Delta) R = R_{13} R_{12}\,, \nonumber\\ 
\forall_{h \in H}: \Delta^\prime(h) & = R \Delta(h) R^{-1}\,,
\end{align}
where $R_{12} = R \otimes {\bf 1}$, $R_{13} = R_{(1)} \otimes {\bf 1} \otimes R_{(2)}$, $R_{23} = {\bf 1} \otimes R$. The third formula in (\ref{eq:11.17}) provides an isomorphism between the representations $(\rho,V_1 \otimes V_2)$, $(\rho^\prime,V_2 \otimes V_1)$. Furthermore, it can be easily verified that $R$ satisfies the quantum Yang-Baxter equation
\begin{align}\label{eq:11.18}
R_{12} R_{13} R_{23} = R_{23} R_{13} R_{12}\,.
\end{align}
This equation means the consistency of an exchange of representations in e.g.\! a scattering process of some particles. In this context (a representation of) $R$ is known as the universal $R$-matrix. If $R$ additionally satisfies $R_{12} R_{21} = 1$, where $R_{21} = R_{(2)} \otimes R_{(1)} \otimes {\bf 1}$, then $H$ is called a triangular bialgebra. 

A bialgebra can be further generalized to a Hopf algebra if we endow it with a linear map called the antipode. The antipode $S: A \rightarrow A$ is defined as satisfying the relation
\begin{align}\label{eq:11.08}
\nabla \circ (S \otimes {\rm id}) \circ \Delta = 
\nabla \circ ({\rm id} \otimes S) \circ \Delta = \eta \circ \epsilon\,.
\end{align}
It can be seen as a generalized group inversion and thus a Hopf algebra may be interpreted as a deformed group, where generally only certain linear combinations of elements are invertible. In the physical context there are two important types of the Hopf algebras. Let us first consider a Lie algebra $\mathfrak{g}$. It has the multiplication given by its Lie bracket and the natural unit $\eta(\lambda) = \lambda {\bf 1}$. $\mathfrak{g}$ can be turned into a (trivial) Hopf algebra if we introduce
\begin{align}\label{eq:11.09}
\Delta(x) = x \otimes {\bf 1} + {\bf 1} \otimes x\,, \qquad 
\forall_{x \neq {\bf 1}}: \epsilon(x) = 0\,,\ \epsilon({\bf 1}) = 1\,, \qquad 
S(x) = -x\,,
\end{align}
where $x \in \mathfrak{g}$. Strictly speaking, the maps (\ref{eq:11.09}) have to be extended to the universal enveloping algebra $U(\mathfrak{g})$ of the algebra $\mathfrak{g}$ since Lie algebras generally are not associative and do not have the unit element. The extension can be done in the unique way and then the full $U(\mathfrak{g})$ becomes a Hopf algebra. On the other hand, a Hopf algebra can also be obtained from the space $C(G)$ of continuous functions on a finite group $G$. In this case the multiplication is the pointwise product of functions $(f_1\cdot f_2)(g) = f_1(g) f_2(g)$ and the unit $\eta(\lambda) = \lambda {\bf 1}$ is given by the constant function ${\bf 1}(g) = 1$, $\forall g \in G$. We turn $C(G)$ into a Hopf algebra by defining
\begin{align}\label{eq:11.10}
\Delta(f)(g_1\otimes g_2) = f(g_1 g_2)\,, \qquad \epsilon(f) = f(e)\,, \qquad 
S(f)(g) = f(g^{-1})\,,
\end{align}
where $e$ denotes the unit element of $G$. For a given $G$ and the corresponding $\mathfrak{g}$ the Hopf algebras $C(G)$ and $U(\mathfrak{g})$ can actually be shown \cite{Tjin:1992as} to be (weakly) dual. 

In order to introduce more complex structures we need to know how algebras and coalgebras act on each other. A left action of a (bi)algebra $H$ on algebra $A$ is defined as a map $\alpha_L: H \otimes A \rightarrow A$, $\alpha_L(h \otimes a) \equiv h \triangleright a$ such that
\begin{align}\label{eq:11.11}
(g \cdot h) \triangleright a = g \triangleright (h \triangleright a)\,, \qquad 
{\bf 1}_H \triangleright a = a\,.
\end{align}
Similarly one may introduce a right action $\alpha_R: A \otimes H \rightarrow A$, $\alpha_R(a \otimes h) \equiv a \triangleleft h$. Usually we are interested in the actions which respect the algebraic structure of $A$. Such a covariant action is determined by the coproduct of $H$ and satisfies
\begin{align}\label{eq:11.12}
h \triangleright (a \cdot b) = 
(h_{(1)} \triangleright a) \cdot (h_{(2)} \triangleright b)\,, \qquad 
h \triangleright {\bf 1}_A = \epsilon_H(h)\, {\bf 1}_A\,.
\end{align}
The coalgebraic counterpart of the action is called the coaction. A left coaction of a bi-/coalgebra $A$ on algebra $H$ is a map $\beta_L: H \rightarrow A \otimes H$, $\beta_L(h) = \sum_i (h^{(1)})^i \otimes (h^{(2)})^i \equiv h^{(1)} \otimes h^{(2)}$ (where we again introduced the Sweedler notation and notice that $(h^{(1)})^i \in A$, $(h^{(2)})^i \in H$) such that
\begin{align}\label{eq:11.13}
({\rm id}_A \otimes \beta_L) \circ \beta_L = 
(\Delta_A \otimes {\rm id}_H) \circ \beta_L\,, \qquad 
(\epsilon_A \otimes {\rm id}_H) \circ \beta_L = {\rm id}_H\,.
\end{align}
A coaction which respects the algebraic structure of $H$ has to satisfy
\begin{align}\label{eq:11.14}
\beta_L(h \cdot g) = \beta_L(h) \beta_L(g)\,, \qquad \beta_L({\bf 1}_H) = {\bf 1}_A \otimes {\bf 1}_H\,,
\end{align}
where $\beta_L(h) \beta_L(g) = (h^{(1)} \cdot g^{(1)}) \otimes (h^{(2)} \cdot g^{(2)})$. A right coaction may be defined analogously. It is easy to see that the natural action of an algebra on itself is given by the multiplication and the natural coaction inside a coalgebra by the comultiplication. 

Let us consider a pair of Hopf algebras $H$, $A$, with a right action of $H$ on $A$ and a left coaction of $A$ on $H$ (or a right coaction and left action). We can combine them into a particular type of the Hopf algebra which is useful in the description of deformed relativistic symmetries. Namely, a (right-left) bicrossproduct Hopf algebra $A \bcp H$ (where the black triangle denotes the coaction) is the tensor product algebra $H \otimes A$ equipped with
\begin{flalign}\label{eq:11.15}
& (h \otimes a) \cdot (g \otimes b) = 
h g_{(1)} \otimes (a \triangleleft g_{(2)}) b\,, \quad 
{\bf 1} = {\bf 1}_H \otimes {\bf 1}_A\,, \nonumber\\ 
& \Delta(h \otimes a) = \lb h_{(1)} \otimes (h_{(2)})^{(1)} a_{(1)}\rb \otimes 
\lb(h_{(2)})^{(2)} \otimes a_{(2)}\rb \,, \quad 
\epsilon(h \otimes a) = \epsilon_H(h) \epsilon_A(a)\,, \nonumber\\ 
& S(h \otimes a) = \lb{\bf 1}_H \otimes S_A(h^{(1)} a)\rb \cdot 
\lb S_H(h^{(2)}) \otimes {\bf 1}_A\rb\,,
\end{flalign}
assuming that certain additional compatibility conditions are satisfied \cite{Majid:1995fy}. 

On the other hand, if we have a Hopf algebra $H$ and the dual Hopf algebra with the opposite coproduct $H^{*o}$ (cf.\! (\ref{eq:11.16})) then we can combine them into a type of the Hopf algebra that is particularly useful in the context of three-dimensional gravity. Namely, the quantum double of $H$ is a quasitriangular Hopf algebra given by the tensor product ${\cal D}(H) = H \otimes H^{*o}$ equipped with
\begin{flalign}\label{eq:11.19}
& (h \otimes e) \cdot (g \otimes f) = h g_{(2)} \otimes f e_{(2)} \left<g_{(1)}, e_{(1)}\right> \left<S(g_{(3)}), e_{(3)}\right>\,, \qquad 
{\bf 1} = {\bf 1} \otimes {\bf 1}\,, \nonumber\\ 
& \Delta(h \otimes e) = \lb h_{(1)} \otimes e_{(1)}\rb \otimes \
\lb h_{(2)} \otimes e_{(2)}\rb\,, \qquad 
\epsilon(h \otimes e) = \epsilon(h) \epsilon(e)\,, \nonumber\\ 
& S(h \otimes e) = \lb{\bf 1} \otimes S^{-1}(e)\rb \cdot 
\lb S(h) \otimes {\bf 1}\rb\,,
\end{flalign}
where $h_{(1)} \otimes h_{(2)} \otimes h_{(3)} \equiv (h_{(1)})_{(1)} \otimes (h_{(1)})_{(2)} \otimes h_{(2)} = h_{(1)} \otimes (h_{(2)})_{(1)} \otimes (h_{(2)})_{(2)}$ and $\left<.,.\right>$ denotes the duality pairing. Choosing a basis $\{e_\mu\}$ of $H$ and a dual basis $\{f^\mu\}$ of $H^{*o}$ we can write the universal $R$-matrix of ${\cal D}(H)$ as
\begin{align}\label{eq:11.19a}
R = \lb e_\mu \otimes {\bf 1}\rb \otimes \lb{\bf 1} \otimes f^\mu \rb\,.
\end{align}
In particular, for the Hopf algebra (\ref{eq:11.10}) the dual is the so-called group algebra $\mathbbm{C}[G]$ and the quantum double is ${\cal D}(G) \equiv {\cal D}(C(G)) = C(G) \otimes \mathbbm{C}[G]$. 

We can now explain that, briefly speaking, a quantum group (or algebra) is a deformation of a trivial Hopf algebra of the type either (\ref{eq:11.09}) or (\ref{eq:11.10}). In the approach using universal enveloping algebras (\ref{eq:11.09}) the deformation is controlled by a parameter $q$, which is usually expressed as the exponential of the Planck constant and thus provides a transition between the classical and quantum regimes. A simple example of such a quantized algebra is $U_q(\mathfrak{sl}(2))$, which is a deformation of $U(\mathfrak{sl}(2))$ and in terms of its generators $J_0,J_+,J_-$ is given by
\begin{flalign}\label{eq:11.20}
& [J_0,J_\pm] = \pm 2J_\pm\,, \qquad [J_+,J_-] = 
\frac{q^{2J_0} - q^{-2J_0}}{q - q^{-1}} \nonumber\\ 
& \Delta(J_0) = J_0 \otimes {\bf 1} + {\bf 1} \otimes J_0\,, \qquad 
\Delta(J_\pm) = J_\pm \otimes q^{J_0} + q^{-J_0} \otimes J_\pm\,, \nonumber\\ 
& \epsilon(J_0) = \epsilon(J_\pm) = 0 \,, \qquad S(J_0) = -J_0\,, \qquad 
S(J_\pm) = -q^{\pm 1} J_\pm
\end{flalign}
(where $q^{J_0}$ is understood as a series in $J_0$). The Hopf algebra $U(\mathfrak{sl}(2))$ can be recovered by taking the classical limit $q \rightarrow 1$. Let us also remark that $U_q(\mathfrak{sl}(2))$ is a quasitriangular Hopf algebra, with the universal $R$-matrix
\begin{align}\label{eq:11.21}
R = q^{1/2\, J_0 \otimes J_0} \sum_{n=0}^{\infty} \frac{((1 - q^{-2})\, 
q^{J_0/2} J_+ \otimes q^{-J_0/2} J_-)^n}{[n]_{q^{-2}}!}\,,
\end{align}
where $[n]_{q}! \equiv \Pi_{m=1}^{n} \frac{1 - q^m}{1 - q}$.

\section{$\kappa$-Poincar\'{e} algebra}
The $\kappa$-Poincar\'{e} Hopf algebra was first obtained \cite{Lukierski:1991qa,Lukierski:1992ny} as the contraction of the quantum anti-de Sitter algebra $U_q(\mathfrak{so}(3,2))$ such that one takes the limit of the de Sitter radius $R \rightarrow \infty$ and the (real) deformation parameter $q \rightarrow 1$ but keeps fixed the ratio $R\, \log q = \kappa^{-1}$, $\kappa > 0$. The new deformation parameter $\kappa$ has the dimension of inverse length and thus may be given by the Planck length. The structure of the $\kappa$-Poincar\'{e} algebra can be presented in different ways, depending on the choice of a basis of algebra generators. In particular, we may put it in the form \cite{Majid:1994by} of the bicrossproduct Hopf algebra $U(\mathfrak{so}(3,1)) \bcp {\cal T}$, where ${\cal T}$ is the deformed enveloping algebra of translations. 

In the bicrossproduct basis, where $M_a$, $N_a$, $K_\mu$, $a = 1,2,3$, $\mu = 0,1,2,3$ are, respectively, the generators of rotations, boosts and translations, the Lorentz subalgebra and some of the commutators involving translations are undeformed
\begin{align}\label{eq:12.01}
[M_a,M_b] & = i \epsilon_{abc} M^c\,, & 
[M_a,N_b] & = i \epsilon_{abc} N^c\,, & 
[N_a,N_b] & = -i \epsilon_{abc} M^c\,, \nonumber\\ 
[M_a,K_0] & = 0\,, & [M_a,K_b] & = i \epsilon_{abc} K^c\,, & 
[K_\mu,K_\nu] & = 0
\end{align}
and the deformation of the ordinary Poincar\'{e} algebra occurs only between translations and boosts, to wit
\begin{align}\label{eq:12.02}
[N_a,K_0] & = i K_a\,, \nonumber\\ 
[N_a,K_b] & = i\delta_{ab} \lb\frac{\kappa}{2} \lb 1 - e^{-2 K_0/\kappa}\rb + \frac{1}{2\kappa} K_c K^c\rb - \frac{i}{\kappa} K_a K_b\,.
\end{align}
The Lorentz sector of the coalgebra has trivial coproducts and antipodes for rotations and deformed ones for boosts
\begin{align}\label{eq:12.03}
\Delta M_a & = M_a \otimes {\bf 1} + {\bf 1} \otimes M_a\,, \qquad 
S(M_a) = -M_a\,, \nonumber\\ 
\Delta N_a & = N_a \otimes {\bf 1} + e^{-K_0/\kappa} \otimes N_a + 
\frac{1}{\kappa} \epsilon_{abc} K^b \otimes M^c\,, \nonumber\\ 
S(N_a) & = -e^{K_0/\kappa} N_a + 
\frac{1}{\kappa} \epsilon_{abc} e^{K_0/\kappa} K^b M^c\,,
\end{align}
while for translations we have
\begin{align}\label{eq:12.04}
\Delta K_0 = K_0 \otimes {\bf 1} + {\bf 1} \otimes K_0\,, \qquad 
\Delta K_a = K_a \otimes {\bf 1} + e^{-K_0/\kappa} \otimes K_a
\end{align}
and
\begin{align}\label{eq:12.05}
S(K_0) = -K_0\,, \qquad S(K_a) = -e^{K_0/\kappa} K_a\,.
\end{align}
Finally, all counits are trivially equal to $0$. In the classical limit $\kappa \rightarrow \infty$ we recover the Poincar\'{e} algebra together with the trivial coalgebra. 

Let us note that the mass Casimir element of the above algebra (which commutes with all the generators) has the form \cite{Lukierski:1992ny}
\begin{align}\label{eq:12.06}
C_1(K_0,\{K_a\}) = 4\kappa^2 \sinh^2\frac{K_0}{2\kappa} - e^{K_0/\kappa} K_aK^a
\end{align}
and in the limit $\kappa \rightarrow \infty$ it becomes the ordinary expression $C_0(K_0,\{K_a\}) = K_0^2 - K_aK^a$. On the other hand, any function of (\ref{eq:12.06}) with the correct classical limit is also a Casimir element, e.g.\! the function
\begin{align}\label{eq:12.07}
C_0(K_0,\{K_a\}) = C_1(K_0,\{K_a\}) \lb 1 + 
\frac{1}{4\kappa^2}\, C_1(K_0,\{K_a\})\rb\,.
\end{align}
This ambiguity will be relevant in the next Chapter. 

Furthermore, the Euclidean counterpart of the $\kappa$-Poincar\'{e} algebra can be obtained \cite{Lukierski:1994qt} using the transformation $\kappa \mapsto i \kappa$, $K_0 \mapsto i K_0$, $N_a \mapsto i N_a$, which converts the Lorentz algebra into the Euclidean algebra, while the deformed commutators (\ref{eq:12.02}) become
\begin{align}\label{eq:12.08}
[N_a,K_0] & = -i K_a\,, \nonumber\\ 
[N_a,K_b] & = i\delta_{ab} \lb\frac{\kappa}{2} \lb 1 - e^{-2 K_0/\kappa}\rb - 
\frac{1}{2\kappa} K_c K^c\rb + \frac{i}{\kappa} K_a K_b\,,
\end{align}
the coproducts and antipodes (\ref{eq:12.03}) become
\begin{align}\label{eq:12.09}
\Delta N_a & = N_a \otimes {\bf 1} + e^{-K_0/\kappa} \otimes N_a - 
\frac{1}{\kappa} \epsilon_{abc} K^b \otimes M^c\,, \nonumber\\ 
S(N_a) & = -e^{K_0/\kappa} N_a - 
\frac{1}{\kappa} \epsilon_{abc} e^{K_0/\kappa} K^b M^c
\end{align}
and the other Hopf algebraic structures remain unchanged. 

\section{$\kappa$-Minkowski phase space}
The Hopf algebra presented in the previous Section can be straightforwardly generalized \cite{Lukierski:1994qs} to any number of spacetime dimensions. Meanwhile, the dual of the subalgebra of translations of the ($n\!+\!1$-dimensional) $\kappa$-Poincar\'{e} algebra is naturally interpreted as the algebra of spacetime coordinates. In the bicrossproduct basis it can be shown \cite{Majid:1994by} that the latter is covariant under the action of the full algebra. Such a spacetime is $n\!+\!1$-dimensional noncommutative $\kappa$-Minkowski space, whose time $X_0$ and spatial coordinates $X_a$, $a = 1,...,n$ satisfy the commutation relations
\begin{align}\label{eq:13.01}
[X_0,X_a] = \frac{i}{\kappa}\, X_a\,, \qquad [X_a,X_b] = 0\,.
\end{align}
As a vector space, $\kappa$-Minkowski space is isomorphic to the ordinary commutative Minkowski space in $n\!+\!1$ dimensions, which we recover in the classical limit $\kappa \rightarrow \infty$. The Lie algebra spanned by $X_0,X_a$ is often denoted $\mathfrak{an}(n)$ since it has $n$ Abelian and nilpotent generators $X_a$. 

The $\mathfrak{an}(n)$ algebra generates the Lie group ${\rm AN}(n)$, whose elements can be written as the ordered exponentials of algebra elements, with a given ordering being equivalent to the choice of coordinates on the group. In particular, in the time-to-the-right ordering a group element has the form
\begin{align}\label{eq:13.02}
g = e^{-i k^a X_a} e^{i k_0 X_0}
\end{align}
and the (bicrossproduct) coordinates $k_0,k_a \in \mathbbm{R}$ are associated with the basis (\ref{eq:12.01})-(\ref{eq:12.05}) of the $\kappa$-Poincar\'{e} algebra. If we treat such exponentials as plane waves on $n\!+\!1$-dimensional $\kappa$-Minkowski space then ${\rm AN}(n)$ can be interpreted as the corresponding momentum space, with the structure of the algebra of translations and related with spacetime via a noncommutative Fourier transform \cite{Freidel:2006fe}. The product of two group elements $g = e^{-i k^a X_a} e^{i k_0 X_0}$, $h = e^{-i l^a X_a} e^{i l_0 X_0}$ is given by
\begin{align}\label{eq:13.03}
g\, h = e^{-i (k^a \oplus l^a) X_a} e^{i (k_0 \oplus l_0) X_0}\,,
\end{align}
where the non-Abelian addition of coordinates
\begin{align}\label{eq:13.04}
k_0 \oplus l_0 = k_0 + l_0\,, \qquad k^a \oplus l^a = k^a + e^{-k_0/\kappa} l^a
\end{align}
is determined by the coproduct (\ref{eq:12.04}). Meanwhile, the inverse of a group element $g = e^{-i k^a X_a} e^{i k_0 X_0}$ is
\begin{align}\label{eq:13.05}
g^{-1} = e^{-i (\ominus k^a) X_a} e^{i (\ominus k_0) X_0}\,,
\end{align}
with the deformed reflection of coordinates
\begin{align}\label{eq:13.06}
\ominus k_0 = -k_0\,, \qquad \ominus k^a = -e^{k_0/\kappa} k^a
\end{align}
resulting from the antipode (\ref{eq:12.05}). 

The ${\rm AN}(n)$ group is isomorphic to a subgroup of the Lorentz group ${\rm SO}(n+1,1)$. In particular, generators of the $\mathfrak{an}(n)$ algebra may be given by
\begin{align}\label{eq:13.07}
X_0 = \frac{1}{\kappa}\, J_{0,n+1}\,, \qquad 
X_a = \frac{1}{\kappa} \lb J_{0,a} + J_{n+1,a}\rb\,,
\end{align}
where $J_{0,n+1}$, $J_{0,a}$ are the boost generators and $J_{n+1,a}$ is the generator of rotations (with the appropriate signs). Let us remark that Lorentz transformations generated by such homogenous combinations of rotations and boosts as $J_{0,a} + J_{n+1,a}$ are called null rotations, or parabolic Lorentz transformations, and their orbits in $n\!+\!2$-dimensional Minkowski space are parabolae lying in the null planes. The algebra (\ref{eq:13.07}) may be written in the $(n+2) \times (n+2)$ matrix representation
\begin{align}\label{eq:13.08}
X_0 = -\frac{i}{\kappa} \lb
\begin{array}{ccc}
0\! & {\bf 0}\! & 1 \\ 
{\bf 0}\! & {\bf 0}\! & {\bf 0} \\ 
1\! & {\bf 0}\! & 0 
\end{array}
\rb\,, \qquad 
X_a = \frac{i}{\kappa} \lb
\begin{array}{ccc}
0\! & {\bf e}_a^{\rm T}\! & 0 \\ 
{\bf e}_a\! & {\bf 0}\! & {\bf e}_a \\ 
0\! & -{\bf e}_a^{\rm T}\! & 0 
\end{array}
\rb\,,
\end{align}
where ${\bf e}_a$ is the $n$-vector $(0,\ldots,1,\ldots,0)$ with $1$ at the $a$'th position. In such a representation nilpotent generators satisfy the relation $X_a^3 = 0$ (independently of $n$). For the chosen ordering (\ref{eq:13.02}) an element of ${\rm AN}(n)$ is represented by
\begin{align}\label{eq:13.09}
g = \lb
\renewcommand{\arraystretch}{1.5}\begin{array}{ccc}
\cosh\frac{k_0}{\kappa} + e^{k_0/\kappa} \frac{k_ak^a}{2\kappa^2}\! 
& \tfrac{1}{\kappa} {\bf k}^{\rm T}\! & \sinh\frac{k_0}{\kappa} + 
e^{k_0/\kappa} \frac{k_ak^a}{2\kappa^2} \\ 
e^{k_0/\kappa} \tfrac{1}{\kappa} {\bf k}\! & \mathbbm{1}\! & 
e^{k_0/\kappa} \tfrac{1}{\kappa} {\bf k} \\ 
\sinh\frac{k_0}{\kappa} - e^{k_0/\kappa} \frac{k_ak^a}{2\kappa^2}\! & 
-\tfrac{1}{\kappa} {\bf k}^{\rm T}\! & \cosh\frac{k_0}{\kappa} - 
e^{k_0/\kappa} \frac{k_ak^a}{2\kappa^2} 
\end{array}
\rb\,,
\end{align}
where $\mathbbm{1}$ denotes $n \times n$ identity matrix and ${\bf k}$ is the $n$-vector $(k_1,\ldots,k_n)$. 

Since $g$ may be treated as an element of the Lorentz group it has the natural action on vectors in $n\!+\!2$-dimensional Minkowski space that is the matrix multiplication. Therefore let us choose a single spacelike vector $v_L = (0,\ldots,0,\kappa)$ and define the map $m_L: g \rightarrow g \cdot v_L$, $\forall g \in {\rm AN}(n)$ from the ${\rm AN}(n)$ group to Minkowski space. Then $m_L$ determines embedding coordinates $(p_0,\{p_a\},p_{-1}) \equiv g \cdot v_L$ on the ${\rm AN}(n)$ group, which are given by \cite{Kowalski:2002dy,Kowalski:2003de}
\begin{align}\label{eq:13.10}
p_0 & = \kappa \sinh\frac{k_0}{\kappa} + 
\frac{1}{2\kappa}\, e^{k_0/\kappa} k_ak^a\,, \nonumber\\ 
p_a & = e^{k_0/\kappa} k_a\,, \nonumber\\ 
p_{-1} & = \kappa \cosh\frac{k_0}{\kappa} - 
\frac{1}{2\kappa}\, e^{k_0/\kappa} k_ak^a\,.
\end{align}
We observe that such coordinates, as the functions of $k_0,k_a$, satisfy two conditions $-p_0^2 + p_a p^a + p_{-1}^2 = \kappa^2$ and $p_0 + p_{-1} > 0$. The first of them defines a $(n,1)$-hyperboloid, which is equivalent to the well-known embedding of $n\!+\!1$-dimensional de Sitter space in $n\!+\!2$-dimensional Minkowski space. The other condition restricts the coordinates to a half of the hyperboloid with the boundary $p_0 + p_{-1} = 0$, $p_a p^a = \kappa^2$. Thus half of de Sitter space is the (Lorentzian) manifold of the ${\rm AN}(n)$ group, i.e.\! the $n\!+\!1$-dimensional $\kappa$-Minkowski momentum space. It can also be seen as elliptic de Sitter space, which is de Sitter space divided by the equivalence relation $p_{0,a,-1} = -p_{0,a,-1}$. However, since there are $n+2$ group coordinates (\ref{eq:13.10}) one of them is superfluous. In the classical limit $\kappa \rightarrow \infty$ they flatten to bicrossproduct coordinates, $p_0 \rightarrow k_0$, $p_a \rightarrow k_a$, with the exception of diverging $p_{-1} \rightarrow \infty$. Therefore it is $p_{-1}$ that is the auxiliary coordinate, constrained by the hyperboloid condition. Let us note that one can also map ${\rm AN}(n)$ to the other half of de Sitter space if the action of a group element $g$ in the map $m_L$ is given by $(g \cdot {\cal N}) \cdot v_L$ instead of $g \cdot v_L$, where the matrix
\begin{align}\label{eq:13.11}
{\cal N} = \lb
\begin{array}{ccc}
-1\! & {\bf 0}\! & 0 \\ 
{\bf 0}\! & {\mathbbm 1}\! & {\bf 0} \\ 
0\! & {\bf 0}\! & -1 
\end{array}
\rb\,.
\end{align}
The existence of the above two maps from the ${\rm AN}(n)$ group to the $(n,1)$-hyperboloid is related to the (local) Iwasawa decomposition of the Lorentz group ${\rm SO}(n+1,1) = {\rm AN}(n)\, {\rm SO}(n,1) \cup {\rm AN}(n)\, {\cal N}\, {\rm SO}(n,1)$, see \cite{Vilenkin:1992rs}, and the quotient ${\rm SO}(n+1,1) / {\rm SO}(n,1)$ being equivalent to $n\!+\!1$-dimensional de Sitter space. 

One may suppose that the Euclidean version of the ${\rm AN}(n)$ manifold could be obtained in a similar manner to the above Lorentzian case. Indeed, following the suggestion made in \cite{Kowalski:2013le}, as we did in \cite{Arzano:2014de}, let us now take a timelike vector $v_E = (\kappa,0,\ldots,0)$ and define the map $m_E: g \rightarrow g \cdot v_E$, $\forall g \in {\rm AN}(n)$. It introduces a different set of embedding coordinates $(p_{-1},\{p_a\},p_0) \equiv g \cdot v_E$ on the ${\rm AN}(n)$ group, which are given by
\begin{align}\label{eq:13.12}
p_0 & = \kappa \sinh\frac{k_0}{\kappa} - 
\frac{1}{2\kappa}\, e^{k_0/\kappa} k_ak^a\,, \nonumber\\ 
p_a & = e^{k_0/\kappa} k_a\,, \nonumber\\ 
p_{-1} & = \kappa \cosh\frac{k_0}{\kappa} + 
\frac{1}{2\kappa}\, e^{k_0/\kappa} k_ak^a\,.
\end{align}
The coordinates again satisfy two conditions $p_0^2 + p_a p^a - p_{-1}^2 = -\kappa^2$ and $p_{-1} > 0$. The first one defines a $(1,n)$-hyperboloid, which has two sheets, corresponding to the embedding of two copies of $n\!+\!1$-dimensional Euclidean anti-de Sitter space in $n\!+\!2$-dimensional Minkowski space.\footnote{The Euclidean counterpart of anti-de Sitter space is given by the hyperbolic space in the same sense as the sphere is the Euclidean counterpart of de Sitter space.} The other condition restricts the coordinates to just one copy of Euclidean anti-de Sitter space, i.e.\! the sheet of the hyperboloid with $p_{-1} = \sqrt{p_0^2 + p_a p^a + \kappa^2} > \kappa$. We observe that $p_{-1}$ has to be the auxiliary coordinate and this is confirmed by taking the infrared limit of (\ref{eq:13.12}), which gives $p_0 \rightarrow k_0$, $p_a \rightarrow k_a$ but $p_{-1} \rightarrow \infty$. This also explains why we denoted $p_0$ and $p_{-1}$ in a reverse order than in the Lorentzian case. Thus Euclidean anti-de Sitter space can be regarded as the Euclidean manifold of the ${\rm AN}(n)$ group, or the $n\!+\!1$-dimensional $\kappa$-Minkowski momentum space. Finally, similarly as in the Lorentzian case, to map ${\rm AN}(n)$ to the other sheet of the $(1,n)$-hyperboloid we may use the map $m_E$ with the modified group action $(g \cdot {\cal N}) \cdot v_E$. The above two maps from the ${\rm AN}(n)$ group to the hyperboloid correspond to another Iwasawa decomposition of the Lorentz group ${\rm SO}(n+1,1) = {\rm AN}(n)\, {\rm SO}(n+1) \cup {\rm AN}(n)\, {\cal N}\, {\rm SO}(n+1)$ \cite{Vilenkin:1992rs}, while the quotient ${\rm SO}(n+1,1) / {\rm SO}(n+1)$ is equivalent to the $(1,n)$-hyperboloid. 

The Lorentzian and Euclidean manifolds of ${\rm AN}(n)$ should be connected by means of an appropriate transformation similar to the Wick rotation. In the previous Section we mentioned the associated transformation which allows to obtain the Euclidean version of the $\kappa$-Poincar\'{e} algebra (\ref{eq:12.08}), (\ref{eq:12.09}). For the ${\rm AN}(n)$ group, in bicrossproduct coordinates it corresponds to
\begin{align}\label{eq:13.13}
k_0 \mapsto i k_0\,, \qquad \kappa \mapsto i \kappa\,,
\end{align}
while in embedding coordinates the analogous transformation has the form
\begin{align}\label{eq:13.14}
p_0 \mapsto i p_0\,, \qquad p_{-1} \mapsto i p_{-1}\,, \qquad 
\kappa \mapsto i \kappa\,.
\end{align}
Applying both (\ref{eq:13.13}) and (\ref{eq:13.14}) to the expressions for Lorentzian coordinates (\ref{eq:13.10}) we obtain their Euclidean counterparts (\ref{eq:13.12}). 

To conclude, let us also write down the coproducts and antipodes of the translation generators of the $\kappa$-Poincar\'{e} algebra in the so-called classical basis, which corresponds to momentum coordinates (\ref{eq:13.10}). In the Lorentzian case they have the form \cite{Kowalski:2002da}
\begin{align}\label{eq:13.15}
\Delta P_0 & = P_0 \otimes \frac{1}{\kappa} (P_0 + P_{-1}) + \kappa (P_0 + P_{-1})^{-1} \otimes P_0 + (P_0 + P_{-1})^{-1} P_a \otimes P^a\,, \nonumber\\ 
\Delta P_a & = P_a \otimes \frac{1}{\kappa} (P_0 + P_{-1}) + 
{\bf 1} \otimes P_a\,, \nonumber\\ 
S(P_0) & = -P_0 + (P_0 + P_{-1})^{-1} P_a P^a\,, \qquad 
S(P_a) = -\kappa (P_0 + P_{-1})^{-1} P_a\,,
\end{align}
where $P_{-1} = (\kappa^2 + P_0^2 - P_a P^a)^{\frac{1}{2}}$, while in the Euclidean case we found \cite{Arzano:2014de}
\begin{align}\label{eq:13.16}
\Delta P_0 & = P_0 \otimes \frac{1}{\kappa} (P_0 + P_{-1}) + \kappa (P_0 + P_{-1})^{-1} \otimes P_0 - (P_0 + P_{-1})^{-1} P_a \otimes P^a\,, \nonumber\\ 
\Delta P_a & = P_a \otimes \frac{1}{\kappa} (P_0 + P_{-1}) + 
{\bf 1} \otimes P_a\,, \nonumber\\ 
S(P_0) & = -P_0 - (P_0 + P_{-1})^{-1} P_a P^a\,, \qquad 
S(P_a) = -\kappa (P_0 + P_{-1})^{-1} P_a\,,
\end{align}
where $P_{-1} = (\kappa^2 + P_0^2 + P_a P^a)^{\frac{1}{2}}$. Thus, in contrast to the expressions in the bicrossproduct basis (\ref{eq:12.04}), (\ref{eq:12.05}), there is a difference between the Lorentzian and Euclidean case.

\newpage
\thispagestyle{empty}
~

\chapter{Dimensional flow of $\kappa$-Minkowski space}
One of the basic assumptions of the quantum gravity research is that in the ultraviolet regime of the theory spacetime itself should become quantized. Trying to understand what does it mean we should take into account the following two issues. Namely, whether the relativistic symmetries will be violated and the number of spacetime dimensions will change. The latter suggestion is probably more difficult to explain. However, it is closely related to the issue of the perturbative nonrenormalizability of general relativity. Spacetime dimension contributes to the degree of divergence of terms of the perturbative expansion and running to the appropriate value in the ultraviolet regime it could make the expansion finite. Furthermore, the dimensional flow may be a symptom of the specific small-distance structure of spacetime. To explore this scale dependence of dimension we need a different notion than the standard topological dimension. One of the more sensitive definitions is the spectral dimension, which we will introduce below. It has been extensively studied for quantum spacetimes and the flow of the spectral dimension was indeed discovered in a variety of approaches to the quantization of gravity, including causal dynamical triangulations \cite{Ambjorn:2005tt}, Ho\v{r}ava-Lifshitz gravity \cite{Horava:2009st}, asymptotic safety scenario \cite{Lauscher:2005fy} and loop quantum gravity and spin foam models \cite{Modesto:2009fm}. The most typical pattern is of the dimension running from the topological dimension, i.e.\! 4 dimensions in the infrared limit to 2 dimensions in the ultraviolet limit. Analogous results in most of the above models were also obtained in the context of three-dimensional quantum gravity \cite{Horava:2009st,Lauscher:2005fy,Benedetti:2009se,Caravelli:2009fs}. 

The dimensional flow is often easier to explore than the potential violations of relativistic symmetries. Meanwhile, it is not difficult to observe that there is some mutual relation between both phenomena. At first sight it may even seem that a change in the number of dimensions implies the breakdown of relativistic invariance, which is the actual situation in Ho\v{r}ava-Lifshitz gravity. However, the dimensional flow may also arise when the symmetries are not broken but only deformed \cite{Amelino:2014pe}. On the other hand, from a given scale dependence of the spectral dimension one can in principle reconstruct the corresponding modified dispersion relation in momentum space \cite{Sotiriou:2011fn}. Such dispersion relations can be either associated with a breaking or deformations of relativistic symmetries. From this perspective it is worth to calculate the spectral dimension of $\kappa$-Minkowski space, which was first done in \cite{Benedetti:2009fe} and then improved by us in \cite{Arzano:2014de}, as will discuss below. 

The spectral dimension is defined for the Riemannian manifolds but can also be applied to spacetime if we use the Euclideanized version of the latter. It is introduced in the following way. On a given manifold $M$ of $d$ topological dimensions and with a metric $h$ we consider a fictitious diffusion process that is governed by the heat equation with a Laplacian $\Delta$ and auxiliary time parameter $\sigma$
\begin{align}\label{eq:20.01}
\frac{\partial}{\partial \sigma} K(x,x_0; \sigma) = \Delta K(x,x_0; \sigma)\,, 
\quad K(x,x_0; 0) = \frac{\delta(x - x_0)}{\sqrt{\det h(x)}}\,,
\end{align}
where we also specified the initial condition. In general the form of $\Delta$ may be different from the ordinary $\Delta = h^{ab} \nabla_a \nabla_b$, $a,b = 1,\ldots,d$. Let us assume that $M$ is flat, as in the case of Minkowski space. Then the solution of (\ref{eq:20.01}), also known as the heat kernel, in the momentum representation is given by
\begin{align}\label{eq:20.02}
K(x,x_0; \sigma) = \frac{1}{(2\pi)^d} \int\! d^dp\ e^{i p (x - x_0)} 
e^{-\sigma {\cal L}(p)}\,,
\end{align}
where ${\cal L}(p)$ denotes the momentum space version of the Laplacian $\Delta$. The heat kernel can be characterized by its trace, called the (average) return probability
\begin{align}\label{eq:20.03}
P(\sigma) \equiv K(x,x; \sigma) = \frac{1}{(2\pi)^d} \int\! d^dp\ 
e^{-\sigma {\cal L}(p)}\,,
\end{align}
which measures the probability of the diffusion returning to the same point on $M$. The spectral dimension of the manifold $M$ is
\begin{align}\label{eq:20.04}
d_S(\sigma) = -2\, \frac{d\log P(\sigma)}{d\log\sigma}\,.
\end{align}
It can be described as the effective topological dimension for which the standard diffusion process in $d$-dimensional Euclidean space (with $\Delta = \partial_a \partial^a$) approximates the $\Delta$-governed diffusion on $M$, at a given value of the scale $\sigma$. With large $\sigma$ we are probing the infrared structure of $M$, while small $\sigma$ correspond to its ultraviolet structure. In particular, the ordinary diffusion in Euclidean space is characterized by $d_S(\sigma) = d$ at all scales. Let us note that such a definition of the dimension is also sensitive (for sufficiently large $\sigma$) to the finite size of $M$ and the curvature of $h$ and one has to subtract their potential contribution from (\ref{eq:20.04}). 

The application of the spectral dimension in certain quantum gravity models may require \cite{Calcagni:2013pn} modifications of the diffusion equation (\ref{eq:20.01}) to make the return probability $P(\sigma)$ positive semidefinite. However, in our calculations for $\kappa$-Minkowski space we do not encounter such a situation. We also avoid the complexities associated with the noncommutative geometry by using $P(\sigma)$ in the momentum representation (\ref{eq:20.03}). On the other hand, as we shown in Section 1.3, (the Euclidean version of) the corresponding momentum space ${\rm AN}(d-1)$ is a curved manifold and this has to be accounted for. Thus, as we did in \cite{Arzano:2014de}, we calculate $P(\sigma)$ as the integral over embedding coordinates (\ref{eq:13.12}) of ${\rm AN}(d-1)$ in Minkowski space. In $d$ topological dimensions we can write it as
\begin{align}\label{eq:20.05}
P(\sigma) = \frac{1}{(2\pi)^d} \int\! d^{d+1}p\ \delta\lb p_{-1}^2 - (p_0^2 + 
p_a p^a + \kappa^2)\rb \Theta(p_{-1} - \kappa)\, 
e^{-\sigma {\cal L}(p_0,\{p_a\})}\,,
\end{align}
where the Dirac delta and Heaviside function constrain coordinates to Euclidean anti-de Sitter space. When we integrate out the auxiliary coordinate $p_{-1}$ the return probability simplifies to
\begin{align}\label{eq:20.06}
P(\sigma) = \frac{1}{(2\pi)^d} \int\! d^dp\ \frac{1}{2 \sqrt{p_0^2 + p_a p^a + \kappa^2}}\, e^{-\sigma {\cal L}(p_0,\{p_a\})}
\end{align}
and $(2 \sqrt{p_0^2 + p_a p^a + \kappa^2})^{-1} d^dp$ is actually the group invariant measure on ${\rm AN}(d-1)$. What we still need is the form of the Laplacian ${\cal L}(p_0,\{p_a\})$ but this remains an open issue for $\kappa$-Minkowski space, due to the ambiguity of the $\kappa$-Poincar\'{e} Casimir (\ref{eq:12.06}). Therefore we will calculate the spectral dimension for several of different Laplacians proposed in the literature, in both 3+1 and 2+1 topological dimensions. 

A first possible choice is the Laplacian determined by the bicovariant (i.e.\! covariant under the $\kappa$-Poincar\'{e} algebra) differential calculus on $\kappa$-Minkowski space \cite{Sitarz:1995ne,Gonera:1996de}, which corresponds to the Casimir element (\ref{eq:12.07}). Its Euclidean version has the form
\begin{align}\label{eq:20.07}
{\cal L}_0(p_0,\{p_a\}) = p_0^2 + p_a p^a = p_{-1}^2 - \kappa^2
\end{align}
(this explains why $p_0,p_a$ coordinates are often called classical). For this Laplacian the spectral dimension was already computed numerically in \cite{Benedetti:2009se,Benedetti:2009fe}, where the return probability was derived with the help of the Wick rotation applied to a less convenient basis of momentum generators. We use the formula (\ref{eq:20.06}) and in 3+1 topological dimensions obtain the analytical expression for the spectral dimension
\begin{align}\label{eq:20.08}
d_S(\sigma) = \frac{2\kappa \sqrt{\sigma} (2\kappa^2 \sigma - 3) - 
\sqrt{\pi}\, e^{\kappa^2 \sigma} (4\kappa^4 \sigma^2 - 4\kappa^2 \sigma + 3)\, 
{\rm erfc}(\kappa \sqrt{\sigma})}{-2\kappa \sqrt{\sigma} + 
\sqrt{\pi}\, e^{\kappa^2 \sigma} (2\kappa^2 \sigma - 1)\, 
{\rm erfc}(\kappa \sqrt{\sigma})}\,,
\end{align}
where ${\rm erfc}(.)$ denotes the complementary error function. The most significant features of $d_S$ are its values in the infrared regime, where it should coincide with the topological dimension, and in the ultraviolet regime. Taking $\kappa$ into account we observe that small scales of $\kappa$-Minkowski space are probed with the parameter $\sigma$ when $\kappa^2 \sigma \ll 1$ and large scales when $\kappa^2 \sigma \gg 1$. We can calculate the corresponding limits of (\ref{eq:20.08}), which give
\begin{align}\label{eq:20.09}
\lim_{\sigma \to 0} d_S(\sigma) = 3\,, \quad 
\lim_{\sigma \to \infty} d_S(\sigma) = 4\,.
\end{align}
As one can see in the plot in Fig.~2.1, the dimension is indeed decreasing monotonically from the large-scale, topological dimension to the small-scale value. 

\begin{figure}[ht]\label{fig:dsc1}
\includegraphics[width=0.47\textwidth]{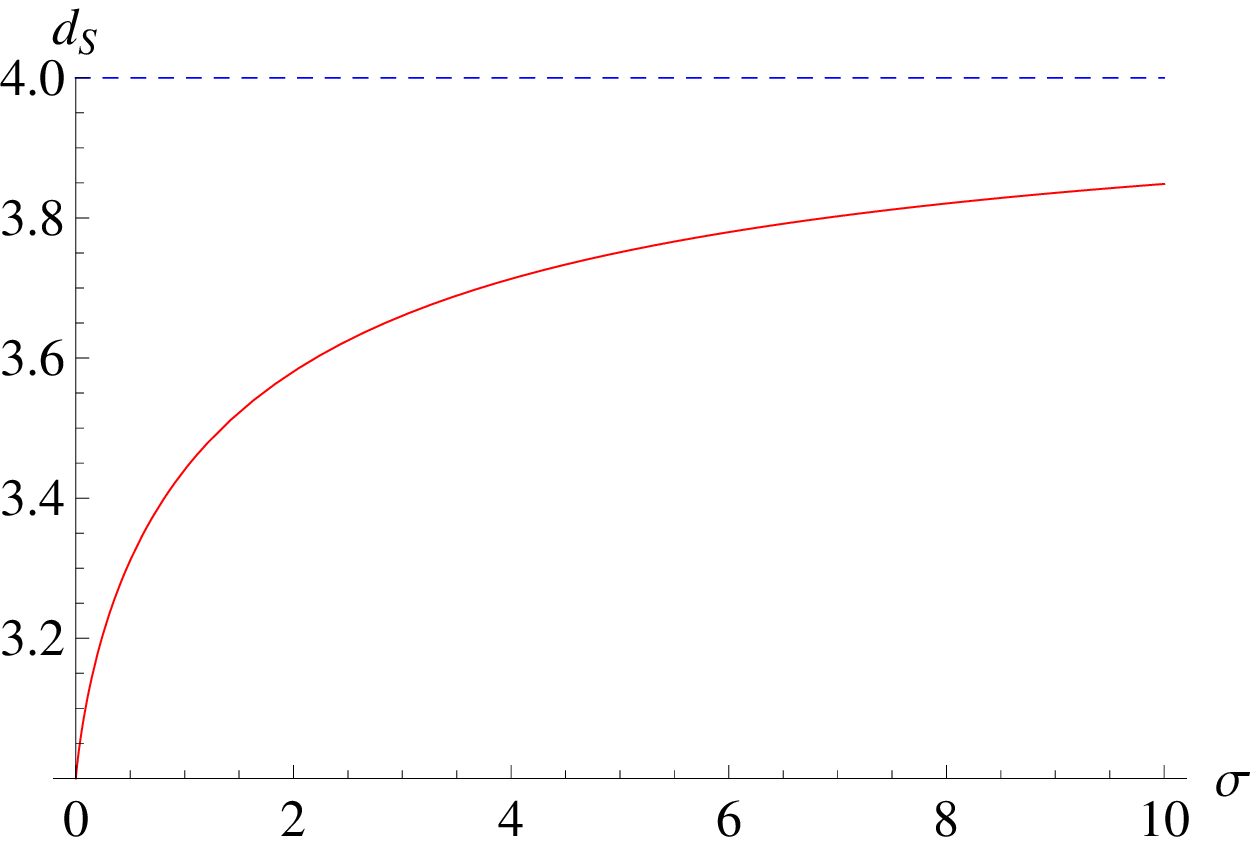}
\hspace{0.05\textwidth}
\includegraphics[width=0.47\textwidth]{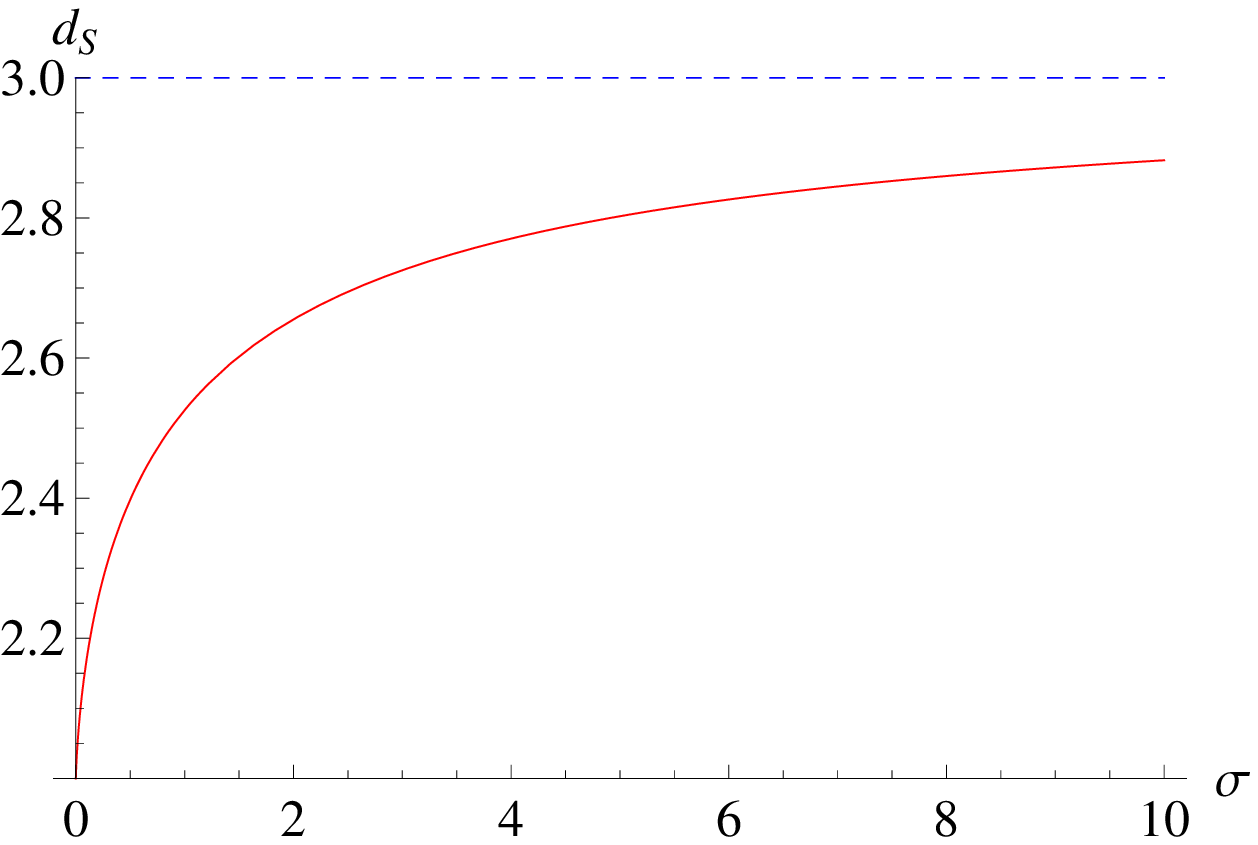}
\caption{Spectral dimension $d_S(\sigma)$ for the Laplacian ${\cal L}_0$ in 3+1 dimensions (left) and in 2+1 dimensions (right) with $\kappa = 1$}
\end{figure}

In 2+1 topological dimensions we can also find the explicit form of the spectral dimension
\begin{align}\label{eq:20.10}
d_S(\sigma) = 2 + \frac{\kappa^2 \sigma\, U(\frac{3}{2},1,\kappa^2 \sigma)}
{U(\frac{1}{2},0,\kappa^2 \sigma)}\,,
\end{align}
where $U(.,.,.)$ denotes a Tricomi confluent hypergeometric function. The ultraviolet and infrared limits of this expression are given by
\begin{align}\label{eq:20.11}
\lim_{\sigma \to 0} d_S(\sigma) = 2\,, \qquad 
\lim_{\sigma \to \infty} d_S(\sigma) = 3\,.
\end{align}
The plot in Fig.~2.1 shows the dimensional reduction similar to the case of 3+1 dimensions. Repeating the numerical calculations of \cite{Benedetti:2009fe} we can show that in both cases they perfectly agree with our analytical results. We then observe that for the Laplacian ${\cal L}_0$ in both 3+1 and 2+1 dimensions the overall behaviour of the spectral dimension is similar to many other approaches to quantum gravity that were mentioned at the beginning of the current Chapter. However, in 3+1 dimensions the small-scale value of the dimension (\ref{eq:20.08}) is different from the usual $d_S = 2$, while in 2+1 dimensions the ultraviolet limit of (\ref{eq:20.10}) agrees with the value of $d_S$ found in causal dynamical triangulations or Ho\v{r}ava-Lifshitz gravity. Let us also note that the dimensional reduction may result from the small-scale fractal structure of a given space, which is the actual situation in causal dynamical triangulations. For $\kappa$-Minkowski space such an interpretation was already suggested in \cite{Benedetti:2009fe} and relations between fractal and noncommutative geometries were explored in e.g.\! \cite{Arzano:2011ft}. A similar reason may be the fuzziness of points in noncommutative spacetime. 

As another possibility we consider the Laplacian defined by the mass Casimir (\ref{eq:12.06}), whose Euclidean version has the form
\begin{align}\label{eq:20.12}
{\cal L}_1(p_0,\{p_a\}) = 2\kappa \lb\sqrt{p_0^2 + p_a p^a + \kappa^2} - \kappa\rb = 2\kappa \lb p_{-1} - \kappa\rb\,.
\end{align}
In 3+1 topological dimensions we find that the spectral dimension is given by the simple rational function
\begin{align}\label{eq:20.13}
d_S(\sigma) = \frac{8\kappa^2 \sigma + 6}{2\kappa^2 \sigma + 1}\,.
\end{align}
Its ultraviolet and infrared limits are
\begin{align}\label{eq:20.14}
\lim_{\sigma \to 0} d_S(\sigma) = 6\,, \qquad 
\lim_{\sigma \to \infty} d_S(\sigma) = 4\,.
\end{align}
Thus in this case we observe the increased number of dimensions at small scales. The plot in Fig.~2.2 shows that the dimension is growing monotonically. 

\begin{figure}[ht]\label{fig:dsc0}
\includegraphics[width=0.47\textwidth]{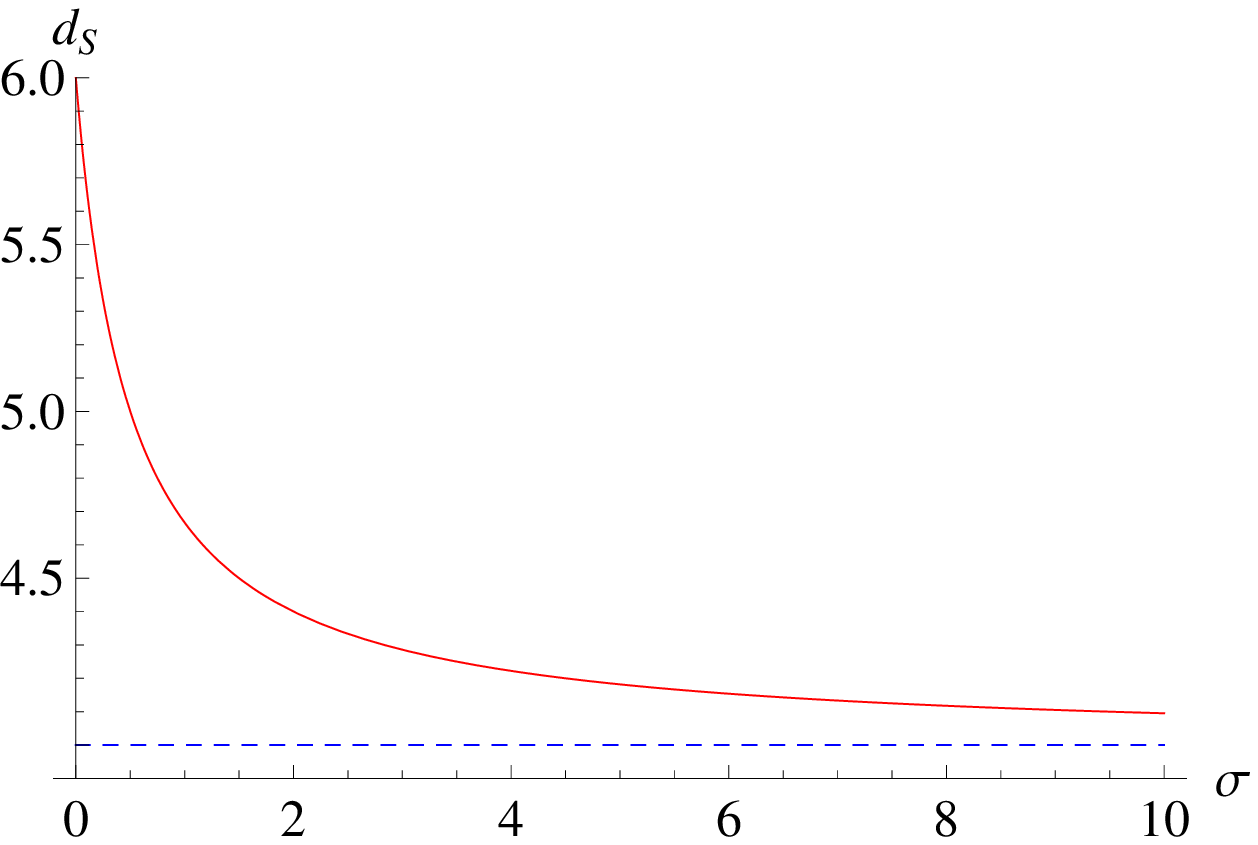}
\hspace{0.05\textwidth}
\includegraphics[width=0.47\textwidth]{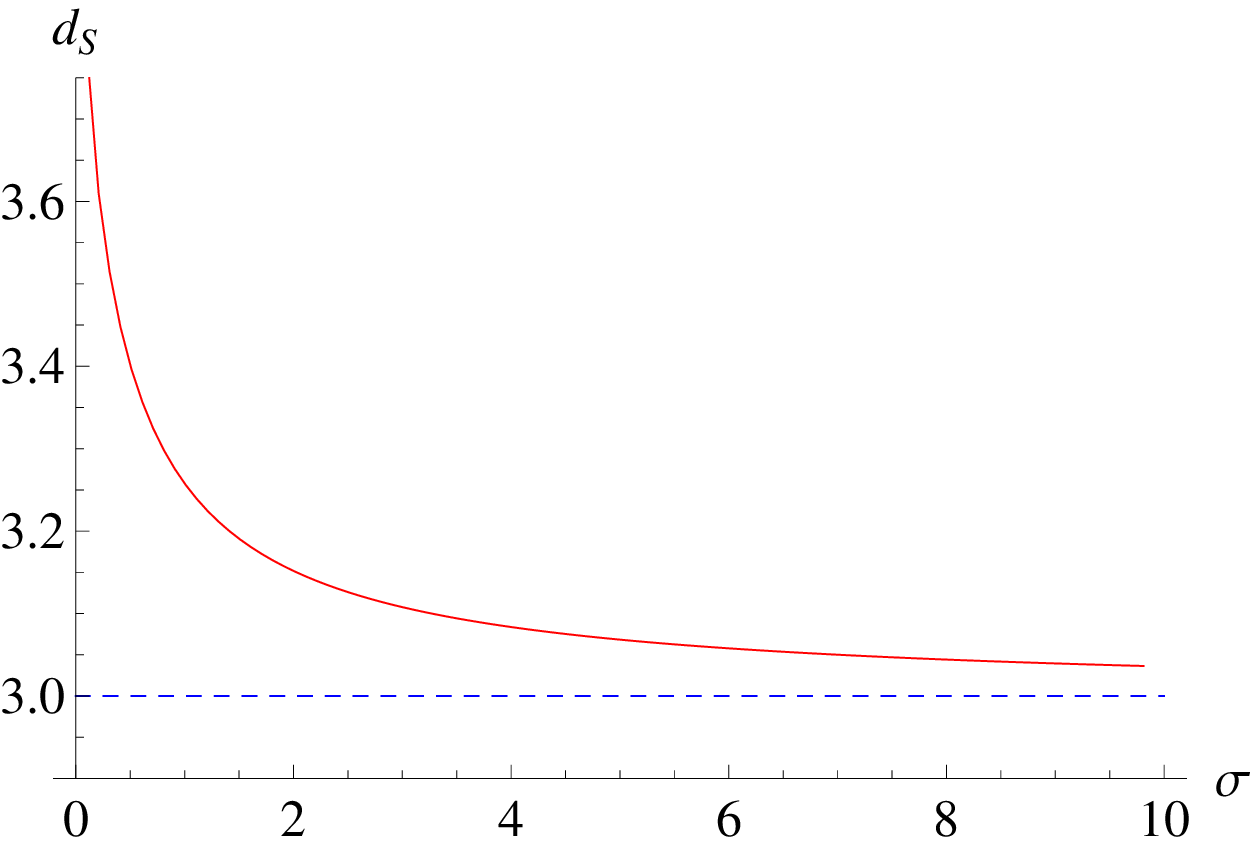}
\caption{Spectral dimension $d_S(\sigma)$ for the Laplacian ${\cal L}_1$ in 3+1 dimensions (left) and in 2+1 dimensions (right) with $\kappa = 1$}
\end{figure}

In 2+1 dimensions we find that the return probability can only be evaluated numerically. The resulting spectral dimension is presented in the plot in Fig.~2.2. Its ultraviolet and infrared limits are approximately
\begin{align}\label{eq:20.15}
\lim_{\sigma \to 0} d_S(\sigma) \approx 4\,, \qquad 
\lim_{\sigma \to \infty} d_S(\sigma) \approx 3\,,
\end{align}
which means that for the Laplacian ${\cal L}_1$ in both 3+1 and 2+1 dimensions we obtain the growing spectral dimension. In the terminology of statistical physics such a pattern is known as the superdiffusion. We will comment on this when discussing the results for the last Laplacian. Let us also note that for the Laplacian ${\cal L}_0$ the only reasone of the spectral dimension's flow is the nontrivial integration measure in the return probability (\ref{eq:20.06}), while for ${\cal L}_1$, as well as ${\cal L}_d$ considered below, it is also an effect of the deformed Laplacian. 

Finally, we may take the Laplacian coming from the framework of relative locality that we mentioned in the Preface. In this geometric approach the Laplacian on momentum space is identified as the square of the geodesic distance from the origin. Therefore in the case of the $\kappa$-Poincar\'{e} algebra it is given \cite{Gubitosi:2013re} by the geodesic distance in de Sitter space. Its Euclidean version should be the (squared) distance in Euclidean anti-de Siter space and have the form
\begin{align}\label{eq:20.16}
{\cal L}_d(p_0,\{p_a\}) = \kappa^2 {\rm arccosh}^2
\lb \frac{1}{\kappa} \sqrt{p_0^2 + p_a p^a + \kappa^2}\rb = 
\kappa^2 {\rm arccosh}^2\frac{p_{-1}}{\kappa}\,.
\end{align}
With such a Laplacian we have to evaluate the return probability numerically in both 3+1 and 2+1 topological dimensions. The resulting spectral dimension can be seen in the plots in Fig.~2.3. In 3+1 dimensions $d_S$ has the approximate ultraviolet and infrared limits
\begin{align}\label{eq:20.17}
\lim_{\sigma \to 0} d_S(\sigma) \approx \infty\,, \qquad 
\lim_{\sigma \to \infty} d_S(\sigma) \approx 4\,.
\end{align}
Similarly in 2+1 dimensions
\begin{align}\label{eq:20.18}
\lim_{\sigma \to 0} d_S(\sigma) \approx \infty\,, \qquad 
\lim_{\sigma \to \infty} d_S(\sigma) \approx 3\,.
\end{align}
Thus for the Laplacian ${\cal L}_d$ in both 3+1 and 2+1 dimensions the spectral dimension is apparently diverging at small scales. Such a behaviour seems to imply the breakdown of (classical) geometry. It was interpreted in this way in \cite{Arzano:2013bn}, where a model of the quantized black hole was found to lead to a divergence of the spectral dimension at a small, but finite, value of $\sigma$. Below that minimal scale the diffusion turns out to be ill defined, which can be seen as the discreteness of spacetime. On the other hand, the diverging dimension is also a feature of the phase $B$ in the space of coupling constants of causal dynamical triangulations \cite{Ambjorn:2010cy}. That phase is dominated by the quantum configurations in which, simply speaking, every point in spacetime lies close to all others. The dimensional flow for the Laplacian ${\cal L}_d$ could also indicate a similar type of the small-scale structure of $\kappa$-Minkowski space. Meanwhile, in the case of the Laplacian ${\cal L}_1$ the corresponding structure could be not very different. 

\begin{figure}[ht]\label{fig:dscd}
\includegraphics[width=0.47\textwidth]{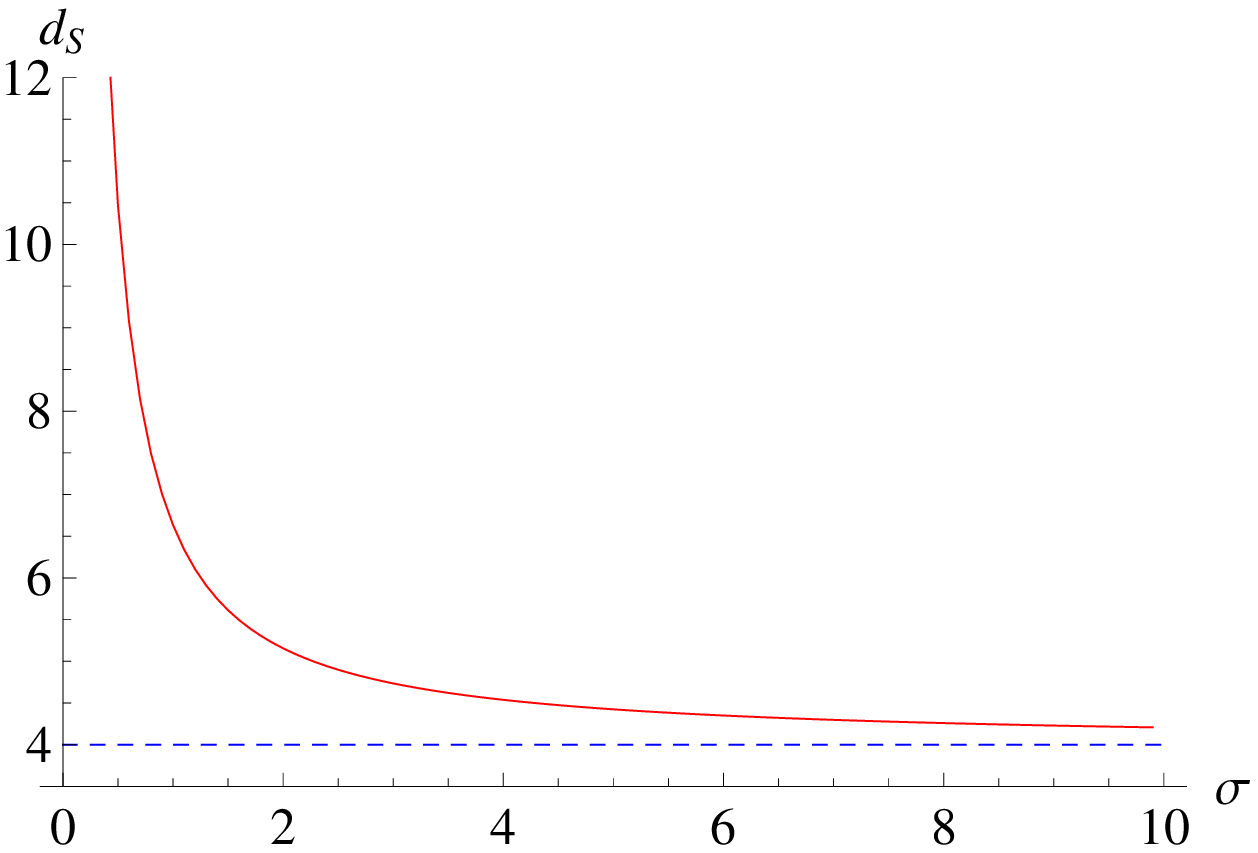}
\hspace{0.05\textwidth}
\includegraphics[width=0.47\textwidth]{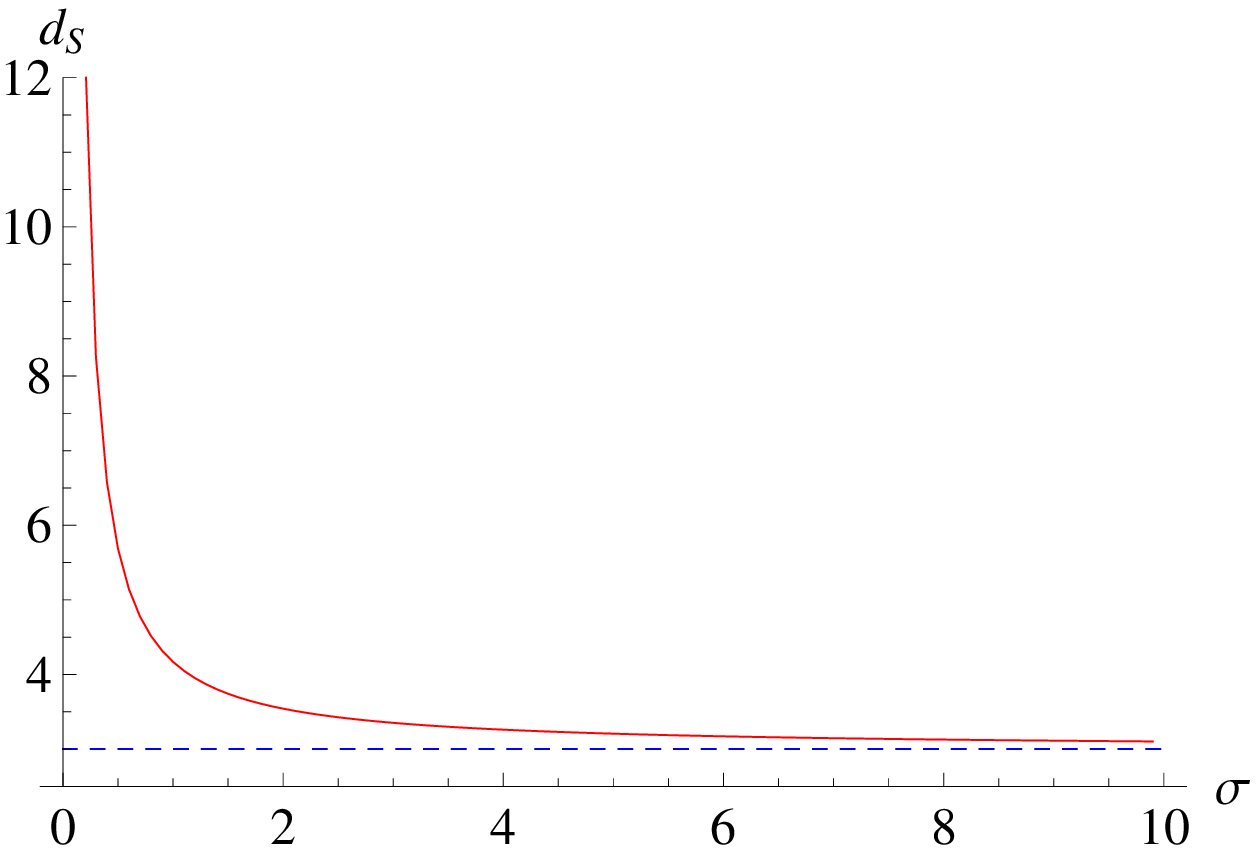}
\caption{Spectral dimension $d_S(\sigma)$ for the Laplacian ${\cal L}_d$ in 3+1 dimensions (left) and in 2+1 dimensions (right) with $\kappa = 1$}
\end{figure}

To conclude, we note that for various Laplacians on $\kappa$-Minkowski space one obtains a very different behaviour of the dimension. What remains to be explained is whether one of them is the physical Laplacian. Otherwise the spectral dimension would characterize a given field model on $\kappa$-Minkowski space rather than quantum spacetime itself. Furthermore, the application of the spectral dimension to a noncommutative spacetime may be not as straightforward as it is usually assumed, which would change the reported results. Let us also observe that none of the studied Laplacians results in a dimensional flow that is similar to the one found in \cite{Alesci:2012ay} in the context of the quantum field theory of particles coupled to three-dimensional gravity. In this model going to small scales one first encounters the superdiffusion and then in the ultraviolet limit the dimension falls to zero. On the other hand, for the Laplacian ${\cal L}_0$ in 2+1 dimensions we found the same behaviour of the dimension as in some fundamental quantum gravity models. It suggests that the $\kappa$-Poincar\'{e} algebra does not arise in the same semiclassical regime as the one considered in \cite{Alesci:2012ay}. Incidentally, it is worth to mention that diffusion processes can also be used in the calculations of other important quantities in quantum gravity, in particular the vacuum energy density and entanglement entropy. However \cite{Nesterov:2011gy}, they both turn out to be necessarily divergent for any form of the Laplacian, while we found that the case of $\kappa$-Minkowski space does not change the situation.

\newpage
\thispagestyle{empty}
~

\chapter{Point particle in 3d gravity}
In order to try to tackle the problems associated with the quantization of gravity we may study such physical models that preserve the fundamental features of general relativity but are devoid of at least some of its complexities. One of the natural choices is to consider gravity in 2+1 spacetime dimensions. It can not give us a lot of insight into the dynamics of the four-dimensional quantum theory since even classical solutions in three dimensions are generally quite different than in four and do not have a good Newtonian limit \cite{Deser:1984tc}. On the other hand, the conceptual aspects of the transition between the classical and quantum regime are essentially identical in both cases and in this context three-dimensional gravity can be really helpful (for a review see \cite{Carlip:2005qe}), being a much simpler theory. 
 
We first observe that, in any number of spacetime dimensions, solving Einstein field equations
\begin{align}\label{eq:30.01}
R_{\alpha\beta} - \frac{1}{2} R\, g_{\alpha\beta} = 8\pi G\, T_{\alpha\beta}
\end{align}
we can express the Ricci curvature tensor $R_{\alpha\beta}$ in terms of the energy-momentum tensor $T_{\alpha\beta}$. In more than three dimensions the Riemann curvature tensor is not completely determined by the Ricci tensor, since the former has more independent components, and the remaining freedom may lead to a non-flat metric $g_{\alpha\beta}$ even for vacuum $T_{\alpha\beta} = 0$. In the three-dimensional case this is no longer true and hence spacetime is always flat, or has the constant curvature (with the appropriate sign) when we turn on the cosmological constant. In other words, gravity in 2+1 dimensions does not have local degrees of freedom. Thus in this theory there are no local interactions and no gravitational waves. 

Nevertheless, it turns out that three-dimensional gravity (in the absence of matter fields) still has some residual dynamics. The latter is given by the topological degrees of freedom, which can be introduced in several ways. Firstly, one may consider nontrivial topologies of spacetime \cite{Thurston:1997ty}. Secondly, in the case of negative cosmological constant there exist black hole solutions \cite{Banados:1992be}. Finally, we may couple particles to the gravitational field. The purpose of the current Chapter is to discuss the latter situation, in the context of two different formulations of three-dimensional gravity. 

Let us note that another peculiar feature of gravity in 2+1 dimensions is the fact that the Newton's constant $G$ has the dimension of inverse mass, as can be seen from (\ref{eq:30.01}). Thus $G$ provides a natural deformation scale in the space of energy-momenta and the Planck mass is actually the classical quantity $\frac{1}{4G}$. As we will see, this leads to a nontrivial geometry of momentum space already in the classical theory.

\section{Dreibein formalism}
A point particle solution of three-dimensional gravity with vanishing cosmological constant was first obtained in \cite{Staruszkiewicz:1963ge} and subsequently generalized to the case of multiple particles in \cite{Deser:1984tc}. The metric of spacetime with a single static particle of mass $m$ in cylindrical coordinates has the form
\begin{align}\label{eq:31.01}
ds^2 = -dt^2 + dr^2 + (1 - 4Gm)^2 r^2 d\phi^2\,.
\end{align}
It describes the geometry of a cone, with the range of $\phi$ reduced by the deficit angle $\delta = 8\pi Gm$, $\delta \in (0,2\pi)$ (after the transformation $\phi \rightarrow (1 - 4Gm)\, \phi$). Apart from the cone's vertex at $r = 0$ spacetime is locally isometric to Minkowski space, as one can verify by a straightforward calculation of the Riemann curvature. Therefore the parallel transport of a vector along an arbitrary loop around the vertex depends only on the loop's winding number and a possible curvature singularity at $r = 0$. We will see below that this parallel transport is described by a rotationlike Lorentz transformation (i.e.\! a transformation conjugate to a rotation), which is naturally interpreted \cite{Regge:1961gs} as a consequence of the conical singularity created by the particle at $r = 0$. 

Let us now follow \cite{Matschull:1997qy} and change variables from the metric $g_{\alpha\beta}$ to the dreibein $e_\alpha^{\ \mu}$ and spin connection $\omega_\alpha^{\ \mu}$, $\alpha = t,r,\phi$, $\mu = 0,1,2$, defined by
\begin{align}\label{eq:31.02}
e_\alpha^{\ \mu} e_\beta^{\ \nu} \eta_{\mu\nu} = g_{\alpha\beta}\,, \qquad 
\omega_\alpha^{\ \mu} = \frac{1}{2} \epsilon^\mu_{\ \sigma\nu} 
\lb e_\beta^{\ \nu} \partial_\alpha e^{\beta\sigma} + 
e_\beta^{\ \nu} \Gamma_{\ \alpha\gamma}^\beta e^{\gamma\sigma}\rb\,,
\end{align}
where $\eta_{\mu\nu}$ denotes the Minkowski metric and $\Gamma_{\ \alpha\gamma}^\beta$ are the Christoffel symbols. Since the local isometry algebra of spacetime is the (three-dimensional) Poincar\'{e} algebra $e_\alpha^{\ \mu}$, $\omega_\alpha^{\ \mu}$, $\mu = 0,1,2$ may be treated as coordinates of the respective algebra elements $e_\alpha$ and $\omega_\alpha$. Outside the particle's worldline at $r = 0$ they have to satisfy the vacuum Einstein equations, whose general solution is given by
\begin{align}\label{eq:31.03}
e_\alpha = L^{-1} \partial_\alpha {\bf q}\, L\,, \qquad 
\omega_\alpha = L^{-1} \partial_\alpha L\,,
\end{align}
with a Lorentz transformation $L$ and translation ${\bf q}$ determining an embedding of a neighbourhood of the particle into Minkowski space. Due to the presence of a curvature singularity we have to introduce a cut in this neighbourhood, which may be done along the plane $\phi = 2\pi$. Imposing the continuity of the dreibein and spin connection across the cut, via $e_\alpha(\phi = 0) = e_\alpha(\phi = 2\pi)$, $\omega_\alpha(\phi = 0) = \omega_\alpha(\phi = 2\pi)$, we find the conditions
\begin{align}\label{eq:31.04}
L_+ = P^{-1} L_-\,, \qquad {\bf q}_+ = P^{-1} \lb {\bf q}_- - {\bf a}\rb P\,,
\end{align}
where $-,+$ subscripts denote, respectively, the values of $L$, ${\bf q}$ at $\phi = 0$ and $\phi = 2\pi$, while $P$ is a constant Lorentz transformation and ${{\bf a}}$ a constant translation. Briefly speaking, conical spacetime is constructed from Minkowski space by removing a wedge whose edge is the particle's worldline and identifying the faces of the wedge by $P$. 

To obtain the proper solution one still has to remove or regularize the curvature singularity. One of the possibilities \cite{Matschull:1997qy} is to change the topology of spacetime by replacing the worldline at $r = 0$ with a cylindrical boundary, which is enforced by the condition $e_\phi(r = 0) = 0$. It then follows from (\ref{eq:31.03}) that ${\bf q}(r = 0) \equiv {\bf x} = {\bf x}(t)$ and from (\ref{eq:31.04}) that
\begin{align}\label{eq:31.05}
{\bf a} = {\bf x}(t) - P\, {\bf x}(t) P^{-1}\,.
\end{align}
Notice that in the appropriate representation (see the next Section) any element $P$ of the three-dimensional Lorentz group can be decomposed into a term proportional to the identity transformation and an element ${\bf p}$ of the Lorentz algebra (which is isomorphic to the group of translations), i.e.
\begin{align}\label{eq:31.06}
P = p_3\, \mathbbm{1} + {\bf p}\,,
\end{align}
where $p_3$ is a number constrained by ${\bf p}$. Taking the time derivative of (\ref{eq:31.05}) we find that $\dot{{\bf x}}(t)$ commutes with $P$ and hence it is proportional to the algebra element ${\bf p}$. Thus in the most general form ${\bf x}(t)$ can be written as
\begin{align}\label{eq:31.07}
{\bf x}(t) = {\bf x}_0 + {\bf p}\, \tau(t)\,,
\end{align}
with a constant translation ${\bf x}_0$ and an arbitrary function $\tau$. It is naturally interpreted as the position of a particle with the momentum ${\bf p}$ and the eigentime proportional to $\tau(t)$, passing through the point ${\bf x}_0$. 

The remaining detail of the description is the particle's mass. In conical spacetime the energy-momentum, which is equivalent to the curvature, can be measured by a parallel transport around the singularity. Then the transport operator, or holonomy of the spin connection $\omega_\alpha$, is actually given by $P$, as one can see from (\ref{eq:31.04}). Therefore \cite{Matschull:1997qy} it is a Lorentz group element $P$ rather than algebra element ${\bf p}$ that is the particle's momentum. In a frame in which the particle is at rest the holonomy $P$ has to be a rotation by the deficit angle $\delta = 8\pi Gm$, which means that the scalar product of a given spatial vector before and after the parallel transport should be proportional to $\cos\delta$. It can be shown that this leads to the mass shell condition
\begin{align}\label{eq:31.08}
p_3 = \cos(4\pi G m)
\end{align}
For a moving particle the holonomy $P$ is given by the conjugation of the static holonomy with a boost. Then (\ref{eq:31.08}) is preserved but the deficit angle becomes wider than $8\pi Gm$, since it is determined by the particle's total energy, while the regularized worldline tilts in the direction of motion. The rotationlike holonomies span the full Lorentz group, which is thus the extended momentum space of the particle. As a manifold it is equivalent to 2+1-dimensional anti-de Sitter space. 

To conclude the current Section let us remark that for a particle endowed with spin $s \in \mathbbm{R}$ the metric (\ref{eq:31.01}) generalizes to \cite{Deser:1984tc}
\begin{align}\label{eq:31.09}
ds^2 = -\lb dt + 4Gs\, d\phi\rb^2 + dr^2 + \lb 1 - 4Gm\rb^2 r^2 d\phi^2\,.
\end{align}
Such a geometry of spacetime is called the spinning cone and can be constructed from Minkowski space by cutting out a wedge characterized by the deficit angle $8\pi Gm$ and identifying the wedge's faces with the time offset $8\pi Gs$. The resulting helical structure of time may lead to the problematic existence of closed timelike curves. Furthermore, as we will see below, the solution (\ref{eq:31.09}) has a non-vanishing torsion.

\section{Chern-Simons formalism}
Due to its peculiar nature gravity in 2+1 dimensions can also be formulated in a different way, introduced in \cite{Achucarro:1986as,Witten:1988dm}, as a Chern-Simons gauge theory, which is a type of the topological field theory. This is related to different areas of mathematical physics, such as moduli spaces of flat connections, knot theory and quantum groups. 

The gauge group of three-dimensional gravity is a local isometry group of spacetime. In the case of vanishing cosmological constant it is the Poincar\'{e} group ${\rm ISO}(2,1)$, whose algebra $\mathfrak{iso}(2,1)$ is spanned by the generators of Lorentz transformations $J_\mu$ and translation generators $P_\mu$, $\mu = 0,1,2$ which satisfy the commutation relations
\begin{align}\label{eq:32.01}
[J_\mu,J_\nu] = \epsilon_{\mu\nu\sigma} J^\sigma\,, \qquad 
[J_\mu,P_\nu] = \epsilon_{\mu\nu\sigma} P^\sigma\,, \qquad 
[P_\mu,P_\nu] = 0\,.
\end{align}
The indices will be raised or lowered by the Minkowski metric $\eta_{\mu\nu}$ with signature $(1,-1,-1)$ and the convention for the Levi-Civita symbol is $\epsilon_{012} = 1$. Since the ${\rm SL}(2,\mathbbm{R})$ group is a double cover of the Lorentz (sub)group ${\rm SO}(2,1)$ the latter can be given in the corresponding matrix representation for the $\mathfrak{so}(2,1)$ algebra
\begin{align}\label{eq:32.14}
J_0 = \frac{1}{2} \lb
\begin{array}{cc}
0\! & 1 \\ 
-1\! & 0
\end{array}
\rb\,, \qquad 
J_1 = \frac{1}{2} \lb
\begin{array}{cc}
0\! & 1 \\ 
1\! & 0
\end{array}
\rb\,, \qquad 
J_2 = \frac{1}{2} \lb
\begin{array}{cc}
1\! & 0 \\ 
0\! & -1
\end{array}
\rb
\end{align}
and its generators satisfy the relation
\begin{align}\label{eq:32.15}
J_\mu J_\nu = -\frac{1}{4}\, \eta_{\mu\nu}\, \mathbbm{1} + 
\frac{1}{2}\, \epsilon_{\mu\nu\sigma} J^\sigma\,.
\end{align}
Furthermore, there exists \cite{Meusburger:2008qr} the useful identification of the translation generators as $P_\mu = \theta J_\mu$, with a formal parameter $\theta$ such that $\theta^2 = 0$. The Poincar\'{e} group has the so-called semidirect product structure ${\rm ISO}(2,1) \simeq \mathfrak{so}(2,1)^* \rsp {\rm SO}(2,1)$, where the dual algebra $\mathfrak{so}(2,1)^* \simeq \mathbbm{R}^3$ is the group of translations and the semidirect product $\rsp$ means that we have a right action of ${\rm SO}(2,1)$ on $\mathfrak{so}(2,1)^*$. Thus a group element $\gamma \in {\rm ISO}(2,1)$ can be written in the factorized form \cite{Meusburger:2008qr}
\begin{align}\label{eq:32.16}
\gamma = \mathfrak{j}\, \mathfrak{p} = (\iota_3 \mathbbm{1} + \iota^\mu J_\mu) 
(1 + \xi^\nu P_\nu)\,,
\end{align}
where $\mathfrak{j} \in {\rm SO}(2,1)$, $\xi \equiv \xi^\mu P_\mu$, $\xi \in \mathfrak{so}(2,1)^*$ and coordinates on ${\rm SO}(2,1)$ satisfy the constraint $\iota_3^2 + \frac{1}{4} \iota_\mu \iota^\mu = 1$. Then the multiplication in ${\rm ISO}(2,1)$ is given by
\begin{align}\label{eq:32.17}
\gamma_{(1)} \gamma_{(2)} = \mathfrak{j}_{(1)} \mathfrak{j}_{(2)} \lb 1 + 
{\rm Ad}(\mathfrak{j}_{(2)}^{-1})\, \xi_{(1)} + \xi_{(2)}\rb\,,
\end{align}
with the conjugation action ${\rm Ad}(\mathfrak{j})\, \xi \equiv \mathfrak{j}\, \xi\, \mathfrak{j}^{-1}$.\footnote{The standard factorization of the Poincar\'{e} group, with the left conjugation action of $\mathfrak{j}_{(1)}$ on $\xi_{(2)}$, is connected to (\ref{eq:32.16}) by $\xi \mapsto {\rm Ad}(\mathfrak{j})\, \xi$.} 

On the algebra (\ref{eq:32.01}) there exists the natural scalar product
\begin{align}\label{eq:32.02}
\left<J_\mu P_\nu\right> = \eta_{\mu\nu}\,, \qquad 
\left<J_\mu J_\nu\right> = \left<P_\mu P_\nu\right> = 0\,.
\end{align}
Precisely speaking, we have the two-dimensional space of such products but only (\ref{eq:32.02}) allows us to obtain the correct gravitational action \cite{Witten:1988dm}. In order to express three-dimensional gravity as a gauge theory we combine the dreibein and spin connection one-forms into a gauge field, which is the Cartan connection
\begin{align}\label{eq:32.03}
A = \omega^\mu J_\mu + e^\mu P_\mu
\end{align}
and whose curvature, i.e.\! field strength is given by $F = dA + [A,A]$. Thus the action of pure gravity is the Chern-Simons action
\begin{align}\label{eq:32.04}
S_g = \frac{k}{4\pi} \int \left(\left<dA \wedge A\right> + 
\frac{1}{3} \left<A \wedge [A,A]\right>\right)
\end{align}
with the integration over the whole spacetime and the coupling constant $k = \frac{1}{4G}$. If we assume that spacetime can be decomposed into time $\mathbbm{R}$ and space ${\cal S}$ it is convenient to split the connection $A$ accordingly into the temporal and spatial parts, $A = A_t dt + A_S$, where $A_t$ is a function and $A_S$ a one-form. Then (\ref{eq:32.04}) can be rewritten as
\begin{align}\label{eq:32.05}
S_g = \frac{k}{4\pi} \int\! dt \int_{\cal S} \left<\dot A_S \wedge A_S\right> + 
\frac{k}{2\pi} \int\! dt \int_{\cal S} \left<A_t\, F_S\right>\,,
\end{align}
where the spatial curvature $F_S = dA_S + [A_S,A_S]$. 

Point particles can be added to the theory as punctures, i.e.\! pointlike topological defects on ${\cal S}$. For a single particle at rest the natural action has the form \cite{Witten:1989ql,Sousa:1990os}
\begin{align}\label{eq:32.06}
S_p = -\int\! dt \int_{\cal S} \left<{\cal C}\, A_t\right> \delta^2(\vec{y})\, 
dy^1 \wedge dy^2\,,
\end{align}
where $y^1,y^2$ are coordinates on ${\cal S}$ in which the particle is located at the origin. Mass $m$ and spin $s$ of the particle are encoded in the algebra element ${\cal C} = m\, J_0 + s\, P_0$ and we denote ${\cal C}_J \equiv m\, J_0$, ${\cal C}_P \equiv s\, P_0$ (notice that $J_0$ is the rotation generator). To obtain a moving particle one applies a gauge transformation of the connection $A \mapsto h^{-1} A\, h + h^{-1} dh$, $h \in {\rm ISO}(2,1)$ to the action (\ref{eq:32.06}), which then becomes the gauge invariant expression
\begin{align}\label{eq:32.07}
S_p = -\int\! dt\ \left<{\cal C}\, h^{-1} \dot h\right> - 
\int\! dt \int_{\cal S} \left<A_t\, h\, {\cal C}\, h^{-1}\right> 
\delta^2(\vec{y})\, dy^1 \wedge dy^2\,.
\end{align}
The second term describes the coupling of the puncture to the gravitational field, while the first one can be easily converted to the usual form of a free (spinning) particle action. Namely, factorizing the group element $h = (1 + \zeta)\, u$ and denoting the particle's momentum ${\bf q} = q^\mu J_\mu \equiv u\, {\cal C}_J u^{-1}$ and position ${\bf x} = x^\mu P_\mu \equiv u \zeta u^{-1}$ we obtain the familiar expression
\begin{align}\label{eq:32.07a}
S_p = \int\! dt\ \lb q_\mu \dot x^\mu + s\, (u^{-1} \dot u)_0\rb\,.
\end{align}
The sum of (\ref{eq:32.05}) and (\ref{eq:32.07}) gives the total action $S = \int\! dt\ L$ with the Lagrangian
\begin{align}\label{eq:32.08}
L = \frac{k}{4\pi} \int_{\cal S} \left<\dot A_S \wedge A_S\right> - 
\left<{\cal C}\, h^{-1} \dot h\right> + \nonumber\\ 
\int_{\cal S} \left<A_t \lb\frac{k}{2\pi} F_S - 
h\, {\cal C}\, h^{-1} \delta^2(\vec{y})\, dy^1 \wedge dy^2\rb\right>\,,
\end{align}
whose last term imposes the following constraint on the curvature
\begin{align}\label{eq:32.09}
\frac{k}{2\pi} F_S = h\, {\cal C}\, h^{-1} \delta^2(\vec{y})\, 
dy^1 \wedge dy^2\,,
\end{align}
while $A_t$ plays the role of a Lagrange multiplier. From the definition (\ref{eq:32.03}) it follows that $F_S = R_S + T_S$ and the spatial Riemann curvature and torsion are, respectively,
\begin{align}\label{eq:32.10}
R_S = \frac{2\pi}{k}\, {\bf q}\, \delta^2(\vec{y})\, dy^1 \wedge dy^2\,, \qquad 
T_S = \frac{2\pi}{k}\, {\bf k}\, \delta^2(\vec{y})\, dy^1 \wedge dy^2\,.
\end{align}
Thus the particle's momentum ${\bf q}$ is a source of the curvature singularity at the puncture, while its generalized angular momentum ${\bf k} = {\bf x} \times {\bf q} + s\, \hat{\bf q}$ (where $\hat{\bf q} \equiv u\, P_0 u^{-1}$ is a normalized vector) is a source of the torsion and $F_S$ vanishes everywhere else, which agrees with what we know from the previous Section. 

The constraint (\ref{eq:32.09}) provides a relation between the gravitational gauge field and particle degrees of freedom, which may allow us to eliminate the former in favour of the latter (since it is the particle that carries the degrees of freedom of gravity) and obtain the action describing the effective dynamics of the particle. This was indeed achieved for the corresponding symplectic form (with multiple particles) \cite{Meusburger:2005by}, which in principle also determines the action. However, here we will apply similar methods and derive the effective action independently, repeating what we did in \cite{Arzano:2014by,Kowalski:2014dy}. The starting point \cite{Meusburger:2005by} is to decompose space ${\cal S}$ into the disc centred on the particle ${\cal D}$, with polar coordinates $r \in [0,1]$, $\phi \in [0,2\pi]$, and the empty asymptotic region ${\cal E}$ (corresponding to $r \geq 1$), sharing the circular boundary $\Gamma$ at $r = 1$. Then it follows from (\ref{eq:32.09}) that on the empty region the connection is flat and hence given by
\begin{align}\label{eq:32.11}
A_S^{({\cal E})} = \gamma d\gamma^{-1}\,,
\end{align}
where $\gamma$ is an element of the gauge group ${\rm ISO}(2,1)$. On the disc the constraint (\ref{eq:32.09}) can also be solved, using the identity $dd\phi = 2\pi\, \delta^2(\vec{y})\, dy^1 \wedge dy^2$, and $A_S$ has the form
\begin{align}\label{eq:32.12}
A_S^{({\cal D})} = \bar\gamma\, \frac{1}{k} {\cal C} d\phi\, \bar\gamma^{-1} + 
\bar\gamma d\bar\gamma^{-1}\,, \quad \bar\gamma(r = 0) = h\,,
\end{align}
where an element $\bar\gamma$ characterizes the motion of the particle. By construction, the connection has to be continuous across the boundary, i.e.\! $A_S^{({\cal D})}|_\Gamma = A_S^{({\cal E})}|_\Gamma$. In terms of gauge group elements it leads to the sewing condition
\begin{align}\label{eq:32.13}
\gamma^{-1}|_\Gamma = N e^{\frac{1}{k} {\cal C} \phi}
\bar\gamma^{-1}|_\Gamma\,,
\end{align}
where $N = N(t)$ is an arbitrary element of ${\rm ISO}(2,1)$. This corresponds to the cut (\ref{eq:31.04}) that is introduced in a neighbourhood of the particle in the dreibein formalism. Indeed, $\bar\gamma$ is single-valued on ${\cal D}$ and hence $\gamma$ has a jump at the point $\phi = 2\pi$ when it is identified with $\phi = 0$. 

We may now factorize the connections (\ref{eq:32.11}), (\ref{eq:32.12}) via (\ref{eq:32.16}) and plug them into the Lagrangian (\ref{eq:32.08}), which cancels out the free particle term $<{\cal C}\, h^{-1} \dot h>$. Neglecting total time derivatives we obtain the purely boundary expression
\begin{align}\label{eq:32.18}
L = \frac{k}{2\pi} \int_{\Gamma} \left<\mathfrak{j}^{-1} \dot{\mathfrak{j}}\, 
d\xi - \bar{\mathfrak{j}}^{-1} \dot{\bar{\mathfrak{j}}}\, d\bar\xi + 
\frac{1}{k} {\cal C}_J d\phi \left[\bar{\mathfrak{j}}^{-1} 
\dot{\bar{\mathfrak{j}}}, \bar\xi\right] + \frac{1}{k} {\cal C}_P d\phi\, 
\bar{\mathfrak{j}}^{-1} \dot{\bar{\mathfrak{j}}}\right>\,,
\end{align}
where we took the opposite orientation of $\Gamma$ for contributions coming from the disc ${\cal D}$. In the next step we apply the factorization to the sewing condition (\ref{eq:32.13}) and split it into two parts
\begin{align}\label{eq:32.19}
\mathfrak{j}^{-1} = \mathfrak{n}\, e^{\frac{1}{k} {\cal C}_J \phi} 
\bar{\mathfrak{j}}^{-1}\,, \qquad 
-{\rm Ad}(\mathfrak{n}^{-1})\, \xi = \nu - 
{\rm Ad}(e^{\frac{1}{k} {\cal C}_J \phi})\, \bar{\xi} + 
\tfrac{1}{k} {\cal C}_P \phi\,,
\end{align}
where $N = \mathfrak{n}\, (1 + \nu)$, $\mathfrak{n} \in {\rm SO}(2,1)$, $\nu \in \mathfrak{so}(2,1)^*$. Substituting these conditions into (\ref{eq:32.18}) and rearranging the derivatives we obtain the integral over a total derivative
\begin{align}\label{eq:32.20}
L = \frac{k}{2\pi} \int_\Gamma d \left<e^{-\frac{1}{k} {\cal C}_J \phi} 
\dot{\mathfrak{n}}^{-1} \mathfrak{n}\, e^{\frac{1}{k} {\cal C}_J \phi} 
\bar{\xi} + \mathfrak{n}^{-1} \dot{\mathfrak{n}}\, 
\frac{1}{k} {\cal C}_P \phi\right>\,.
\end{align}
Integrating it over $\phi \in [0,2\pi]$ and evaluating the scalar product we arrive at the final effective particle Lagrangian
\begin{align}\label{eq:32.21}
L = \kappa \lb\dot\Pi^{-1} \Pi\rb_\mu x^\mu + 
s \lb\mathfrak{n}^{-1} \dot{\mathfrak{n}}\rb_0\,,
\end{align}
where we denote $\kappa \equiv \frac{k}{2\pi}$ and the new variables of position $x \equiv \mathfrak{n}\, \bar{\xi}\, \mathfrak{n}^{-1}$, $x = x^\mu \tilde P_\mu$ (notice that $\bar{\xi} = \bar{\xi}(\phi = 0) = \bar{\xi}(\phi = 2\pi)$) and momentum
\begin{align}\label{eq:32.22}
\Pi \equiv \mathfrak{n}\, e^{\frac{2\pi}{k} {\cal C}_J} \mathfrak{n}^{-1} = 
e^{\frac{m}{\kappa} \mathfrak{n} J_0 \mathfrak{n}^{-1}}\,.
\end{align}
Thus the momentum of a particle becomes an element of the Lorentz group (conjugate to the rotation by $\frac{m}{\kappa} = 8\pi Gm$) instead of a Lorentz algebra element ${\bf q}$, as we already discussed in the previous Section. The Lagrangian (\ref{eq:32.21}) agrees with the symplectic form constructed in \cite{Meusburger:2003py} and its spinless part is equivalent to the Lagrangian obtained in the dreibein formalism in \cite{Matschull:1997qy}. Without loss of generality we may also fix the gauge at the boundary $\Gamma$ via $\gamma(\phi = 0) = 1$, which leads to the simple relations $\bar{\mathfrak{j}}(\phi = 0,2\pi) = \mathfrak{n}$ and $\bar{\xi}(\phi = 0,2\pi) = \nu$. Then we have $\Pi = \bar{\mathfrak{j}}\, e^{\frac{m}{\kappa} J_0} \bar{\mathfrak{j}}^{-1}$, $x = \bar{\mathfrak{j}}\, \bar{\xi}\, \bar{\mathfrak{j}}^{-1}$. We also calculate the holonomy of the connection $A_S$ around the loop $\Gamma$, which is the path-ordered exponential
\begin{align}\label{eq:32.23}
{\cal P}\, e^{-\int_\Gamma A_S} = \gamma(\phi = 0)\, \gamma^{-1}(\phi = 2\pi) = \nonumber\\ 
\Pi \lb 1 + \lb{\rm Ad}(\Pi^{-1}) - 1\rb x + {\rm Ad}(\Pi^{-1})\, 
\bar{\mathfrak{j}}\, \tfrac{1}{\kappa} {\cal C}_P \bar{\mathfrak{j}}^{-1}\rb\,.
\end{align}
In particular, the Lorentzian part of (\ref{eq:32.23}) gives $\mathfrak{j}(\phi = 0)\, \mathfrak{j}^{-1}(\phi = 2\pi) = \Pi$, which supports the interpretation that $\Pi$ is the particle's momentum, as can also be shown more generally \cite{Meusburger:2003py}. The other factor in (\ref{eq:32.23}) measures the generalized angular momentum of the particle \cite{Meusburger:2003py}
\begin{align}\label{eq:32.23a}
{\rm I} \equiv \kappa\, (1 - {\rm Ad}(\Pi))\, x + 
\bar{\mathfrak{j}}\, {\cal C}_P \bar{\mathfrak{j}}^{-1}\,,
\end{align}
which is deformed due to the presence of the group momentum. 

Let us now restrict to the spinless case $s = 0$. To explore the particle's dynamics, governed by the Lagrangian (\ref{eq:32.21}), we parametrize the momentum via $m\, \bar{\mathfrak{j}} J_0 \bar{\mathfrak{j}}^{-1} = q^\mu J_\mu$ and $\Pi = p_3 + \frac{1}{\kappa}\, p^\mu J_\mu$. It can be easily shown that $q_\mu q^\mu = m^2$ and
\begin{align}\label{eq:32.24}
p_3 = \cos\frac{|q|}{2\kappa}\,, \qquad 
p^\mu = 2\kappa\, \frac{q^\mu}{|q|}\, \sin\frac{|q|}{2\kappa}\,,
\end{align}
where $|q| \equiv \sqrt{q_\mu q^\mu}$. From these expressions and the constraint on group coordinates $p_3^2 + p_\mu p^\mu/(4\kappa^2) = 1$ we obtain the mass shell condition $p_\mu p^\mu = 4\kappa^2 \sin^2\frac{m}{2\kappa}$, which is equivalent to (\ref{eq:31.08}) from the dreibein formalism. If we take it as a constraint with the Lagrange multiplier $\lambda$ then the action determined by (\ref{eq:32.21}), rewritten in components, has the form
\begin{align}\label{eq:32.25}
S = -\int\! dt\ \lb\lb p_3 \dot p_\mu - \dot p_3 p_\mu - 
\frac{1}{2\kappa}\, \epsilon_{\mu\nu\sigma} \dot p^\nu p^\sigma\rb x^\mu + 
\lambda \lb p_\mu p^\mu - 4\kappa^2 \sin^2\frac{m}{2\kappa}\rb\rb\,,
\end{align}
where $p_3 = \sqrt{1 - p_\mu p^\mu/(4\kappa^2)}$. Hence in the no-gravity, or low-energy limit $\kappa \rightarrow \infty$ (equivalent to $G \rightarrow 0$) we recover the action of a free relativistic particle
\begin{align}\label{eq:32.26}
S = -\int\! dt\ \lb\dot p_\mu x^\mu + \lambda \lb p_\mu p^\mu - m^2\rb\rb\,.
\end{align}
As can be shown after some calculations, a variation of (\ref{eq:32.25}) over $x^\mu$, $p_\mu$ leads to the same equations of motion (up to a certain rescaling of $\lambda$) as the ones of (\ref{eq:32.26}), to wit
\begin{align}\label{eq:32.27}
\dot x^\mu = 2\lambda\, p^\mu\,, \qquad \dot p_\mu = 0\,.
\end{align}
This is actually consistent with the expression for the particle's position (\ref{eq:31.07}) from the previous Section. Thus the motion of a particle coupled to three-dimensional gravity is not affected by the nontrivial structure of its momentum space. Let us also note that the particle's angular momentum (\ref{eq:32.23a}) in components ${\rm I} = I^\mu P_\mu$ is given by
\begin{align}\label{eq:32.27a}
I^\mu = p_3\, \epsilon^{\mu}_{\ \nu\sigma} x^\nu p^\sigma + 
\frac{1}{2\kappa} \lb p_\nu p^\nu x^\mu - x_\nu p^\nu p^\mu\rb + 
s\, \hat p^\mu\,,
\end{align}
which satisfies the deformed relation between momentum and angular momentum $p_\mu I^\mu = 4\kappa^2 \sin^2\frac{m}{2\kappa}\, s$ (in contrast to the usual $p_\mu I^\mu = ms$) and in the limit $\kappa \rightarrow \infty$ leads to the ordinary expression $I^\mu = \epsilon^{\mu}_{\ \nu\sigma} x^\nu p^\sigma + s\, \hat p^\mu$.

\section{Multiple particles}
We will now consider the generalization of our derivation of the effective particle Lagrangian to the case of multiple particles. The Chern-Simons Lagrangian for a system of $n$ particles coupled to three-dimensional gravity has the form \cite{Meusburger:2005by}
\begin{align}\label{eq:33.01}
L_{(n)} = \frac{k}{4\pi} \int_{\cal S} \left<\dot A_S \wedge A_S\right> - 
\sum_{i=1}^{n} \left<{\cal C}_i h_i^{-1} \dot h_i\right> + \nonumber\\ 
\int_{\cal S} \left<A_t \lb\frac{k}{2\pi} F_S - 
\sum_{i=1}^{n} h_i {\cal C}_i h_i^{-1} \delta^2(\vec{y} - \vec{y}_i)\, 
dy^1 \wedge dy^2\rb\right>\,,
\end{align}
where the particles, labelled by $i = 1,\ldots,n$, are located at the points $\vec{y}_i$ and characterized at rest by the algebra elements ${\cal C}_i = m_i J_0 + s_i P_0$. The total deficit angle of the particles should satisfy $\sum_{i=1}^{n} \delta_i < 2\pi$ for an open topology of space ${\cal S}$ and $\sum_{i=1}^{n} \delta_i = 4\pi$ for a closed one (the latter case is topologically possible for $n \geq 3$). In the case of an open topology we also have to impose the appropriate conditions at spatial infinity so that the whole system is equivalent to a single effective particle \cite{Meusburger:2005by}. 

In order to solve the constraint on $F_S$ a decomposition of ${\cal S}$ can be constructed in the following manner \cite{Meusburger:2005by}. We choose a point away from the particles and starting from it draw a separate loop around every particle, dividing ${\cal S}$ into $n$ non-overlapping particle regions ${\cal D}_i$ and the remaining empty region ${\cal E}$, with the boundary $\Gamma$. Since the theory is topological every region ${\cal D}_i$ is equivalent to a disc, on which we introduce polar coordinates $r_i \in [0,1]$, $\phi_i \in [0,2\pi]$ and the connection $A_S$ is found to be
\begin{align}\label{eq:33.02}
A_S^{({\cal D}_i)} = \bar\gamma_i \frac{1}{k} {\cal C}_i d\phi_i 
\bar\gamma_i^{-1} + \bar\gamma_i d\bar\gamma_i^{-1}\,, \quad 
\bar\gamma_i(r_i = 0) = h_i\,.
\end{align}
The empty region ${\cal E}$ can be deformed into a polygon, whose edges $\Gamma_i$ coincide with the boundaries of the consecutive discs. Namely, at the $i$'th vertex of ${\cal E}$ the incoming edge $\Gamma_i$ with the coordinate $\phi_i = 2\pi$ meets the outgoing edge $\Gamma_{i+1}$ with the coordinate $\phi_{i+1} = 0$. Then on each of $\Gamma_i$ we may sew the connection $A_S^{({\cal D}_i)}$ with $A_S^{({\cal E})}$, given by (\ref{eq:32.11}), and follow the same steps as in the case of a single particle. As the result, for every particle we find the effective Lagrangian
\begin{align}\label{eq:33.03}
L_i = \kappa \lb\dot{\bar\Pi}_i^{-1} \bar\Pi_i\rb_\mu \bar x_i^\mu + 
s_i \lb\mathfrak{n}_i^{-1} \dot{\mathfrak{n}}_i\rb_0\,,
\end{align}
where we denote $\bar\Pi_i \equiv \mathfrak{n}_i e^{\frac{1}{\kappa} m_i J_0} 
\mathfrak{n}_i^{-1}$, $\bar x_i \equiv \mathfrak{n}_i \bar\xi_i \mathfrak{n}_i^{-1}$, while $\bar\xi_i = \bar\xi_i(\phi_i = 0,2\pi)$. 

To uncover the relations between individual particles we use the continuity of $\gamma$ on every $i$'th vertex of ${\cal E}$, $i \neq 1$ (at the 1'st vertex $\gamma$ has a jump, corresponding to the total deficit angle of the system), which is enforced by the conditions $\gamma(\phi_{i+1} = 0) = \gamma(\phi_i = 2\pi)$. Similarly to the single particle case we may also fix the gauge at the first vertex via $\gamma(\phi_1 = 0) = 1$. In this way we obtain the following sequence of conditions
\begin{align}\label{eq:33.04}
\mathfrak{n}_1\, \bar{\mathfrak{j}}_1^{-1} = 1\,, \qquad 
\mathfrak{n}_2\, \bar{\mathfrak{j}}_2^{-1} = \Pi_1\,, \qquad 
\mathfrak{n}_3\, \bar{\mathfrak{j}}_3^{-1} = \Pi_1 \Pi_2\,, \qquad \ldots\,,
\end{align}
where $\Pi_i = \bar{\mathfrak{j}}_i e^{\frac{1}{\kappa} m_i J_0} \bar{\mathfrak{j}}_i^{-1}$. (We do not need here the conditions involving $\bar\xi_i$'s.) Applying (\ref{eq:33.04}) to the individual Lagrangians (\ref{eq:33.03}) we eliminate variables $\mathfrak{n}_i$ in favour of $\bar{\mathfrak{j}}_i$ and then take the sum over all particles. The final effective $n$-particle Lagrangian can be written as the iterative expression
\begin{align}\label{eq:33.05}
L_{(n)} = L_{(n-1)} + \kappa \lb\dot\Pi_n^{-1} \Pi_n\rb_\mu x_n^\mu + 
s_n \lb\bar{\mathfrak{j}}_n^{-1} \dot{\bar{\mathfrak{j}}}_n\rb_0 + \nonumber\\ 
\kappa \lb\Pi_n^{-1} \partial_0(\Pi_{n-1}^{-1} \ldots \Pi_1^{-1}) 
\Pi_1 \ldots \Pi_n \rb_\mu x_n^\mu -  \nonumber\\ 
\kappa \lb\partial_0(\Pi_{n-1}^{-1} \ldots \Pi_1^{-1}) \Pi_1 \ldots 
\Pi_{n-1} \rb_\mu x_n^\mu + \nonumber\\ 
s_n \lb\bar{\mathfrak{j}}_n^{-1} \Pi_{n-1}^{-1} \ldots \Pi_1^{-1} 
\partial_0(\Pi_1 \ldots \Pi_{n-1}) \bar{\mathfrak{j}}_n\rb_0\,,
\end{align}
where $x_i = \bar{\mathfrak{j}}_i \bar\xi_i \bar{\mathfrak{j}}_i^{-1}$. For example, in the 2-particle case it amounts to
\begin{align}\label{eq:33.06}
L_{(2)} = \kappa \lb\dot\Pi_1^{-1} \Pi_1\rb_\mu x_1^\mu + 
\kappa \lb\dot\Pi_2^{-1} \Pi_2\rb_\mu x_2^\mu + 
\kappa \lb\Pi_2^{-1} \dot\Pi_1^{-1} \Pi_1 \Pi_2 - 
\dot\Pi_1^{-1} \Pi_1\rb_\mu x_2^\mu + \nonumber\\ 
s_1 \lb\bar{\mathfrak{j}}_1^{-1} \dot{\bar{\mathfrak{j}}}_1\rb_0 + 
s_2 \lb\bar{\mathfrak{j}}_2^{-1} \dot{\bar{\mathfrak{j}}}_2\rb_0 + 
s_2 \lb\bar{\mathfrak{j}}_2^{-1} \Pi_1^{-1} \dot\Pi_1 
\bar{\mathfrak{j}}_2\rb_0\,.
\end{align}
Thus the Lagrangian (\ref{eq:33.05}) is the sum of free particle Lagrangians and terms describing the topological interaction between individual particles. We also observe that the Lorentzian part of the holonomy of the connection $A_S$ along a given edge $\Gamma_i$ is given by $\mathfrak{j}(\phi_i = 0)\, \mathfrak{j}^{-1}(\phi_{i+1} = 2\pi)$ (cf.\! (\ref{eq:32.23})) and hence the holonomy around the whole boundary $\Gamma$ is 
\begin{align}\label{eq:33.06a}
\mathfrak{j}(\phi_1 = 0)\, \mathfrak{j}^{-1}(\phi_n = 2\pi) = \Pi_1 \ldots \Pi_n 
\equiv \Pi\,,
\end{align}
which is naturally interpreted as the total momentum of the system. However, since space ${\cal S}$ is two-dimensional and $\Pi_i$ are elements of a non-Abelian group, $\Pi$ is invariant not under a usual permutation of a pair of (holonomies characterizing) particles $(\Pi_i,\Pi_{i+1}) \rightarrow (\Pi_{i+1},\Pi_i)$ but under a so-called braiding $(\Pi_i,\Pi_{i+1}) \rightarrow (\Pi_{i+1},\Pi_{i+1}^{-1} \Pi_i \Pi_{i+1})$ or $(\Pi_i,\Pi_{i+1}) \rightarrow (\Pi_i \Pi_{i+1} \Pi_i^{-1},\Pi_i)$ \cite{Carlip:1989ey} (for consequences for the quantum statistics see Chapter 6). For the same reasons the interacting terms in (\ref{eq:33.05}) depend on the particle ordering. 

To conclude the discussion of the Chern-Simons formalism let us remark \cite{Meusburger:2005by} that in a similar way to the punctures on space ${\cal S}$, which represent point particles, one can also include in the action (\ref{eq:33.01}) the contribution of a nontrivial spatial topology, represented by the handles\footnote{A handle is a torus attached to the surface.} on ${\cal S}$.

\newpage
\thispagestyle{empty}
~

\chapter{$\kappa$-deformed Carroll particle}
When the cosmological constant $\Lambda$ is non-zero, three-dimensional gravity is described \cite{Witten:1988dm} by the Chern-Simons action (\ref{eq:32.04}) with the gauge group being the de Sitter group ${\rm SO}(3,1)$ for $\Lambda > 0$ or anti-de Sitter group ${\rm SO}(2,2)$ for $\Lambda < 0$. From Section 1.3 we know that there exists a correspondence between 2+1-dimensional de Sitter space and the ${\rm AN}(2)$ group, which is the $\kappa$-Minkowski momentum space. Furthermore, it was shown \cite{Meusburger:2009gy} that the structure of the Chern-Simons theory with the de Sitter gauge group is associated with the $\kappa$-de Sitter algebra (i.e.\! the de Sitter counterpart of the $\kappa$-Poincar\'{e} algebra) and \cite{Osei:2012oy} the $\kappa$-Poincar\'{e} algebra arises as the symmetry algebra in the quantized theory, although in a unphysical regime. Therefore it may be worth to further explore the case of the ${\rm SO}(3,1)$ gauge group, which was our motivation in \cite{Kowalski:2014dy}. 

The de Sitter algebra $\mathfrak{so}(3,1)$ has the generators of rotations $J_\mu$ and boosts $P_\mu$, $\mu = 0,1,2$, with the commutators
\begin{align}\label{eq:40.01}
[J_\mu,J_\nu] = \epsilon_{\mu\nu\sigma} J^\sigma\,, \qquad 
[J_\mu,P_\nu] = \epsilon_{\mu\nu\sigma} P^\sigma\,, \qquad 
[P_\mu,P_\nu] = -\Lambda\, \epsilon_{\mu\nu\sigma} J^\sigma\,.
\end{align}
To uncover the Iwasawa decomposition of ${\rm SO}(3,1)$ that is related to (\ref{eq:13.10}) we may introduce new generators
\begin{align}\label{eq:40.02}
S_\mu \equiv P_\mu + \sqrt{\Lambda}\, \epsilon_{\mu 0\nu} J^\nu\,.
\end{align}
Then the algebra (\ref{eq:40.01}) acquires the form
\begin{align}\label{eq:40.03}
[J_\mu,J_\nu] & = \epsilon_{\mu\nu\sigma} J^\sigma\,, \qquad 
[J_\mu,S_\nu] = \epsilon_{\mu\nu\sigma} S^\sigma + 
\sqrt{\Lambda}\, (\eta_{\nu 0} J_\mu - \eta_{\mu\nu} J_0)\,, \nonumber\\ 
[S_\mu,S_\nu] & = \sqrt{\Lambda}\, (\eta_{\mu 0} S_\nu - \eta_{\nu 0} S_\mu)\,,
\end{align}
splitting into two subalgebras. $J_\mu$'s span the Lorentz algebra $\mathfrak{so}(2,1)$, while rewriting the last commutator we find $[S_0,S_a] = \sqrt{\Lambda}\, S_a$, $[S_a,S_b] = 0$, $a,b = 1,2$ and thus $S_\mu$'s span the $\mathfrak{an}(2)$ algebra. Similarly to the translation generators of the Poincar\'{e} algebra (\ref{eq:32.01}) we have \cite{Meusburger:2008qr} the identification of the boost generators as $P_\mu = i\sqrt{\Lambda}\, J_\mu$. Then the algebra (\ref{eq:40.03}) may be given in the matrix representation (\ref{eq:32.14}),
\begin{align}\label{eq:40.03a}
S_0 = \frac{1}{2} \lb
\begin{array}{cc}
-\sqrt{\Lambda}\! & 0 \\ 
0\! & \sqrt{\Lambda}
\end{array}
\rb\,, \qquad 
S_1 = \lb
\begin{array}{cc}
0\! & 0 \\ 
\sqrt{\Lambda}\! & 0
\end{array}
\rb\,, \qquad 
S_2 = \lb
\begin{array}{cc}
0\! & 0 \\ 
i\sqrt{\Lambda}\! & 0
\end{array}
\rb
\end{align}
(in the more general framework of the quaternionic representations of local isometry algebras in three-dimensional gravity \cite{Meusburger:2008qr}) in which its generators satisfy the relations (\ref{eq:32.15}) and
\begin{align}\label{eq:40.04}
S_\mu S_\nu = \frac{1}{4} \Lambda\, \eta_{\mu 0} \eta_{\nu 0}\, \mathbbm{1} + \frac{1}{2} \sqrt{\Lambda} \lb\eta_{\mu 0} S_\nu - \eta_{\nu 0} S_\mu\rb\,, \nonumber\\ 
J_\mu S_\nu = -\frac{i}{4} \sqrt{\Lambda}\, \eta_{\mu\nu}\, \mathbbm{1} + 
\sqrt{\Lambda} \lb -\frac{1}{4}\, \epsilon_{0\mu\nu}\, \mathbbm{1} + 
\frac{1}{2}\, \eta_{\nu 0} J_\mu - \frac{1}{2}\, \eta_{\mu\nu} J_0\rb + 
\frac{1}{2}\, \epsilon_{\mu\nu\sigma} S^\sigma\,.
\end{align}
Thus for group elements $\gamma \in {\rm SO}(3,1)$ a counterpart of the factorization (\ref{eq:32.16}) is \cite{Meusburger:2009gy}
\begin{align}\label{eq:40.05}
\gamma = \mathfrak{j}\, \mathfrak{s} = (\iota_3 \mathbbm{1} + \iota^\mu J_\mu) 
(\xi_3 \mathbbm{1} + \xi^\nu S_\nu)\,,
\end{align}
where $\mathfrak{j} \in {\rm SO}(2,1)$, $\mathfrak{s} \in {\rm AN}(2)$ and coordinates satisfy the constraints $\iota_3^2 + \frac{1}{4} \iota_\mu \iota^\mu = 1$, $\xi_3^2 - \frac{\Lambda}{4} \xi_0^2 = 1$. More precisely \cite{Osei:2012oy}, the ${\rm SO}(3,1)$ has the so-called double cross product structure ${\rm SO}(3,1) \simeq {\rm AN}(2) \dcp {\rm SO}(2,1)$, with the respective left and right action of one subgroup on the other. 

The correct scalar product on the algebra (\ref{eq:40.03}), analogous to (\ref{eq:32.02}), is given by
\begin{align}\label{eq:40.06}
\left<J_\mu S_\nu\right> = \eta_{\mu\nu}\,, \qquad 
\left<J_\mu J_\nu\right> = \left<S_\mu S_\nu\right> = 0\,.
\end{align}
Then a particle coupled to three-dimensional gravity with $\Lambda > 0$ is described by the Lagrangian (\ref{eq:32.08}) with the ${\rm SO}(3,1)$ gauge group, the scalar product (\ref{eq:40.06}) and the algebra element characterizing the particle at rest ${\cal C} = {\cal C}_J + {\cal C}_S$, ${\cal C}_J = m\, J_0$, ${\cal C}_S = s\, S_0$ (notice that $S_0 = P_0$) \cite{Meusburger:2009gy}. The Cartan connection $A_S = \omega^\mu J_\mu + e^\mu P_\mu$ remains flat outside the particle's singularity (in contrast to the spin connection $\omega^\mu$) and therefore we may follow Section 3.2 and decompose space into the particle disc and the asymptotic region, with the common boundary $\Gamma$. Using the group factorization (\ref{eq:40.05}) the Lagrangian can be converted to the boundary form
\begin{align}\label{eq:40.07}
L = \frac{k}{2\pi} \int_\Gamma \left<d\mathfrak{s}\, \mathfrak{s}^{-1} 
\mathfrak{j}^{-1} \dot{\mathfrak{j}} - d\bar{\mathfrak{s}}\, 
\bar{\mathfrak{s}}^{-1} \bar{\mathfrak{j}}^{-1} \dot{\bar{\mathfrak{j}}} + 
\frac{1}{k} {\cal C} d\phi \lb\bar{\mathfrak{s}}^{-1} \bar{\mathfrak{j}}^{-1} 
\dot{\bar{\mathfrak{j}}}\, \bar{\mathfrak{s}} + \bar{\mathfrak{s}}^{-1} 
\dot{\bar{\mathfrak{s}}}\rb\right>\,.
\end{align}
From the sewing condition (\ref{eq:32.13}) we find the expression for $d\mathfrak{s}\, \mathfrak{s}^{-1}$ which does not depend on $\mathfrak{s}$ and substituting it into (\ref{eq:40.07}) we obtain
\begin{align}\label{eq:40.08}
L = \frac{k}{2\pi} \int_\Gamma \left<\partial_0\lb\bar{\mathfrak{j}}^{-1} 
\mathfrak{j}\rb \mathfrak{j}^{-1} \bar{\mathfrak{j}} 
\lb d\bar{\mathfrak{s}}\, \bar{\mathfrak{s}}^{-1} - 
\bar{\mathfrak{s}}\, \frac{1}{k} {\cal C} d\phi\, \bar{\mathfrak{s}}^{-1}\rb + 
\frac{1}{k} {\cal C} d\phi\, \bar{\mathfrak{s}}^{-1} 
\dot{\bar{\mathfrak{s}}}\right>\,.
\end{align}
Unfortunately, it is rather difficult to derive the effective particle action from (\ref{eq:40.08}) in the way we did it for the Poincar\'{e} gauge group, even for small $\Lambda$, due to the complicated double product structure. 

On the other hand, if we take the limit $\Lambda \rightarrow 0$ then it is equivalent to the contraction of the de Sitter group to the Poincar\'{e} group and we obviously recover the Chern-Simons action for a particle in flat spacetime. In other words, in the contraction limit the ${\rm AN}(2)$ component of the factorization ${\rm AN}(2) \dcp {\rm SO}(2,1)$ is flattened out to the group of translations and (\ref{eq:40.05}) simplifies to (\ref{eq:32.16}). By analogy, we may consider \cite{Kowalski:2014dy} an alternative contraction of the gauge group, in which it is the Lorentz component of the de Sitter group that becomes Abelian, i.e.\! flattened out. To this end we first rescale the algebra generators to $\tilde J_\mu \equiv \sqrt{\Lambda}\, J_\mu$, $\tilde P_\mu \equiv 1/\sqrt{\Lambda}\, P_\mu$ and accordingly define $\tilde S_\mu \equiv \tilde P_\mu + \epsilon_{\mu 0\nu} \tilde J^\nu$. Then the commutators (\ref{eq:40.03}) are transformed into
\begin{align}\label{eq:40.09}
[\tilde J_\mu,\tilde J_\nu] & = \sqrt{\Lambda}\, 
\epsilon_{\mu\nu\sigma} \tilde J^\sigma\,, \qquad 
[\tilde J_\mu,\tilde S_\nu] = \sqrt{\Lambda}\, 
\epsilon_{\mu\nu\sigma} \tilde S^\sigma + \eta_{\nu 0} \tilde J_\mu - \eta_{\mu\nu} \tilde J_0\,, \nonumber\\ 
[\tilde S_\mu,\tilde S_\nu] & = \eta_{\mu 0} \tilde S_\nu - 
\eta_{\nu 0} \tilde S_\mu\,.
\end{align}
Taking the contraction limit $\Lambda \rightarrow 0$ we obtain the following algebra
\begin{align}\label{eq:40.10}
[\tilde J_\mu,\tilde J_\nu] = 0\,, \qquad 
[\tilde J_\mu,\tilde S_\nu] = \eta_{\nu 0} \tilde J_\mu - 
\eta_{\mu\nu} \tilde J_0\,, \qquad 
[\tilde S_\mu,\tilde S_\nu] = \eta_{\mu 0} \tilde S_\nu - 
\eta_{\nu 0} \tilde S_\mu\,.
\end{align}
The last commutator is still that of the $\mathfrak{an}(2)$ algebra, while the first one may naturally be treated as the commutator of the dual algebra $\mathfrak{an}(2)^*$ (see below). The full algebra (\ref{eq:40.10}) can also be seen as a modified version of the 2+1-dimensional Carroll algebra, as we will now discuss. 

Namely, the $d\!+\!1$-dimensional Carroll group ${\rm Carr}(d+1)$, generated by the corresponding algebra, can be regarded \cite{Duval:2014ce} as a subgroup of the Poincar\'{e} group ${\rm ISO}(d+1,1)$ that is the dual counterpart to the Galilei group ${\rm Gal}(d+1)$. On the other hand, ${\rm Carr}(d+1)$ was first introduced \cite{Bacry:1968ps} as the contraction of ${\rm ISO}(d,1)$ obtained by taking the limit of vanishing speed of light, in contrast to the Galilean case (where the speed of light goes to infinity). Such a limit corresponds to the situation in which all lightcones shrink to null worldlines and thus it may be called ultralocal. Consequently, it represents the hypothetical asymptotic silence scenario, in which spacetime points become causally disconnected, and therefore a perturbative expansion around the Carrollian limit is a potentially useful description of the strong curvature regime of general relativity \cite{Dautcourt:1998oy}. There is also some support for the analogous state of asymptotic silence in the context of quantum gravity, particularly in loop quantum cosmology \cite{Mielczarek:2012ay}. 

The Carroll algebra in 2+1 dimensions has the commutators
\begin{align}\label{eq:40.11}
[M,N_a] & = \epsilon_{0ab} N^b\,, & [N_a,N_b] & = 0\,, & 
[M,T_a] & = \epsilon_{0ab} T^b\,, & [N_a,T_b] & = \delta_{ab} T_0\,, \nonumber\\ 
[M,T_0] & = 0\,, & [N_a,T_0] & = 0\,, & [T_0,T_a] & = 0\,, & 
[T_a,T_b] & = 0\,,
\end{align}
where $M$, $N_a$, $T_a$, $T_0$, $a = 1,2$ generate, respectively, rotations, boosts and spatial and temporal translations. The difference with respect to the Poincar\'{e} algebra lies in the boost sector since the Carrollian boosts act only in the time direction. If we now denote the generators in (\ref{eq:40.10}) as $\tilde J_0 \equiv -T_0$, $\tilde J_a \equiv T_a$, $\tilde S_0 \equiv M$, $\tilde S_a \equiv N_a$ then it may be presented in the following form
\begin{align}\label{eq:40.12}
[M,N_a] & = N_a\,, & [N_a,N_b] & = 0\,, & 
[M,T_a] & = -T_a\,, & [N_a,T_b] & = \delta_{ab} T_0\,, \nonumber\\ 
[M,T_0] & = 0\,, & [N_a,T_0] & = 0\,, & 
[T_0,T_a] & = 0\,, & [T_a,T_b] & = 0\,.
\end{align}
Comparing (\ref{eq:40.12}) with (\ref{eq:40.11}) we note that indeed $J_\mu$'s, $\tilde S_0$ and $\tilde S_a$'s fulfil the roles of the generators of translations, rotations and Carrollian boosts but with the modified first and third commutator. We will see below that this similarity of the algebra (\ref{eq:40.10}) to (\ref{eq:40.11}) has the actual physical consequences. 

The group generated by the algebra (\ref{eq:40.10}) is the semidirect product ${\rm AN}(2) \lsp \mathfrak{an}(2)^*$, $\mathfrak{an}(2)^* \simeq \mathbbm{R}^3$. Namely, an element $\gamma \in {\rm AN}(2) \lsp \mathfrak{an}(2)^*$ factorizes into (cf.\! (\ref{eq:40.05}), (\ref{eq:32.16}))
\begin{align}\label{eq:40.13}
\gamma = \mathfrak{i}\, \mathfrak{s} = 
(1 + \iota^\mu \tilde J_\mu) (\xi_3 \mathbbm{1} + \xi^\nu \tilde S_\nu)\,,
\end{align}
where $\iota \equiv \iota^\mu \tilde J_\mu$, $\iota \in \mathfrak{an}(2)^*$, $\mathfrak{s} \in {\rm AN}(2)$ and ${\rm AN}(2)$ coordinates satisfy $\xi_3^2 - \frac{1}{4} \xi_0^2 = 1$. The group multiplication is given by
\begin{align}\label{eq:40.14}
\gamma_{(1)} \gamma_{(2)} = \lb 1 + \iota_{(1)} + {\rm Ad}(\mathfrak{s}_{(1)})\, 
\iota_{(2)}\rb \mathfrak{s}_{(1)} \mathfrak{s}_{(2)}
\end{align}
and the relation between the generators $\tilde J_\mu$ and $\tilde S_\mu$ in (\ref{eq:40.04}) becomes
\begin{align}\label{eq:40.14a}
\tilde J_\mu \tilde S_\nu = \frac{1}{2}\lb \eta_{\nu 0} \tilde J_\mu - 
\eta_{\mu\nu} \tilde J_0\rb\,.
\end{align}

Let us mention that in \cite{Meusburger:2006pe} there was derived (although in a complicated notation) the symplectic structure for particles coupled to the Chern-Simons theory with the gauge group of the form $G \lsp \mathfrak{g}^*$, where $G$ is an arbitrary Lie group. Instead, we will calculate the effective particle action for the ${\rm AN}(2) \lsp \mathfrak{an}(2)^*$ group, in a simple manner analogous to Section 3.2 and repeating what we did in \cite{Kowalski:2014dy}. To this end we need two remaining ingredients. Firstly, the scalar product (\ref{eq:40.06}) in terms of the generators $\tilde J_\mu$, $\tilde S_\mu$ still has the form
\begin{align}\label{eq:40.15}
\left<\tilde J_\mu \tilde S_\nu\right> = \eta_{\mu\nu}\,, \qquad 
\left<\tilde J_\mu \tilde J_\nu\right> = \left<\tilde S_\mu \tilde S_\nu\right> 
= 0\,.
\end{align}
Secondly, since we want ${\rm AN}(2)$ to become the particle's momentum space and $\mathfrak{an}(2)^*$ its position space we have to exchange the algebra elements encoding particle's mass and spin so that ${\cal C}_{\tilde J} = s\, \tilde J_0$ and ${\cal C}_{\tilde S} = m\, \tilde S_0$. Then from (\ref{eq:32.10}) it follows that either the interpretation of geometrical variables will change or mass will become a source of torsion and spin a source of curvature (which could be seen as the so-called semidualization of the theory, see \cite{Osei:2012oy}). This issue remains to be explored. 

Taking all the above into account we may return to the Lagrangian (\ref{eq:40.08}). We plug the factorization (\ref{eq:40.13}) into the sewing condition (\ref{eq:32.13}) and find the relation
\begin{align}\label{eq:40.16}
\mathfrak{s} = \bar{\mathfrak{s}}\, e^{-\frac{1}{k} {\cal C}_{\tilde S} \phi} 
\mathfrak{v}^{-1}\,,
\end{align}
where $N = (1 + n)\, \mathfrak{v}$, $\mathfrak{v} \in {\rm AN}(2)$, $n \in \mathfrak{an}(2)^*$. Then we can also find the second relation
\begin{align}\label{eq:40.17}
\bar{\mathfrak{i}}^{-1} \mathfrak{i} = e^{-\frac{1}{k} {\cal C}_{\tilde J} \phi} 
\mathfrak{s}\, (1 - n)\, \mathfrak{s}^{-1}\,,
\end{align}
where $\mathfrak{s}$ is determined by (\ref{eq:40.16}). Substituting (\ref{eq:40.17}) into (\ref{eq:40.08}) and rearranging the derivatives we arrive at
\begin{align}\label{eq:40.18}
L = \frac{k}{2\pi} \int_\Gamma d\left<\partial_0\lb\bar{\mathfrak{s}}\, 
e^{-\frac{1}{k} {\cal C}_{\tilde S} \phi} \mathfrak{v}^{-1} n\, \mathfrak{v}\rb 
e^{\frac{1}{k} {\cal C}_{\tilde S} \phi} \bar{\mathfrak{s}}^{-1} + 
\frac{1}{k} {\cal C}_{\tilde J} d\phi\, \bar{\mathfrak{s}}^{-1} 
\dot{\bar{\mathfrak{s}}}\right>\,.
\end{align}
We integrate it over $\phi \in [0,2\pi]$, evaluate the scalar product and eventually obtain the particle Lagrangian
\begin{align}\label{eq:40.19}
L = \kappa \lb\dot\Pi\, \Pi^{-1}\rb_\mu x^\mu + 
s \lb\bar{\mathfrak{s}}^{-1} \dot{\bar{\mathfrak{s}}}\rb_0\,.
\end{align}
Its form is analogous to the Lagrangian (\ref{eq:32.21}) of a gravitating particle in flat spacetime but the momentum $\Pi$ is here an element of the ${\rm AN}(2)$ group
\begin{align}\label{eq:40.20}
\Pi \equiv \bar{\mathfrak{s}}\, e^{\frac{m}{\kappa} \tilde S_0} 
\bar{\mathfrak{s}}^{-1}
\end{align}
and the position $x \equiv \bar{\mathfrak{s}}\, \mathfrak{v}^{-1} n\, \mathfrak{v}\, \bar{\mathfrak{s}}^{-1}$, $x = x^\mu \tilde J_\mu$ (where $\bar{\mathfrak{s}} = \bar{\mathfrak{s}}(\phi = 0) = \bar{\mathfrak{s}}(\phi = 2\pi)$). Since, as we discussed in Section 1.3, the ${\rm AN}(2)$ manifold is equivalent to 2+1-dimensional elliptic de Sitter space we obtained the momentum space with a positive curvature, in contrast to a negative curvature in the case of (\ref{eq:32.21}). We may also fix the gauge at the boundary $\Gamma$ via $\gamma(\phi = 0) = 1$, which leads to $\bar\iota(\phi = 0,2\pi) = n$ and $\bar{\mathfrak{s}}(\phi = 0,2\pi) = \mathfrak{v}$. Then we have $\Pi = \mathfrak{v}\, e^{\frac{m}{\kappa} \tilde S_0} \mathfrak{v}^{-1}$, $x = n$. Furthermore, we find that the holonomy of the connection $A_S$ along $\Gamma$ is given by (cf.\! (\ref{eq:32.23}))
\begin{align}\label{eq:40.21}
{\cal P}\, e^{-\int_\Gamma A_S} = \gamma(\phi = 0)\, \gamma^{-1}(\phi = 2\pi) 
= \lb 1 + \lb 1 - {\rm Ad}(\Pi)\rb x + \tfrac{1}{\kappa} {\cal C}_{\tilde J}\rb 
\Pi\,.
\end{align}
In particular, $\mathfrak{s}(\phi = 0)\, \mathfrak{s}^{-1}(\phi = 2\pi) = \Pi$ and thus $\Pi$ is indeed the particle's momentum. If we treat $\Pi$ as an element of $SO(2,1)$, according to the isomorphism (\ref{eq:13.07}), then it is a Lorentz transformation conjugate to the boost by $\frac{m}{\kappa}$. In the gravitational Lagrangian (\ref{eq:32.21}) such a holonomy would describe \cite{Baez:2007ey} a tachyon with (imaginary) mass $im$. However, in the case of (\ref{eq:40.19}) the gauge group is not the Poincar\'{e} group and the interpretation of the obtained particle model is not obvious. The Lagrangian (\ref{eq:40.19}) rewritten below in components will actually turn out to be rather peculiar. Meanwhile, by analogy with (\ref{eq:32.23a}), we interpret
\begin{align}\label{eq:40.21a}
{\rm I} \equiv \kappa\, ({\rm Ad}(\Pi^{-1}) - 1)\, x + {\cal C}_{\tilde J}\,.
\end{align}
as the particle's generalized angular momentum. 

Let us also observe that if we calculate a variation $\delta\mathfrak{v} = \varepsilon\, \mathfrak{v}$, $\delta\mathfrak{v}^{-1} = -\mathfrak{v}^{-1} \varepsilon$ of the spin term in (\ref{eq:40.19}) then we obtain the total time derivative $\delta \left<{\cal C}_{\tilde J} \mathfrak{v}^{-1} \dot{\mathfrak{v}}\right> = \partial_0 \left<{\cal C}_{\tilde J} \varepsilon\right>$ (since ${\cal C}_{\tilde J}$ commutes with $\mathfrak{v}$) and hence the latter does not contribute to the equations of motion. Therefore in what follows we will restrict to the spinless case $s = 0$. For the further discussion of the particle's properties we may switch to the plane-wave parametrization of ${\rm AN}(2)$ elements $\mathfrak{s} = e^{\sigma^a \tilde S_a} e^{\sigma^0 \tilde S_0}$, as in (\ref{eq:13.02}), which is connected to the parametrization $\mathfrak{s} = \xi_3 + \xi^\mu \tilde S_\mu$ used above by $\sigma^0 = 2\log(\xi_3 + \frac{1}{2} \xi^0)$, $\sigma^a = (\xi_3 + \frac{1}{2} \xi^0)\, \xi^a$. In particular, we write
\begin{align}\label{eq:40.22}
\Pi = e^{p^a/\kappa\, \tilde S_a} e^{p^0/\kappa\, \tilde S_0}\,, \qquad 
\mathfrak{v} = e^{\upsilon^a\, \tilde S_a} e^{\upsilon^0\, \tilde S_0}\,.
\end{align}
Then from (\ref{eq:40.20}) it follows that
\begin{align}\label{eq:40.23}
p^0 = m\,, \qquad p^a = \kappa \lb 1 - e^{\frac{m}{\kappa}}\rb \upsilon^a
\end{align}
and thus the energy of the particle is fixed to be its rest energy. Imposing this as a constraint with the Lagrange multiplier $\lambda$ we may rewrite the action determined by the Lagrangian (\ref{eq:40.19}) as
\begin{align}\label{eq:40.24}
S = -\int\! dt\ \lb x^0 \dot p_0 + x^a \dot p_a - \kappa^{-1} x^a p_a \dot p_0 + 
\lambda \lb p_0^2 - m^2\rb\rb\,.
\end{align}
Without the constraint it would be the off-shell action of a particle with the $\kappa$-Poincar\'{e} symmetry \cite{Imilkowska:2006dy}. However, the constraint turns it into the action describing a particle with the $\kappa$-deformed Carroll symmetry, as the algebra (\ref{eq:40.12}) was implying. The action of an ordinary Carroll particle \cite{Bergshoeff:2014ds} can be recovered in the limit $\kappa \rightarrow \infty$, which gives 
\begin{align}\label{eq:40.25}
S = -\int\! dt\ \lb x^0 \dot p_0 + x^a \dot p_a + 
\lambda \lb p_0^2 - m^2\rb\rb\,.
\end{align}
The equations of motion following from (\ref{eq:40.24}) are not $\kappa$-deformed but the same as the ones of (\ref{eq:40.25}) and have the form
\begin{align}\label{eq:40.26}
\dot x^0 = 2\lambda\, m\,, \qquad \dot x^a = 0\,, \qquad \dot p_\mu = 0\,,
\end{align}
which shows that the particle is always at rest.\footnote{This explains the name of the Carroll group \cite{Carroll:1871}: ``Now, here, you see, it takes all the running you can do, to keep in the same place."} The action (\ref{eq:40.24}) is invariant under infinitesimal $\kappa$-deformed Carroll transformations, which include ordinary rotations
\begin{align}\label{eq:40.27}
\delta x^a = \rho\, \epsilon^a_{\ b} x^b\,, \qquad 
\delta p_a = \rho\, \epsilon_a^{\ b} p_b\,, \qquad 
\delta x^0 = \delta p_0 = 0\,,
\end{align}
deformed Carrollian boosts
\begin{align}\label{eq:40.28}
\delta x^0 = \lb1 + \kappa^{-1} p_0\rb \lambda_a x^a\,, \qquad 
\delta p_a = -\lambda_a p_0\,, \qquad 
\delta x^a = \delta p_0 = 0\,,
\end{align}
deformed translations
\begin{align}\label{eq:40.29}
\delta x^0 = \alpha^0\,, \qquad 
\delta x^a = e^{p_0/\kappa} \alpha^a\,, \qquad 
\delta p_\mu = 0
\end{align}
and spatial conformal transformations
\begin{align}\label{eq:40.30}
\delta x^a = \eta\, x^a\,, \qquad 
\delta p_a = -\eta\, p_a\,, \qquad 
\delta x^0 = \delta p_0 = 0\,,
\end{align}
where $\rho$, $\lambda_a$, $\alpha^\mu$, $\eta$ are parameters of the respective symmetry transformations. On the other hand, the particle's angular momentum (\ref{eq:40.21a}) in components ${\rm I} = I^0 \tilde J_0 + I^a \tilde J_a$ is given by
\begin{align}\label{eq:40.30a}
I^0 = - x_a p^a + s\,, \quad I^a = \kappa \lb e^{p_0/\kappa} - 1\rb x^a
\end{align}
and in the limit $\kappa \rightarrow \infty$ we obtain $I^0 = - x_a p^a + s$, $I^a = p_0 x^a$, which are not the ordinary expressions (although they satisfy the usual relation $p_\mu I^\mu = ms$). The reason is that the difference between the gauge algebra (\ref{eq:40.12}) and the Carroll algebra (\ref{eq:40.11}) concerns the generator of rotations. 

Let us also briefly consider a system of multiple particles, similarly to what we did in Section 3.3. We again start from the Lagrangian (\ref{eq:33.01}) (where we should take into account the appropriate boundary conditions \cite{Meusburger:2006pe}) and decompose space into the discs with particles and the empty polygon. Following the single particle case we derive the effective Lagrangian for every particle
\begin{align}\label{eq:40.31}
L_i = \kappa \lb\dot{\bar\Pi}_i \bar\Pi_i^{-1}\rb_\mu \bar x_i^\mu + 
s_i \lb\bar{\mathfrak{s}}_i^{-1} \dot{\bar{\mathfrak{s}}}_i\rb_0\,,
\end{align}
where we denote $\bar\Pi_i \equiv \bar{\mathfrak{s}}_i e^{\frac{1}{\kappa} m_i \tilde S_0} \bar{\mathfrak{s}}_i^{-1}$, $\bar x_i \equiv \bar{\mathfrak{s}}_i \mathfrak{v}_i^{-1} n_i \mathfrak{v}_i \bar{\mathfrak{s}}_i^{-1}$, while $\bar{\mathfrak{s}}_i = \bar{\mathfrak{s}}_i(\phi_i = 0,2\pi)$. Imposing the continuity conditions $\gamma(\phi_{i+1} = 0) = \gamma(\phi_i = 2\pi)$ and partially fixing the gauge $\gamma(\phi_1 = 0) = 1$ we find the sequence of conditions
\begin{align}\label{eq:40.32}
\mathfrak{v}_1\, \bar{\mathfrak{s}}_1^{-1} = 1\,, \qquad 
\mathfrak{v}_2\, \bar{\mathfrak{s}}_2^{-1} = \Pi_1\,, \qquad 
\mathfrak{v}_3\, \bar{\mathfrak{s}}_3^{-1} = \Pi_2 \Pi_1\,, \qquad \ldots\,,
\end{align}
where $\Pi_i = \mathfrak{v}_i e^{\frac{1}{\kappa} m_i \tilde S_0} \mathfrak{v}_i^{-1}$. Finally, we use (\ref{eq:40.32}) to replace variables $\bar{\mathfrak{s}}_i$ in (\ref{eq:40.31}) with $\mathfrak{v}_i$ and summing over $i$ we obtain the effective $n$-particle Lagrangian
\begin{align}\label{eq:40.33}
L_{(n)} = L_{(n-1)} + \kappa \lb\dot\Pi_n \Pi_n^{-1}\rb_\mu x_n^\mu + 
s_n \lb\mathfrak{v}_n^{-1} \dot{\mathfrak{v}}_n\rb_0 + \nonumber\\ 
\kappa \lb\Pi_n \partial_0(\Pi_{n-1} \ldots \Pi_1) \Pi_1^{-1} \ldots 
\Pi_n^{-1}\rb_\mu x_n^\mu - \nonumber\\ 
\kappa \lb\partial_0(\Pi_{n-1} \ldots \Pi_1) \Pi_1^{-1} \ldots 
\Pi_{n-1}^{-1}\rb_\mu x_n^\mu + \nonumber\\ 
s_n \lb\mathfrak{v}_n^{-1} \Pi_{n-1} \ldots \Pi_1 
\partial_0(\Pi_1^{-1} \ldots \Pi_{n-1}^{-1}) \mathfrak{v}_n\rb_0\,,
\end{align}
where $x_i = n_i$. In particular, for a pair of particles the Lagrangian (\ref{eq:40.33}) has the form
\begin{align}\label{eq:40.34}
L_{(2)} = \kappa \lb\dot\Pi_1 \Pi_1^{-1}\rb_\mu x_1^\mu + 
\kappa \lb\dot\Pi_2 \Pi_2^{-1}\rb_\mu x_2^\mu + 
\kappa \lb\Pi_2 \dot\Pi_1 \Pi_1^{-1} \Pi_2^{-1} - 
\dot\Pi_1 \Pi_1^{-1}\rb_\mu x_2^\mu + \nonumber\\ 
s_1 \lb\mathfrak{v}_1^{-1} \dot{\mathfrak{v}}_1\rb_0 + 
s_2 \lb\mathfrak{v}_2^{-1} \dot{\mathfrak{v}}_2\rb_0 + 
s_2 \lb\mathfrak{v}_2^{-1} \Pi_1 \dot\Pi_1^{-1} \mathfrak{v}_2\rb_0\,.
\end{align}
Similarly as in Section 3.3 the total momentum of the particles is given by the non-Abelian holonomy
\begin{align}\label{eq:40.34a}
\mathfrak{s}(\phi_1 = 0)\, \mathfrak{s}^{-1}(\phi_n = 2\pi) = \Pi_n \ldots \Pi_1 
\equiv \Pi\,,
\end{align}
which is invariant under a braiding of individual holonomies. Let us also note that in the multiparticle Lagrangian (\ref{eq:40.33}) spin terms are no longer independent from the other ones, in contrast to (\ref{eq:40.19}). More properties of these particles remain to be studied. 

To conclude the current Chapter let us try to understand the relevance of our results for gravity in a higher number of spacetime dimensions. Namely, three-dimensional gravity in the Chern-Simons formulation is an example of the so-called BF theory. General relativity in four dimensions can also be expressed as the appropriate BF theory but with the correction term that breaks the full gauge symmetry of the topological field down to the Lorentz symmetry of gravity, see e.g.\! \cite{Freidel:2003ge,Freidel:2005qs}. If we couple particles to such a theory then in the topological limit they can be described in the same way as in three-dimensional gravity \cite{Freidel:2006pd,Kowalski:2008ey}. On the other hand, in contrast to (\ref{eq:32.21}), the Lagrangian (\ref{eq:40.19}) can be naturally generalized to $d\!+\!1$ dimensions, $d > 2$ if we replace the group ${\rm AN}(2)$ with ${\rm AN}(d)$. Since there always exists the (local) Iwasawa decomposition of the de Sitter group into the Lorentz and ${\rm AN}(d)$ groups such a Lagrangian could in principle arise as a contraction of the BF theory with the ($d\!+\!2$-dimensional) de Sitter gauge group. However, only in 2+1 dimensions the number of ${\rm AN}(d)$ generators is the same as for both the translations and the Lorentz group, which allows us to obtain the algebra (\ref{eq:40.12}).

\chapter{Conical defects in higher dimensions}
In Section 3.1 we explained that point particle solutions of three-dimensional gravity (with vanishing cosmological constant) have the geometry of conical defects in flat spacetime and that the extended momentum space of a gravitating particle is a curved manifold. Such topological defects of codimension 2 can be naturally generalized to higher dimensional spacetimes, where they have momenta with similar properties and therefore may be worth to study from our perspective, which gave us the motivation for \cite{Arzano:2015se}. In particular, to obtain a conical defect in 3+1 dimensions we simply replace the pointlike curvature singularity with a singular straight line. The linear conical defects are known as cosmic strings and may actually represent real physical objects, with some finite dimensions. It was first suggested \cite{Kibble:1976ts} that they could form during a spontaneous gauge symmetry breaking in the early universe. Consequently, they were considered as one of the possible sources of primordial density fluctuations \cite{Hindmarsh:1995cs} and observations of the cosmic microwave background set the specific bounds on their contribution, see e.g.\! \cite{Bevis:2010cd}. Open and closed cosmic strings can also be created in inflation models constructed in the framework of string theory \cite{Sakellariadou:2009cs}. On the other hand, G.\! 't~Hooft in \cite{Hooft:2008ay} employed straight strings as elementary ingredients of his model of piecewise flat gravity. 

The exact 3+1-dimensional counterpart of a point particle in 2+1 spacetime dimensions is an infinitely long and thin, straight cosmic string. Such ideal conical defects can be straightforwardly generalized to any number of dimensions if a string is replaced by a brane, i.e.\! a hyperplane of codimension 2. By analogy, we will call them cosmic branes. As we know from Section 3.1, the geometry of a conical defect is obtained by cutting out a wedge from Minkowski space and identifying the wedge's faces by a Lorentz transformation conjugate to the rotation by the deficit angle characterizing the defect. Let us remark that this construction can be generalized \cite{Puntigam:1996vs} to include identifications of the faces of a cut (which not necessarily has the form of a wedge) by a general Poincar\'{e} transformation, which results in different types of topological defects that one can classify as dislocations or disclinations if we apply the terminology used in the condensed matter physics. An example in 2+1 dimensions is given by the spinning particle (\ref{eq:31.09}). However, below we will restrict to the standard defects.

\section{Massive defects}
Generalizing the case of 2+1 dimensions (\ref{eq:31.01}) and 3+1 dimensions \cite{Meent:2011ps} with vanishing cosmological constant, we may write \cite{Arzano:2015se} in cylindrical coordinates the metric of a single static conical defect in 4+1-dimensional spacetime (or any other dimension)
\begin{align}\label{eq:51.01}
ds^2 = -dt^2 + dz^2 + dw^2 + dr^2 + \lb 1 - 4\mu\rb^2 r^2 d\phi^2\,.
\end{align}
It describes an infinite, flat cosmic brane given by the $zw$-plane, which is the two-dimensional vertex of a defect with the deficit angle $\delta = 8\pi \mu$, $\mu \in (0,\frac{1}{4})$. As we discussed in Section 3.1, in 2+1 dimensions the deficit angle is proportional to the product $\mu \equiv Gm$, which can be regarded as the dimensionless rest energy of the particle. In $n\!+\!1$ dimensions the Newton's constant $G$ has the dimension of inverse mass times length to the power of $n - 2$. Therefore, by analogy, we may introduce the dimensionless rest energy density $\mu = G\rho$, where $\rho$ is mass per unit of volume of the defect's hyperplane (e.g.\! mass per the string's length for 3+1 dimensions). Similarly to the three-dimensional case we find that the Riemann curvature vanishes outside the defect's world-volume. Thus the parallel transport of a vector along an arbitrary loop around $r = 0$ is completely determined by the conical singularity. The result of the parallel transport around a given loop $\gamma(\lambda)$ can be described by the holonomy of the Levi-Civita connection, which is the path-ordered exponential
\begin{align}\label{eq:51.02}
h(\gamma)^\alpha_{\ \beta} = {\cal P} \exp\lb -\int_\gamma 
\Gamma^\alpha_{\ \zeta\beta} \frac{d\gamma^\zeta}{d\lambda} d\lambda\rb\,,
\end{align}
where $\Gamma^\alpha_{\ \zeta\beta}$ are Christoffel symbols for the metric (\ref{eq:51.01}). Notice that in four-dimensional space it is possible to take a loop around a plane since to close a path around a given hyperplane we need at least two directions orthogonal to it. 

Let us remark that one can measure both momentum and position of a conical defect by using a generalization of the Lorentz holonomy (\ref{eq:51.02}) to the Poincar\'{e} holonomy \cite{Hooft:2008ay,Meent:2011ps}. The latter is given by the parallel transport of the whole coordinate frame instead of an individual vector, which is possible due to the flatness of conical spacetime. If the defect is not located at the frame's origin the resulting Poincar\'{e} holonomy is the combination of a Lorentz transformation and a translation corresponding to the defect's displacement. However, since we are interested in the properties of momentum space we will restrict to the Lorentz holonomies. We first calculate the holonomy (\ref{eq:51.02}) for the static metric (\ref{eq:51.01}). This turns out to be the easiest if we take a circular loop parametrized by $r = (1 - 4 \mu)^{-1}$, $\phi = 2\pi \lambda$ and $t,z,w = 0$. Then the path ordering in (\ref{eq:51.02}) is trivial, the coordinate basis is given by Cartesian coordinates $x = r \cos\phi$, $y = r \sin\phi$, $t,z,w$ and we obtain
\begin{align}\label{eq:51.03}
h \equiv R(\delta) = \lb
\begin{array}{cccc}
1\! & 0\! & 0\! & {\bf 0} \\ 
0\! & \cos\delta\! & -\sin\delta\! & {\bf 0} \\ 
0\! & \sin\delta\! & \cos\delta\! & {\bf 0} \\ 
{\bf 0}\! & {\bf 0}\! & {\bf 0}\! & \mathbbm{1}
\end{array}
\rb
\end{align}
(where $\mathbbm{1}$ denotes $2 \times 2$ identity matrix). Thus, as expected, the holonomy of a massive defect is an elliptic Lorentz transformation i.e.\! a rotation by the deficit angle $\delta = 8\pi\mu$ around the origin in the $xy$-plane, which leaves invariant the defect's world-volume. This is a transformation that identifies the faces of the wedge of a defect at rest. 

The metric of a moving defect can be obtained by expressing the static metric (\ref{eq:51.01}) in Cartesian coordinates and performing a boost in a direction lying in the $xy$-plane. Obviously (\ref{eq:51.01}) is invariant under boosts in directions parallel to the brane. For simplicity let us consider a boost in the $x$ direction
\begin{align}\label{eq:51.04}
t \mapsto t \cosh\chi + x \sinh\chi\,, \qquad 
x \mapsto x \cosh\chi + t \sinh\chi\,,
\end{align}
with the rapidity parameter $\chi$. Then making a transformation to lightcone coordinates $u = (x - t)/\sqrt{2}$, $v = (x + t)/\sqrt{2}$ we obtain the metric
\begin{align}\label{eq:51.05}
ds^2 = 2 du dv + dy^2 + dz^2 + dw^2 - \nonumber\\ 
\lb 1 - (1 - 4\mu)^2\rb \frac{(e^\chi (u dy - y du) + 
e^{-\chi} (v dy - y dv))^2} {e^{2\chi} u^2 + e^{-2\chi} v^2 + 2 (uv + y^2)}\,,
\end{align}
which describes a cosmic brane travelling with the velocity $V = \tanh\chi$ in the $x$ direction. After some calculations it can be shown \cite{Meent:2011ps} that the deficit angle $\delta^\prime$ of such a moving defect and the deficit angle $\delta$ of the defect at rest are connected by the relation
\begin{align}\label{eq:51.06}
\tan(\delta^\prime/2) = \tan(\delta/2) \cosh\chi
\end{align}
and thus $\delta^\prime$ becomes wider than $\delta$. Let us observe that in the limit of small $\delta$, $\delta^\prime$ (\ref{eq:51.06}) simplifies to
\begin{align}\label{eq:51.07}
\delta^\prime = \delta \cosh\chi\,,
\end{align}
which has the same form as the familiar expression for energy of an ordinary relativistic particle. Namely, the defect's rest energy density is ${\cal E}_0 = \mu \sim \delta$, while the boost factor $\cosh\chi = (1 - V^2)^{-1/2}$ and hence from (\ref{eq:51.07}) we find that the total energy density $(1 - V^2)^{-1/2} {\cal E}_0 = {\cal E} \sim \delta^\prime$. 

One may note that due to the nontrivial parallel transport in spacetime with a conical defect there appears an ambiguity in the direction of a boost in different inertial frames. The problem can be resolved if the frame in which we define the boost is always chosen in a consistent way. On the other hand, one can avoid the whole issue that conical spacetime is not asymptotically Minkowski space by boosting the holonomy (\ref{eq:51.03}) instead of the metric. Namely, a boost $B(\chi)$ acting on the rotation $R(\delta)$ by the conjugation $B^{-1}(\chi) R(\delta) B(\chi)$ gives the Lorentz group element
\begin{align}\label{eq:51.08}
R(\delta,\chi) = \lb
\begin{array}{cccc}
1 + 2\sinh^2\chi \sin^2\tfrac{\delta}{2}\! & -\sinh(2\chi) 
\sin^2\tfrac{\delta}{2}\! & -\sinh\chi \sin\delta\! & {\bf 0} \\ 
\sinh(2\chi) \sin^2\tfrac{\delta}{2}\! & 1 - 2\cosh^2\chi 
\sin^2\tfrac{\delta}{2}\! & -\cosh\chi \sin\delta\! & {\bf 0} \\ 
-\sinh\chi \sin\delta\! & \cosh\chi \sin\delta\! & \cos\delta\! & 
{\bf 0} \\ 
{\bf 0}\! & {\bf 0}\! & {\bf 0}\! & \mathbbm{1}
\end{array}
\rb\,.
\end{align}
The obtained holonomy is an element of the conjugation class of rotations by the angle $\delta$, which completely characterizes the defect with the static metric (\ref{eq:51.01}).

\section{Massless defects}
Conical defects that we discussed so far were timelike objects, while now we will consider \cite{Arzano:2015se} lightlike (i.e.\! massless) defects. The metric of such a defect can be obtained as the theoretical limit of a moving massive defect (\ref{eq:51.04}) boosted to the speed of light. To this end we may follow the case of a cosmic string \cite{Meent:2013gs} and use the prescription that was first introduced by Aichelburg and Sexl \cite{Aichelburg:1971oe} to derive the gravitational field of a photon from the Schwarzschild solution and later applied to other singular sources, particularly in the study of impulsive gravitational waves \cite{Lousto:1991gs,Barrabes:2002as}. The method consists in taking the limit of the rapidity $\chi \rightarrow \infty$ while the laboratory energy density of the defect $\varrho$ is kept fixed. By a straightforward calculation it can be shown that $\varrho = 8\pi\mu \cosh\chi$ and with the help of the distributional identity
\begin{align}\label{eq:52.01}
\lim_{b \rightarrow 0} \frac{b}{a^2 + b^2} = \pi \delta(a)
\end{align}
we find that in the Aichelburg-Sexl limit the metric (\ref{eq:51.05}) becomes
\begin{align}\label{eq:52.02}
ds^2 = 2 du dv + dy^2 + dz^2 + dw^2 - \sqrt{2}\varrho |y| \delta(u) du^2\,.
\end{align}
The delta function above may seem problematic but distributional solutions of Einstein equations are actually well known and understood, see \cite{Steinbauer:2006ty}. The metric (\ref{eq:52.02}) describes a massless cosmic brane given by the $zw$-plane which moves along the null direction $v$ in the surrounding flat spacetime. Therefore, similarly to massive defects, such a massless brane can be completely characterized (up to a translation) by the holonomy (\ref{eq:51.02}) of an arbitrary loop around the curvature singularity at $u,y = 0$. In lightcone coordinates it is convenient to choose the square loop with $u,y \in [-1,1]$ \cite{Meent:2013gs}. Then the path ordering is trivial and after the resulting holonomy is transformed to Cartesian coordinates we eventually obtain
\begin{align}\label{eq:52.03}
h \equiv P(\varrho) = \lb
\begin{array}{cccc}
1 + \tfrac{\varrho^2}{2}\! & -\tfrac{\varrho^2}{2}\! & -\varrho\! & {\bf 0} \\ 
\tfrac{\varrho^2}{2}\! & 1 - \tfrac{\varrho^2}{2}\! & -\varrho\! & {\bf 0} \\ 
-\varrho\! & \varrho\! & 1\! & {\bf 0} \\ 
{\bf 0}\! & {\bf 0}\! & {\bf 0}\! & \mathbbm{1}
\end{array}
\rb\,.
\end{align}
It is a parabolic Lorentz transformation i.e.\! a null rotation by the (deficit) angle $\varrho$ in the $uy$-plane, which has the invariant $vzw$-hyperplane. 

The metric of a massless defect can also be derived in an alternative way, introduced for a cosmic string \cite{Meent:2013gs}, which does not lead to the ambiguities associated with performing a boost in conical spacetime. As we mentioned in the previous Section, one may directly consider a boost of the holonomy of a static defect, given by the elliptic Lorentz group element (\ref{eq:51.03}). If we take the Aichelburg-Sexl limit of the boosted holonomy (\ref{eq:51.08}) we find that the result is the null rotation (\ref{eq:52.03}). Then the idea is to reconstruct the metric which is characterized by such a holonomy, following the case of a massive defect (\ref{eq:51.01}). For a massless defect the geometry has to be similar but the deficit angle should be cut out from spacetime in a null hyperplane. To this end we need the lightcone version of cylindrical coordinates, which can be obtained from usual lightcone coordinates by a transformation $q = v + \tfrac{1}{2} y^2 u^{-1}$, $\varphi = y u^{-1}$. The angular-like coordinate $\varphi$ has the infinite range and therefore the cut can not correspond to a simple rescaling of it, as it is the case for $\phi$ in (\ref{eq:51.01}). Thus we rescale $\varphi$ only in the region $u > 0$, writing the metric in the form \cite{Meent:2013gs} 
\begin{align}\label{eq:52.04}
ds^2 = 2 du dq + dz^2 + dw^2 + \lb 1 - f(\varphi) \Theta(u)\rb^2 u^2 
d\varphi^2\,, \quad 
\int_{-\infty}^{\infty}\! d\varphi\ f(\varphi) = \sqrt{2} \varrho\,,
\end{align}
where the smooth function $f(\varphi)$, $f(\varphi) < 1$ with the compact support is used as an analogue of the scaling factor $4\mu$ in (\ref{eq:51.01}). It can be verified that this metric has the same holonomy as (\ref{eq:52.02}) and hence they are equivalent. An additional advantage of (\ref{eq:52.04}) is that, in comparison to (\ref{eq:52.02}), it is more regular at the hypersurface $u = 0$. This also allows us to see that a massless conical defect is not accompanied by gravitational waves, as it was once regarded. 

To obtain an intuitive picture of massless defects let us now consider the case of 3+1 dimensions and describe cosmic strings in terms of spacetime vectors. Any lightlike string can be completely specified by 4 parameters \cite{Hooft:2008ay}. Indeed, one has to choose a parabolic angle $\varrho$, which carries the defect's energy density as well as a null vector ${\bf n}$, which is the defect's direction of propagation and a spacelike vector ${\bf d}$, which is the direction of the defect's spatial extension, satisfying the orthogonality condition ${\bf n} \cdot {\bf d} = 0$ (since a defect is invariant under boosts acting along ${\bf d}$). ${\bf n}$ and ${\bf d}$ can also be arbitrarily rescaled. In total it gives $2 + 2 - 1 + 1 = 4$ independent parameters for the possible defects, or the corresponding holonomies. Thus the space of momenta, i.e.\! holonomies, of a massless string is bigger than the three-dimensional momentum space (constrained by the norm) of an ordinary null particle in 3+1-dimensional spacetime. The reason is that conical defects in higher dimensional spacetimes are extended objects and therefore their holonomies depend on the defect's orientation in space. 

The above space of holonomies can be restricted in two natural ways. Firstly, we may fix the spatial orientation ${\bf d}$. However, then the kinematics of the defect reduces to that of a particle in 2+1 dimensions (since it can only move in a plane orthogonal to ${\bf d}$). Secondly, we may fix the direction of motion ${\bf n}$. It turns out that the latter case is more interesting. In order to discuss it in detail let us choose a complete set of orthogonal null vectors ${\bf n}_{(x)} = (1,1,0,0)$, ${\bf n}_{(y)} = (1,0,1,0)$, ${\bf n}_{(z)} = (1,0,0,1)$ as well as a set of orthonormal spacelike vectors ${\bf d}_{(x)} = (0,1,0,0)$, ${\bf d}_{(y)} = (0,0,1,0)$, ${\bf d}_{(z)} = (0,0,0,1)$. Assume that we fix the direction of motion of a string as ${\bf n} = {\bf n}_{(j)}$, $j = x,y,z$. Then the string's spatial orientation is a linear combination of two vectors ${\bf d}_{(l)}$, $l \neq j$. The space of its holonomies is the subgroup of the Lorentz group ${\rm SO}(3,1)$ formed by null rotations with the generators $X^{(j)}_l = J_{tl} + J_{jl}$. For convenience let us replace (in an arbitrary order) the label $l$ with $a = 1,2$. The holonomy of a defect specified by ${\bf n}_{(j)}$, ${\bf d}_{(a)}$ can be written as the group element
\begin{align}\label{eq:52.05}
h^{(j)}_a = e^{i k_a X^{(j)}_a}
\end{align}
(no summation over $a$), where the parabolic angle $\varrho = k_a$. In the case of a defect with an arbitrary ${\bf d}$ for a given ${\bf n}_{(j)}$ the holonomy is naturally the product
\begin{align}\label{eq:52.06}
h^{(j)}_1 h^{(j)}_2 = e^{i k^a X^{(j)}_a}\,.
\end{align}
By a straightforward calculation we can show that the parabolic angle is given by $\varrho = \sqrt{k_1^2 + k_2^2}$, while the spacelike vector is a linear combination ${\bf d} = k_2 (k_2 - k_1)\, {\bf d}_{(1)} + k_1 (k_1 - k_2)\, {\bf d}_{(2)}$. Thus both the defect's energy density and its spatial orientation are now expressed in terms of the Lorentz group coordinates $k_1,k_2$. This agrees with the counting of degrees of freedom done above since fixing ${\bf n}$ reduces the number of free parameters to 2. 

Meanwhile, from Section 1.3 we know that the Abelian subgroup of the ${\rm AN}(n-1)$ group is formed by parabolic Lorentz transformations. Furthermore, together with (\ref{eq:13.07}) we can write down $n$ independent representations of the $\mathfrak{an}(n-1)$ algebra in the $\mathfrak{so}(n,1)$ algebra, labelled by $b = 1,\ldots,n$,
\begin{align}\label{eq:52.07}
X^{(b)}_0 = \frac{1}{\kappa} J_{0b}\,, \qquad X^{(b)}_a = 
\frac{1}{\kappa} \lb J_{0c} + J_{bc}\rb\,, \quad a = 1,\ldots,n-1\,,
\end{align}
where $c = a$ for $c < b$ and $c = a + 1$ otherwise, i.e.\! $c \neq b$. Thus if we set $\kappa = 1$ then the space of holonomies of a massless string with a given direction of propagation ${\bf n}_{(j)}$, $j = x,y,z$ may be treated as the Abelian sector of the corresponding representation (\ref{eq:52.07}) of the ${\rm AN}(2)$ group. In the context of this relation we should also consider hyperbolic Lorentz transformations i.e.\! boosts generated by $X^{(j)}_0 = J_{tj}$, which are given by
\begin{align}\label{eq:52.08}
g^{(j)}_0 = e^{i k_0 X^{(j)}_0}\,.
\end{align}
They form a one-dimensional group of boosts in the direction of the spatial component of ${\bf n}_{(j)}$, which are complementary with respect to null rotations generated by $X^{(j)}_1$, $X^{(j)}_2$. The action of $g^{(j)}_0$ on a holonomy $h^{(j)}_a$ has the form
\begin{align}\label{eq:52.09}
g^{(j)}_0(k_0)\, h^{(j)}_a(k_a)\, (g^{(j)}_0(k_0))^{-1} = 
h^{(j)}_a(e^{k_0} k_a)\,,
\end{align}
therefore $g^{(j)}_0$ rescales the parabolic angle $k_a$ by the factor of $e^{k_0}$ but the defect's spatial orientation is preserved. Notice that, as in the case of massive defects, a boost has to act by the conjugation to keep the holonomy in the parabolic conjugation class. 

The discussed relation between the momentum space of a massless cosmic string and the ${\rm AN}(2)$ group is actually stronger in the case of massless defects in 4+1-dimensional spacetime, whose holonomies are elements of the ${\rm AN}(3)$ group, corresponding to 3+1-dimensional $\kappa$-Minkowski space. In general, any conical defect can be completely specified by the deficit angle and two vectors which span a plane orthogonal to its world-volume \cite{Meent:2011ps}. On the other hand, by analogy with a cosmic string in 3+1 dimensions, we may choose to characterize a given massless defect in $n\!+\!1$ dimensions by the parabolic angle, i.e.\! its energy density, a null vector which determines the direction of motion and $n - 2$ spacelike vectors which span the spatial hyperplane. Since all these vectors are mutually orthogonal and their scaling is irrelevant we are left with $n (n - 1)/2 + 1$ parameters that should be carried by the defect's holonomy. If we now restrict the space of holonomies by fixing the direction of motion, as we did for cosmic strings, then the number of free parameters is reduced by $n - 1$. In particular, in the case of $n = 4$ (i.e.\! 4+1 spacetime dimensions) a restricted defect is characterized by 4 parameters. Nevertheless, in 4+1 dimensions there are 3 null rotations analogous to (\ref{eq:52.05}) and hence 3 coefficients $k_a$. Consequently, a complementary boost analogous to (\ref{eq:52.08}) becomes necessary in order to describe the defect and the role of the missing parameter is filled by the coefficient $k_0$. Let us also note that for $n = 2$ with the fixed direction of motion the situation is the same as in the 3+1-dimensional case, while for $n > 4$ the number of ${\rm AN}(n-1)$ generators is too small to characterize such a restricted defect. 

To be specific, in our parametrization the holonomy of a massless cosmic brane with a given direction of motion ${\bf n}$ can be written as
\begin{align}\label{eq:52.10}
h\lb k_0,\{k_j\}\rb = 
g_0(k_0)\, h_1(k_1)\, h_2(k_2)\, h_3(k_3)\, (g_0(k_0))^{-1} = \nonumber\\ 
h_1(e^{k_0} k_1)\, h_2(e^{k_0} k_2)\, h_3(e^{k_0} k_3)\,,
\end{align}
where $h_a$, $a = 1,2,3$ and $g_0$ are five-dimensional counterparts of (\ref{eq:52.05}), (\ref{eq:52.08}). After some calculations we can show that the brane's energy density is $\varrho = e^{k_0} \sqrt{k_1^2 + k_2^2 + k_3^2}$, while its spatial plane is spanned by two orthogonal vectors which can be put in the form ${\bf d}_I = n\, {\bf d}_{(1)} + n^\prime {\bf d}_{(2)} + {\bf d}_{(3)}$, ${\bf d}_{II} = n^{\prime\prime} {\bf d}_{(1)} - (1 + n n^{\prime\prime})/n^\prime {\bf d}_{(2)} + {\bf d}_{(3)}$, where the vectors ${\bf d}_{(a)}$, $a = 1,2,3$ form an orthonormal set and the coefficients $n = (k_2 - k_3)/(k_1 - k_2)$, $n^\prime = (k_3 - k_1)/(k_1 - k_2)$, $n^{\prime\prime} = (k_2 (k_2 - k_1) + k_3 (k_3 - k_1))/(k_1 (k_1 - k_3) + k_2 (k_2 - k_3))$. Thus the four parameters $n,n^\prime,n^{\prime\prime}$ and $\varrho$ that characterize the geometrical properties of the brane are expressed in terms of the ${\rm AN}(3)$ group parameters $k_1,k_2,k_3$ and $k_0$.

\section{Defects in de Sitter space}
In this Section we will briefly discuss conical defects in spacetime with positive cosmological constant and particularly the derivation of a lightlike defect. For 2+1-dimensional de Sitter and anti-de Sitter spaces massive point particle solutions were already found in \cite{Deser:1984tr}, while massless ones were obtained more recently \cite{Cai:1999ns}. Here let us consider \cite{Arzano:2015se} the case of 3+1-dimensional de Sitter space, in which the metric of a single static defect in static de Sitter coordinates is given by \cite{Linet:1986tt,Mello:2009ve}
\begin{align}\label{eq:53.01}
ds^2 = -(1 - \lambda r^2) d\tau^2 + (1 - \lambda r^2)^{-1} dr^2 + 
r^2 \lb d\theta^2 + (1 - 4\mu)^2 \sin^2\theta d\phi^2\rb\,,
\end{align}
where we denote $\lambda \equiv \Lambda/3$, and describes a cosmic string with the deficit angle $8\pi\mu$ and rest energy density $\mu$ (although it is rather not a global solution \cite{Linet:1986tt}). Similarly to the ordinary de Sitter metric we may perform a transformation of (\ref{eq:53.01}) to embedding Cartesian coordinates in 4+1-dimensional Minkowski space
\begin{align}\label{eq:53.02}
t = \sqrt{\lambda^{-1} - r^2}\, \sinh\lb\sqrt{\lambda} \tau\rb\,, \qquad 
w = \pm\sqrt{\lambda^{-1} - r^2}\, \cosh\lb\sqrt{\lambda} \tau\rb\,, \nonumber\\ 
x = r \cos\phi \sin\theta\,, \qquad y = r \sin\phi \sin\theta\,, \qquad 
z = r \cos\theta\,,
\end{align}
which satisfy the hyperboloid constraint $-t^2 + x^2 + y^2 + z^2 + w^2 = \lambda^{-1}$. The result turns out to be identical to the metric (\ref{eq:51.01}) of a massive defect in 4+1-dimensional Minkowski space but with the imposed hyperboloid condition. However, since the embedded de Sitter space has one dimension less, in this case $\mu$ represents the linear energy density instead of the surface density. This equivalence of conical metrics implies that also a massless defect solution (\ref{eq:52.02}) from the previous Section corresponds to a de Sitter case. Nevertheless, it may be more instructive to generalize the derivation of the metric of a massless defect in de Sitter space from \cite{Cai:1999ns} and see whether in the end we will obtain the same metric as (\ref{eq:52.02}). 

For convenience we begin by rescaling the radial coordinate in the metric (\ref{eq:53.01}) to $r \rightarrow r/(1 - 4\mu)$ and expanding the latter to the first order in $\mu$, which gives
\begin{align}\label{eq:53.03}
ds^2 \approx ds_{dS}^2 + 8\mu \lb\lambda r^2 d\tau^2 + 
(1 - \lambda r^2)^{-2} dr^2 + r^2 d\theta^2\rb\,,
\end{align}
where the pure de Sitter metric
\begin{align}\label{eq:53.04}
ds_{dS}^2 = -(1 - \lambda r^2) d\tau^2 + (1 - \lambda r^2)^{-1} dr^2 + 
r^2 (d\theta^2 + \sin^2\theta d\phi^2)\,.
\end{align}
Then the idea \cite{Hotta:1993st} is to transform (\ref{eq:53.03}) to embedding coordinates (\ref{eq:53.02}) and perform a boost $t \mapsto t \cosh\chi + x \sinh\chi$, $x \mapsto x \cosh\chi + t \sinh\chi$, obtaining
\begin{align}\label{eq:53.05}
ds^2 = ds_{dS}^2 + \nonumber\\ 
\frac{8\mu}{(T^2 - w^2)^2} \lb (\lambda^{-1} + T^2 - w^2) (w dT - T dw)^2 + \frac{\lambda^{-2} (T dT - w dw)^2}{\lambda^{-1} + T^2 - w^2}\rb + \nonumber\\ 
8\mu\, \frac{\lb z (-T dT + z dz + w dw) + 
(\lambda^{-1} + T^2 - z^2 - w^2) dz\rb^2} {(\lambda^{-1} + T^2 -w^2)
(\lambda^{-1} + T^2 - z^2 - w^2)}\,,
\end{align}
where we denote $T \equiv t \cosh\chi - x \sinh\chi$ and $ds_{dS}^2$ is now expressed in terms of embedding coordinates. We also introduce lightcone coordinates $u = (x - t)/\sqrt{2}$, $v = (x + t)/\sqrt{2}$. Finally, we employ the Aichelburg-Sexl boost, taking the limit $\chi \rightarrow \infty$ at the constant laboratory energy density $\varrho = 8\pi\mu \cosh\chi$. Using the distributional identity
\begin{align}\label{eq:53.06}
\lim_{\chi \rightarrow \infty} f(T^2) \cosh\chi = 
\frac{\delta(u)}{\sqrt{2}} \int_{-\infty}^{+\infty}\! dT\ f(T^2)\,,
\end{align}
we find that (\ref{eq:53.05}) becomes
\begin{align}\label{eq:53.07}
ds^2 = 2 du dv + dy^2 + dz^2 + dw^2 - \sqrt{2}\varrho |y| \delta(u) du^2\,.
\end{align}
Thus the obtained metric is identical to (\ref{eq:52.02}) but the hyperboloid condition means that it now describes a null, circular string on a meridian of the cosmological horizon in de Sitter space, with the curvature singularity at the hypersurface $u,y = 0$, $z^2 + w^2 = \lambda^{-1}$.\footnote{The sphere $x - t = 0$, $y^2 + z^2 + w^2 = \lambda^{-1}$ is the past horizon for an observer on the worldline $y,z,w = 0$, $x = x(t) > \lambda^{-1}$.} The situation seems to be qualitatively different from the case of 2+1 dimensions \cite{Cai:1999ns}, where due to the closed geometry of de Sitter space one obtains a pair of null particles at the opposite points of the horizon (which is then a circle), analogously to the massive particle solution \cite{Deser:1984tr}. However, from the geometric perspective this is naturally the dimensional reduction of a circle to a pair of points. 

Perhaps the above solution is better illustrated in the different coordinates, which can be defined in analogy with the 2+1-dimensional case \cite{Cai:1999ns}. To this end we first transform (\ref{eq:53.07}) using
\begin{align}\label{eq:53.08}
t = \frac{1}{2\eta} \lb\lambda^{-1} - \eta^2 + 
\lb X + \sqrt{\lambda}^{-1}\rb^2 + Y^2 + Z^2\rb\,, \quad 
x = \frac{1}{\sqrt{\lambda} \eta} \lb X + \sqrt{\lambda}^{-1}\rb\,, \nonumber\\ 
y = \frac{1}{\sqrt{\lambda} \eta} Y\,, \quad 
z = \frac{1}{\sqrt{\lambda} \eta} Z\,, \quad 
w = \frac{1}{2\eta} \lb\lambda^{-1} + \eta^2 - 
\lb X + \sqrt{\lambda}^{-1}\rb^2 - Y^2 - Z^2\rb
\end{align}
and perform another transformation to spherical coordinates via $X = \rho \cos\phi \sin\theta$, $Y = \rho \sin\phi \sin\theta$, $Z = \rho \cos\theta$, obtaining the metric in the form
\begin{align}\label{eq:53.09}
ds^2 = \frac{1}{\lambda \eta^2} \lb -d\eta^2 + d\rho^2 + 
\rho^2 (d\theta^2 + \sin^2\theta d\phi^2)\rb - \nonumber\\ 
\frac{\varrho}{\sqrt{\lambda}} |\sin\phi \sin\theta| \lb\delta(\eta - \rho) 
(d\eta - d\rho)^2 + \delta(\eta + \rho) (d\eta + d\rho)^2\rb\,.
\end{align}
The first line is actually the pure de Sitter metric, while the second one describes a lightlike string at the great circle $\phi \in \{0, \pi\}$, $\theta \in [0,\pi]$ of the cosmological horizon $\rho = |\eta|$. In particular, the first term with the delta function corresponds to the string at times $\eta \geq 0$ and the other at times $\eta \leq 0$. 

As we shown above, the form of the metric (\ref{eq:53.07}) is identical to the metric (\ref{eq:52.02}) of a massless cosmic brane in 4+1-dimensional Minkowski space. The derivation of (\ref{eq:53.07}) can in principle be generalized to any number of dimensions. Thus, roughly speaking, a $n\!-\!1$-dimensional massless conical defect in $n\!+\!1$-dimensional de Sitter space can be seen as a geometrical projection of a $n$-dimensional defect in $(n\!+\!1)\!+\!1$-dimensional embedding Minkowski space. This can be easily visualized. Namely, the $n\!-\!2$-dimensional sphere of a defect in de Sitter space is determined by the $n\!-\!1$-dimensional hyperplane of a defect in Minkowski space, which is cutting through the $n$-dimensional sphere of a de Sitter's spatial slice. We conjecture that massless defects in de Sitter space can be indirectly characterized by holonomies of their embedding defects. In particular, we presume that massless strings in 3+1-dimensional de Sitter space can be specified by holonomies belonging to the ${\rm AN}(3)$ group but this remains to be verified.

\chapter{Fock space of particles in 3d gravity}
In the quantum field theory we consider states of arbitrarily many identical particles, which are built starting from the one-particle Hilbert space and form the full Fock space. To construct such multiparticle states, which are covariant under the underlying symmetry algebra of a given theory, we need to know how to exchange algebra representations describing individual particles. As we discussed in Section 1.1, for a system equipped with a nontrivial coalgebra this is not always possible but rather requires the existence of a quasitriangular bialgebra structure, with a universal $R$-matrix (\ref{eq:11.17}). In particular, for the $\kappa$-Poincar\'{e} algebra the $R$-matrix has been found only in the approximate form \cite{Young:2009oe,Young:2010ts} and its actual existence remains an open problem, although a lot of work on the $\kappa$-deformed Fock space has been done \cite{Arzano:2007fs,Daszkiewicz:2008dw,Govindarajan:2008te,Arzano:2009rs}. 

Meanwhile, for the quantized theory of particles coupled to three-dimensional gravity the associated $R$-matrix is known and thus we may try \cite{Arzano:2014by} to construct the corresponding Fock space. Let us first observe that the Hilbert space of an ordinary relativistic particle is given by the space of (square-integrable) complex functions on the particle's mass shell in momentum space, which is an orbit of the Lorentz group in Minkowski (momentum) space. This Hilbert space carries an irreducible representation of the Poincar\'{e} group i.e.\! the isometry group of spacetime. As we discussed in Chapter 3, for a particle coupled to three-dimensional gravity the momentum, i.e.\! holonomy, becomes an element of the conjugation class of a rotation by a deficit angle $\delta = 8\pi Gm$, which is an orbit of the Lorentz group acting on itself. Therefore a one-particle Hilbert space of the quantum field theory of gravitating (spinless) particles should be the space of complex functions on such a conjugation class, together with the scalar product
\begin{align}\label{eq:60.01}
(\psi,\chi) = \int\! d\mu(g)\ \overline{\psi(g)}\, \chi(g)\,,
\end{align}
where $d\mu(g)$ is the invariant measure on the conjugation class, induced by the Haar measure on the full group. Indeed, it can be shown \cite{Hooft:1988ny,Carlip:1989ey} that this is the case. Furthermore \cite{Bais:1998tu,Bais:2002qy}, the above Hilbert space carries an irreducible representation of the quantum double of the three-dimensional Lorentz group, also known as the Lorentz double (see \cite{Koornwinder:1997tp}). The latter is a quasitriangular Hopf algebra that can be seen as a deformation of the group algebra of the Poincar\'{e} group, cf.\! Section 1.1. Thus, roughly speaking, gravity is deforming a usual group of relativistic symmetries of a particle into a quantum group. 

Let us denote one-particle states as kets $\left|g\right>$ labelled by an element $g$ of a given conjugation class (characterized by mass $m$). Then the scalar product (\ref{eq:60.01}) becomes
\begin{align}\label{eq:60.02}
\left<h\mid g\right> = \delta(h^{-1} g)\,.
\end{align}
Notice that the delta function on the group enforces the identity $h^{-1} g = e$, where $e$ is the group's unit element, and hence it gives $h = g$. The Lorentz double determines \cite{Bais:1998tu} the action of Lorentz transformations and translations on one-particle states $\left|g\right>$ (which form its irreducible representations). The natural actions of the Lorentz group on itself are given by the left and right conjugation, $g \triangleright h = g h g^{-1}$ and $h \triangleleft g = g^{-1} h g$. Accordingly, the left and right action of Lorentz transformations on a ket have the form
\begin{align}\label{eq:60.03}
\Lambda_L(g) \left|h\right> \equiv \left|g \triangleright h\right> = 
\left|g h g^{-1}\right>\,, \qquad 
\Lambda_R(g) \left|h\right> \equiv \left|h \triangleleft g\right> = 
\left|g^{-1} h g\right>\,.
\end{align}
The two actions are obviously related by the group inversion, i.e.\! $(.) \triangleright g \equiv g^{-1} \triangleleft (.)$ and we could restrict to just one of them. However, for multiparticle states this choice may have a meaning, as we will see below. Meanwhile, for the standard parametrization of the Lorentz group the action of translations on a ket has the usual form
\begin{align}\label{eq:60.04}
T({\bf x}) \triangleright \left|h\right> \equiv 
e^{i x_\mu p^\mu(h)} \left|h\right>\,,
\end{align}
with the coordinates $x^\mu$, $\mu = 0,1,2$. The generators of translations are naturally interpreted as momentum operators, which act as
\begin{align}\label{eq:60.05}
P^\mu \triangleright \left|h\right> = p^\mu(h) \left|h\right>
\end{align}
and have the eigenvalues $p^\mu(h)$. As we already know from Chapter 1, there is no unique choice of such a basis of the algebra generators. This apparently leads to the ambiguities in the properties of momenta \cite{Arzano:2014gy}. 

Let us turn to the two-particle states. The total momentum of a tensor product of two kets is determined by the coproduct of the translation generators $P^\mu$ in the Lorentz double and gives
\begin{align}\label{eq:60.06}
\Delta P^\mu \triangleright \lb\left|h_1\right> \otimes \left|h_2\right>\rb = 
p^\mu(h_1 h_2) \left|h_1\right> \otimes \left|h_2\right>\,.
\end{align}
However, since the space of holonomies is non-Abelian a two-particle state $\left|h_1\right> \otimes \left|h_2\right>$ and its flipped counterpart $\left|h_2\right> \otimes \left|h_1\right>$ carry different momenta $p^\mu(h_1 h_2) \neq p^\mu(h_2 h_1)$. Thus a usual (bosonic) symmetrized state $\frac{1}{\sqrt{2}}\, \lb\left|h_1\right> \otimes \left|h_2\right> + \left|h_2\right> \otimes \left|h_1\right>\rb$ is not an eigenstate of $P^\mu$. If we visualize the flip of a tensor product as a straightforward exchange of the particles' positions we can observe that indeed it alters the holonomy of a loop going around both particles, which describes their total momentum. Therefore the correct prescription for the exchange of particles \cite{Carlip:1989ey} is that one particle is moved around the other and its holonomy becomes conjugated by the holonomy of the other particle. This is a (nontrivial) braiding of holonomies that we mentioned in Section 3.3 in the classical theory and which appears naturally for topological defects coupled to gauge theories in 2+1 spacetime dimensions \cite{Bais:1980fs,Lo:1993ns}. Depending on the direction of the particle exchange it leads to two types of the braided symmetrization, which correspond to a left-symmetrized state
\begin{align}\label{eq:60.07}
\left|h_1, h_2\right>_L \equiv 
\frac{1}{\sqrt{2}}\, \lb\left|h_1\right> \otimes \left|h_2\right> + 
\left|h_1 h_2 h_1^{-1}\right> \otimes \left|h_1\right>\rb
\end{align}
and a right-symmetrized state
\begin{align}\label{eq:60.08}
\left|h_1, h_2\right>_R \equiv 
\frac{1}{\sqrt{2}}\, \lb\left|h_1\right> \otimes \left|h_2\right> + 
\left|h_2\right> \otimes \left|h_2^{-1} h_1 h_2\right>\rb\,.
\end{align}
Both states have the same, well-defined momentum $p^\mu(h_1 h_2)$. 

Let us also check the result of the action of Lorentz transformations on them. The coproduct of the Lorentz sector of the Lorentz double gives
\begin{align}\label{eq:60.09}
\Delta \Lambda_{L,R}(g) \lb\left|h_1\right> \otimes \left|h_2\right>\rb = 
\Lambda_{L,R}(g) \left|h_1\right> \otimes \Lambda_{L,R}(g) \left|h_2\right>\,.
\end{align}
It follows that both left- and right-symmetrized states are Lorentz covariant, to wit
\begin{align}\label{eq:60.10}
\Delta \Lambda_L(g) \left|h_1, h_2\right>_L & = 
\left|g h_1 g^{-1}, g h_2 g^{-1}\right>_L\,, \nonumber\\ 
\Delta \Lambda_R(g) \left|h_1, h_2\right>_{R} & = 
\left|g^{-1} h_1 g, g^{-1} h_2 g\right>_R\,.
\end{align}
Thus the braided symmetrization of a two-particle state is consistent with the underlying gauge symmetry of the theory. This is obviously a consequence of the fact that the Lorentz double is equipped with the universal $R$-matrix, which determines a deformation of the flip operator $\sigma: \left|h_1\right> \otimes \left|h_2\right> \rightarrow \left|h_2\right> \otimes \left|h_1\right>$, describing the particle exchange. Accordingly, let us introduce \cite{Bais:1998tu,Arzano:2014by} the operational formula for the $R$-matrix associated with the left action of Lorentz transformations
\begin{align}\label{eq:60.11}
R \equiv \sum_{g \in G} \delta(g^{-1}.) {\bf 1} \otimes \Lambda_L(g)\,.
\end{align}
Similarly we may write the universal $R$-element for the right action, which we call the $R^\prime$-matrix,
\begin{align}\label{eq:60.12}
R^\prime \equiv \sum_{g\in G} \Lambda_R(g) \otimes \delta(g^{-1}.) {\bf 1}\,.
\end{align}
Then left- and right-symmetrized two-particle states (\ref{eq:60.07}), (\ref{eq:60.08}) are defined as
\begin{align}\label{eq:60.13}
\left|h_1, h_2\right>_L & \equiv 
\frac{1}{\sqrt{2}}\, \lb{\bf 1} \otimes {\bf 1} + \sigma \circ R\rb 
\left|h_1\right> \otimes \left|h_2\right>\,, \nonumber\\ 
\left|h_1, h_2\right>_R & \equiv 
\frac{1}{\sqrt{2}}\, \lb{\bf 1} \otimes {\bf 1} + \sigma \circ R^\prime\rb 
\left|h_1\right> \otimes \left|h_2\right>\,,
\end{align}
where $\tau_L \equiv \sigma \circ R$, $\tau_R \equiv \sigma \circ R^\prime$ are the braided flip operators. $\tau_L$, $\tau_R$ can be straightforwardly generalized to the case of arbitrarily many particles. Namely, instead of a usual exchange operator
\begin{align}\label{eq:60.14}
\sigma(i) \left|h_1, \ldots, h_i, h_{i+1}, \ldots, h_n\right> = 
\left|h_1, \ldots, h_{i+1}, h_i, \ldots, h_n\right>
\end{align}
we have
\begin{align}\label{eq:60.15}
\tau_L(i) \left|h_1, \ldots, h_i, h_{i+1}, \ldots, h_n\right> & = 
\left|h_1, \ldots, h_i h_{i+1} h^{-1}_i, h_i, \ldots, h_n\right>\,, \nonumber\\ 
\tau_R(i) \left|h_1, \ldots, h_i, h_{i+1}, \ldots, h_n\right> & = 
\left|h_1, \ldots, h_{i+1}, h^{-1}_{i+1} h_i h_{i+1}, \ldots, h_n\right>\,.
\end{align}
In contrast to $\sigma(i)$, such exchange operators do not square to the identity and hence do not belong to a representation of the $n$-dimensional symmetric group (i.e.\! the group of permutations). However, they still satisfy the properties
\begin{align}\label{eq:60.16}
\tau_{L,R}(i)\, \tau_{L,R}(j) & = 
\tau_{L,R}(j)\, \tau_{L,R}(i)\,,\ \ |i-j| \geq 2 \,, \nonumber\\ 
\tau_{L,R}(i)\, \tau_{L,R}(i+1)\, \tau_{L,R}(i) & = 
\tau_{L,R}(i+1)\, \tau_{L,R}(i)\, \tau_{L,R}(i+1)\,,
\end{align}
which define the $n$-dimensional braid group. Using $\tau_{L,R}(i)$ a left- or right-symmetrized $n$-particle state can be built iteratively as \cite{Arzano:2014by}
\begin{align}\label{eq:60.17}
\left|h_{1}, \ldots, h_{n}\right>_{L,R} = \mathcal{S}_{L,R}(n) 
\lb\left|h_{1}\right> \otimes \left|h_{2}, \ldots, h_{n}\right>_{L,R}\rb\,, \nonumber\\ 
\mathcal{S}_{L,R}(n) \equiv \frac{1}{\sqrt{n}} \lb{\bf 1}^{\otimes n} + 
\sum_{i=1}^{n-1} \lb\tau_{L,R}(i) \circ \ldots \circ \tau_{L,R}(1)\rb\rb\,.
\end{align}
The action of Lorentz transformations on a $n$-particle state, $n > 2$ is determined via the coassociativity of the coproduct and it can be easily verified that the states (\ref{eq:60.17}) are Lorentz covariant analogously to (\ref{eq:60.10}). Meanwhile, one may note that acting repeatedly with $\tau_{L,R}(i)$'s on such a multiparticle state we apparently obtain new states with the same particle content. For example, the two-particle state given by
\begin{align}\label{eq:60.18}
\tau_L \left|h_1, h_2\right>_L = \frac{1}{\sqrt{2}} 
\lb\left|h_1 h_2 h_1^{-1}\right> \otimes \left|h_1\right> + 
\left|h_1 h_2 h_1 h_2^{-1} h_1^{-1}\right> \otimes 
\left|h_1 h_2 h_1^{-1}\right>\rb
\end{align}
is a momentum eigenstate with the same eigenvalue as the state (\ref{eq:60.07}). However, it can also be obtained by a Lorentz transformation of a state of the form (\ref{eq:60.07}), to wit
\begin{align}\label{eq:60.19}
\Delta\Lambda_L(h_1) \left|h_2, h_1\right>_L = \tau_L \left|h_1, h_2\right>_L\,.
\end{align}
For higher powers of $\tau_L$ the situation is similar. It remains to be verified whether also arbitrary multiparticle states obtained in such a way as (\ref{eq:60.18}) correspond to Lorentz transformations of the presumed generic states (\ref{eq:60.17}), perhaps mixing the left- and right-symmetrization. Let us note left- and right-symmetrized two-particle states are connected by Lorentz transformations, i.e.\! they satisfy
\begin{align}\label{eq:60.20}
\Delta \Lambda_L(h_1^{-1}) \left|h_1, h_2\right>_L = \left|h_2, h_1\right>_R\,, \qquad 
\Delta \Lambda_R(h_2^{-1}) \left|h_1, h_2\right>_R = \left|h_2, h_1\right>_L\,.
\end{align}
These relations do not straightforwardly generalize to other multiparticle states. 

We now define \cite{Arzano:2014by} the creation operators that have the left action on kets and can be used to built the ladder of left-symmetrized multiparticle states (\ref{eq:60.17}) starting from the vacuum state $\left|0\right>$. Namely, we assume that
\begin{align}\label{eq:60.21}
a_L^\dagger(h) \left|0\right> = \left|h\right>\,, \qquad 
\frac{1}{\sqrt{n}}\, a_L^\dagger(h_1) \left|h_2, \ldots, h_n\right>_L = 
\left|h_1, \ldots, h_n\right>_L\,.
\end{align}
We also introduce the corresponding annihilation operators such that
\begin{align}\label{eq:60.22}
a_L(h) \left|0\right> = 0\,, \qquad 
a_L(h) \left|h_1, \ldots, h_n\right> = 
\delta(h^{-1} h_1) \left|h_2, \ldots, h_n\right>
\end{align}
(where states are not symmetrized). In order to derive the commutation relation between creation and annihilation operators we may use the fact that an arbitrary left-symmetrized $n$-particle state can be written in the form
\begin{align}\label{eq:60.23}
\left|h_1, \ldots, h_n\right>_L = \nonumber\\ 
\sqrt{\frac{(n-1)!}{n!}} \sum_{i=1}^n 
\left|h_1 \ldots h_{i-1} h_i h_{i-1}^{-1} \ldots h_1^{-1}\right> \otimes 
\left|h_1, \ldots, h_{i-1}, \check{h_i}, h_{i+1}, \ldots, h_n\right>_L\,,
\end{align}
where $\check{h_i}$ denotes an absent group element. Then we find that
\begin{align}\label{eq:60.24}
a_L(g)\, a_L^\dagger(h_1) \left|h_2, \ldots, h_n\right>_L = 
\delta(g^{-1} h_1) \left|h_2, \ldots, h_n\right>_L + \nonumber\\ 
\sum_{i=2}^n \delta(g^{-1} h_1 \ldots h_{i-1} h_i h_{i-1}^{-1} \ldots h_1^{-1}) 
\left|h_1, \ldots, h_{i-1}, \check{h_i}, h_{i+1}, \ldots, h_n\right>_L\,.
\end{align}
On the other hand, we have
\begin{align}\label{eq:60.25}
a_L^\dagger(h_1)\, a_L(h_1^{-1} g h_1) 
\left|h_2, \ldots, h_n\right>_L = \nonumber\\ 
\sum_{i=2}^n \delta(g^{-1} h_1 \ldots h_{i-1} h_i h_{i-1}^{-1} \ldots h_1^{-1}) 
\left|h_1, \ldots, h_{i-1}, \check{h_i}, h_{i+1}, \ldots, h_n\right>_L\,.
\end{align}
Taking the difference between (\ref{eq:60.24}) and (\ref{eq:60.25}) we obtain the commutation relation
\begin{align}\label{eq:60.26}
a_L(h_1)\, a_L^\dagger(h_2) - a_L^\dagger(h_2)\, a_L(h_2^{-1} h_1 h_2) = 
\delta(h_1^{-1} h_2)\,,
\end{align}
which is the braided counterpart of the usual relation.

Similarly, for the right-symmetrized states (\ref{eq:60.17}) we define the creation operators with the right action, to wit
\begin{align}\label{eq:60.27}
a_R^\dagger(h) \left|0\right> = \left|h\right>\,, \qquad 
\frac{1}{\sqrt{n}}\, a_R^\dagger(h_n) \left|h_1, \ldots, h_{n-1}\right>_R = 
\left|h_1, \ldots, h_n\right>_R
\end{align}
as well as the corresponding annihilation operators
\begin{align}\label{eq:60.28}
a_R(h) \left|0\right> = 0\,, \qquad 
a_R(h) \left|h_1, \ldots, h_n\right> = 
\delta(h^{-1} h_n) \left|h_1, \ldots, h_{n-1}\right>\,.
\end{align}
An arbitrary right-symmetrized $n$-particle state can be written as
\begin{align}\label{eq:60.29}
\left|h_1, \ldots, h_n\right>_R = \nonumber\\ 
\sqrt{\frac{(n-1)!}{n!}} \sum_{i=1}^n 
\left|h_1, \ldots, h_{i-1}, \check{h_i}, h_{i+1}, \ldots, h_n\right>_R \otimes 
\left|h_n^{-1} \ldots h_{i-1}^{-1} h_i h_{i-1} \ldots h_n\right>\,.
\end{align}
Then we find that
\begin{align}\label{eq:60.30}
a_R(g)\, a_R^\dagger(h_n) \left|h_1, \ldots, h_{n-1}\right>_R = 
\delta(g^{-1} h_n) \left|h_1, \ldots, h_{n-1}\right>_R + \nonumber\\ 
\sum_{i=1}^{n-1} \delta(g^{-1} h_n^{-1} \ldots h_{i+1}^{-1} h_i h_{i+1} \ldots 
h_1) \left|h_1, \ldots, h_{i-1}, \check{h_i}, h_{i+1}, \ldots, h_n\right>_R
\end{align}
and
\begin{align}\label{eq:60.31}
a_R^\dagger(h_n)\, a_R(h_n g h_n^{-1}) 
\left|h_1, \ldots, h_{n-1}\right>_R = \nonumber\\ 
\sum_{i=1}^{n-1} \delta(g^{-1} h_n^{-1} \ldots h_{i+1}^{-1} h_i h_{i+1} \ldots 
h_1) \left|h_1, \ldots, h_{i-1}, \check{h_i}, h_{i+1}, \ldots, h_n\right>_R\,.
\end{align}
The difference between (\ref{eq:60.30}) and (\ref{eq:60.31}) gives the braided commutation relation
\begin{align}\label{eq:60.32}
a_R(h_1)\, a_R^\dagger(h_2) - a_R^\dagger(h_1^{-1} h_2 h_1)\, a_R(h_1) = 
\delta(h_1^{-1} h_2)\,.
\end{align}
It remains to to be verified whether the relations (\ref{eq:60.26}), (\ref{eq:60.32}) can be extended to the consistent algebras of creation and annihilation operators.

\chapter*{Conclusions}
\addcontentsline{toc}{chapter}{Conclusions}
\markboth{CONCLUSIONS}{CONCLUSIONS}
Our work on the spectral dimension of $\kappa$-Minkowski space shows that the desired dimensional reduction occurs for only one of the considered Laplacians on momentum space. From this perspective the latter is probably the most interesting option for the physical Laplacian. Meanwhile, only in the case of 2+1 topological dimensions we obtain the same ultraviolet value of the dimension as in some other approaches to quantum gravity. This is another indication that the $\kappa$-Poincar\'{e} algebra is potentially most useful in the context of quantum gravity in three dimensions. 

The model of $\kappa$-deformed Carroll particles that we derived is one more example of the $\kappa$-deformed (not necessarily $\kappa$-Poincar\'{e}) symmetry arising in the broad framework of three-dimensional gravity. However, we have to understand the physical meaning of the applied contraction of the Chern-Simons theory with the de Sitter gauge group and its relation with the known results. Furthermore, the properties of a multiparticle system remain to be explored. What is interesting is that we obtained a connection with the Carrollian physics regime, which is currently becoming a promising area of research. Thus it would be useful to study the Carrollian limit in the general context of three-dimensional gravity and BF theory. 

Our discussion of massive conical defects in higher dimensional gravity did not lead to any novel observations since such a generalization is completely straightforward. On the other hand, for massless defects we presented the correspondence between the holonomies of defects and (the Abelian sector of) a given ${\rm AN}(n)$ group. This relation may seem superficial but perhaps it is a symptom of the underlying nature of massless defects, which could be uncovered in their quantization or the BF theory formulation. We should also try to explore the holonomies of defects in de Sitter space. 

Finally, our attempt at the Fock space formulation of the quantum field theory coupled to three-dimensional gravity is indeed preliminary and inconclusive. Nevertheless, it illustrates some possible modifications of the quantum theory arising in the presence of (topological) gravitational interactions, which should generalize to higher dimensions. It may also help in constructing the Fock space for theories with the structure of deformed symmetries.

\newpage
\thispagestyle{empty}
~

\chapter*{Author's papers}
\addcontentsline{toc}{chapter}{Author's papers}
\Large Thesis-related:\normalsize
\begin{enumerate}
\item M.~Arzano and T.T., {\it Spacetime defects and group momentum space}, [arXiv:\break 1412.8452 [hep-th]]
\item J.~Kowalski-Glikman and T.T., {\it Deformed Carroll particle from 2+1 gravity}, Phys.\ Lett.\ B {\bf 737}, 267 (2014) [arXiv:1408.0154 [hep-th]]
\item M.~Arzano and T.T., {\it Diffusion on $\kappa$-Minkowski space}, Phys.\ Rev.\ D {\bf 89}, 124024 (2014) [arXiv:1404.4762 [hep-th]]
\item M.~Arzano, J.~Kowalski-Glikman and T.T., {\it Beyond Fock space in three dimensional semiclassical gravity}, Class.\ Quant.\ Grav.\ {\bf 31}, 035013 (2014) [arXiv:1305.\break 6220 [hep-th]]
\end{enumerate}
\Large Other:\normalsize
\begin{enumerate}
\item J.~Ambj\o rn, A.~G\"{o}rlich, J.~Jurkiewicz, R.~Loll, J.~Gizbert-Studnicki and T.T., \textit{The Semiclassical Limit of Causal Dynamical Triangulations},  Nucl.\ Phys.\ B {\bf 849}, 144 (2011) [arXiv:1102.3929 [hep-th]]
\item T.T., B.~Czerny, V.~Karas, T.~Pech\'{a}\v{c}ek, M.~Dov\v{c}iak, R.~Goosmann and M. Niko\l ajuk, \textit{The Flare Model for X-ray Variability of NGC~4258}, A\&A {\bf 530}, A136 (2011) [arXiv:1104.4181 [astro-ph.HE]]
\item T.T., \textit{Analysis of the Semiclassical Solution of CDT}, [arXiv:1102.4643 [hep-th]]
\end{enumerate}

\newpage
\thispagestyle{empty}
~

\chapter*{Acknowledgements}
\addcontentsline{toc}{chapter}{Acknowledgements}
First and foremost, I would like to express the gratitude to the supervisor of my thesis, Prof. Jerzy Kowalski-Glikman. 

I also thank very much Dr. Michele Arzano, my tutor during the research visits at the Utrecht University and the Sapienza University of Rome. 

Moreover, I owe a lot to my colleagues from the research group: Micha\l , Remigiusz, Francesco and Giacomo, as well as many others from the Universities of Krak\'{o}w, Wroc\l aw, Warszawa, Utrecht and Rome. 

Finally, I am grateful to my family for all the support.\\ 

I acknowledge the support by the Foundation for Polish Science International PhD Projects Programme co-financed by the EU European Regional Development Fund.

\newpage
\thispagestyle{empty}
~

\end{document}